\documentclass[11pt,a4paper]{article}
\usepackage[margin=2.7cm]{geometry}
\usepackage{amsmath,amssymb,amsthm}
\usepackage{graphicx}
\usepackage[colorlinks,linkcolor=blue,citecolor=blue,urlcolor=blue]{hyperref}

\newcommand{\capindexscale}{0.8}
\DeclareFontFamily{OML}{cmmcap}{\skewchar\font127 }
\edef\next{\noexpand\DeclareFontShape{OML}{cmmcap}{m}{it}{%
  <-8.5>s*[\capindexscale]cmmi7<8.5-9.5>cmmi9<9.5-11.5>cmmi10<11.5->cmmi12}{}}\next
\DeclareSymbolFont{capindex}{OML}{cmmcap}{m}{it}
\makeatletter
\@tfor\cap@x:=ABCDEFGHIJKLMNOPQRSTUVWXYZ\do{%
  \mathcode\expandafter`\cap@x=\numexpr"7000+256*\symcapindex+\expandafter`\cap@x\relax}
\makeatother

\newtheorem{theorem}{Theorem}
\newtheorem{proposition}{Proposition}
\newtheorem{lemma}{Lemma}
\newtheorem{hypothesis}{Hypothesis}
\newcommand{\pr}{\partial}
\newcommand{\vev}[1]{\langle #1 \rangle}
\newcommand{\cO}{\mathcal{O}}
\newcommand{\cA}{\mathcal{A}}
\newcommand{\cL}{\mathcal{L}}
\newcommand{\vep}{\varepsilon}

\title{An all-orders proof of gradient flow for multiscalar $\phi^4$ theory in four dimensions}
\author{Marten Reehorst$^{1,2}$\\[6pt]
{\small $^1$\,\itshape Department of Mathematics, City St.\ George's, University of London,}\\
{\small\itshape Northampton Square, EC1V 0HB, London, UK}\\[4pt]
{\small $^2$\,\itshape Department of Mathematics, King's College London, Strand, London WC2R 2LS, UK}}
\date{\today}

\begin{document}
\maketitle

\begin{abstract}
We prove that the renormalisation group flow of multiscalar $\phi^4$ theory in four dimensions is a gradient flow to all orders in perturbation theory. The proof uses the Weyl consistency conditions of Osborn and Jack, and it assumes two properties of perturbation theory: that the generating functional with local couplings is finite (also assumed by Osborn and Jack), and that the known expansion of massive Feynman integrals at large momenta holds with a bound on its remainder at each momentum. The consistency conditions give a gradient formula in which the tensor that multiplies the beta function is allowed to have an antisymmetric part, determined by coefficients of the Weyl anomaly. We derive a Ward identity for the two-point function of the marginal operators at constant couplings away from fixed points, with sources for the currents of field rotations and for the mass operator. It re-expresses one of the anti-symmetric contributions in terms of two-point functions and the first moment of the three-point function. We show that, after this rewrite, each of the resulting terms is equivalent to a symmetric contribution to the metric and a contribution to the potential, and we give both explicitly. The metric is positive definite to all orders in perturbation theory. The proof implies in particular the four relations among five- and six-loop coefficients of the field rotation, which Pannell and Stergiou derived as the conditions for a gradient flow at six loops. We check by explicit computation that these relations indeed hold. A companion paper treats $d=4-\varepsilon$ and other critical dimensions with a Ward identity on the sphere.
\end{abstract}

\newpage
\tableofcontents

\section{Introduction}

A renormalisation group flow is a gradient flow if the beta function is the gradient of a function on the space of couplings with respect to a symmetric positive metric. In four dimensions the question whether this holds was formulated by Osborn with the help of local couplings~\cite{Osborn:1991gm}, following an analysis in dimensional regularisation by Jack and Osborn~\cite{Jack:1990eb}. The couplings are promoted to functions of the position, the theory is put on a curved background, and the response of the generating functional to a local Weyl rescaling defines a set of coefficient functions on the space of couplings. Two Weyl rescalings commute, which gives relations among these coefficients, the Weyl consistency conditions. Three ingredients which we use are already in~\cite{Osborn:1991gm}. The first is a gradient formula in which the tensor that multiplies the beta function is not known to be symmetric. The second is the treatment of the mass operator. The third is the change of the beta function by a rotation of the fields, which we describe below. They were revisited and extended by Jack and Osborn~\cite{Jack:2013sha}, who in particular treat the currents and the mass operator together. In this form, see also~\cite{Baume:2014rla,Keren-Zur:2014sva}, they imply, for a general renormalisable theory with marginal couplings $g^I$ and to all orders in perturbation theory,
\begin{equation}
\pr_IA = T_{IJ}\,B^J \, , \qquad T_{IJ} = G_{IJ}+T_{[IJ]} \, .
\label{eq:introT}
\end{equation}
Here $\pr_I=\pr/\pr g^I$ denotes the derivative with respect to the couplings, $T_{[IJ]}$ is the antisymmetric part of $T_{IJ}$,\footnote{Square and round brackets around a pair of indices denote the antisymmetric and the symmetric part, $X_{[IJ]}=\tfrac12(X_{IJ}-X_{JI})$ and $X_{(IJ)}=\tfrac12(X_{IJ}+X_{JI})$.} and $B^I$ is a particular choice of the beta function. With several fields the beta function is defined only up to an infinitesimal rotation of the fields, $\beta^I\to\beta^I-(\omega g)^I$, since couplings which differ by a rotation describe the same theory, and $B^I$ is a choice for which \eqref{eq:introT} holds. In a given scheme it is fixed by the renormalisation of the currents of these rotations, as we recall in section~\ref{sec:setup}. Further, $A$ is a function which reduces to the $a$-anomaly at fixed points, and $G_{IJ}$ is a symmetric tensor which is positive definite in perturbation theory~\cite{Jack:1990eb,Jack:2013sha}. The consistency conditions do not require the tensor $T_{IJ}$ to be symmetric. Contracting \eqref{eq:introT} with a second factor of $B^I$ gives $B^I\pr_IA=G_{IJ}B^IB^J$, because an antisymmetric tensor vanishes when it is contracted twice with the same vector. The function $A$ is therefore monotonic along the flow where $G_{IJ}$ is positive. This is weaker than a gradient flow, which requires a symmetric tensor in \eqref{eq:introT} itself. The contraction $T_{[IJ]}B^J$ with a single factor does not vanish in general, and the antisymmetric part of $T_{IJ}$ is the obstacle.

In this paper we show that, for the theory of $N$ real scalar fields with a general quartic interaction, this contraction is always equal to a symmetric tensor contracted with $B^J$ plus a gradient, which can be absorbed in the metric and in the potential. Our main result is the following. To all orders in perturbation theory there are a function $A'$ and a symmetric tensor $G'_{IJ}$ on the space of couplings with
\begin{equation}
\pr_IA' = G'_{IJ}\,B^J \, .
\label{eq:introG}
\end{equation}
Both are power series in the couplings, and $G'_{IJ}$ is positive definite to all orders in perturbation theory. By this we mean that its term of order zero is a positive definite matrix, so that the corrections of any finite order do not change the signature for sufficiently small couplings.\footnote{The metric then has an inverse as a power series, and \eqref{eq:introG} is equivalent to $B^I=(G'^{-1})^{IJ}\pr_JA'$, which is also sometimes used as the definition of a gradient flow. We write the relation in the form \eqref{eq:introG}, as in~\cite{Osborn:1991gm,Jack:2013sha}, because the consistency conditions give it in this form.} We give $A'$ and $G'_{IJ}$ explicitly in terms of the anomaly coefficients and of correlation functions in \eqref{eq:explicitfull}. The precise statement is Theorem~\ref{thm:main}.

The key ingredient is an identity for the antisymmetric part $T_{[IJ]}$, which we derive from a Ward identity for the two-point function $\vev{\cO_I\cO_J}$ of the marginal operators at constant couplings, away from the fixed point. The tensor $T_{[IJ]}$ is determined by the antisymmetric part of a coefficient $E_{IJ}$ of the Weyl anomaly, which multiplies the term $\pr_\mu\sigma\,\pr^\mu g^I\pr^2g^J$ in the variation of the generating functional.\footnote{This coefficient is called $S_{IJ}$ in~\cite{Jack:2013sha}. We use $S_{IJ}$ for a different, symmetric coefficient of the anomaly in section~\ref{sec:setup}.} We obtain the Ward identity from the conformal Ward identity of the generating functional with local couplings by two derivatives with respect to the couplings. Its part which is odd under the exchange of the two points relates $E_{[IJ]}$ to correlation functions. It implies that $E_{[IJ]}$ is a sum of three kinds of terms,
\begin{equation}
E_{[IJ]} = B^K\,\tau_{K;IJ}+\tfrac12\left(\hat\rho_I\cdot X_J-\hat\rho_J\cdot X_I\right)
+\pr_IB^K\,c_{KJ}-c_{IK}\,\pr_JB^K \, ,
\label{eq:introE}
\end{equation}
whose coefficients $c_{IJ}$, $\tau_{K;IJ}$ and $X_J$ are power series in the couplings. Here $\hat\rho_I$ is a coefficient function associated with the field rotations, which we define in \eqref{eq:Psi}.\footnote{The symbol $\rho_I$ without a hat denotes the corresponding coefficient before the redefinition of the beta function by the field rotation, see \eqref{eq:weylvar}.} For the full definition of the right-hand side see \eqref{eq:formscheme}.

We show that each of the three kinds of terms, contracted with $B^J$, is the sum of a symmetric tensor contracted with $B^J$ and a gradient. We then show that the same holds for $T_{[IJ]}B^J$, and \eqref{eq:introG} follows.

The theorem uses the consistency conditions of~\cite{Osborn:1991gm,Jack:2013sha}, in the form of~\cite{Jack:2013sha}, and it rests on two hypotheses, which we state before the theorem in section~\ref{sec:metric}. The first is that the generating functional with local couplings is finite at each order of perturbation theory, with a local anomaly. This is also assumed in~\cite{Osborn:1991gm,Jack:2013sha}, where the consistency conditions are derived from it, so that our result has the same status as these conditions. The second concerns the first moment of the three-point function of the marginal operators, which enters the Ward identity. We define it with a mass as infrared regulator and show in appendix~\ref{app:ir} that it has a massless limit. For this we use the expansion of renormalised Feynman integrals at momenta which are large compared to the mass~\cite{Smirnov:1990rz}. Our second hypothesis is that this expansion holds with a bound on its remainder at each non-zero momentum. In~\cite{Smirnov:1990rz} the bound is proved in detail for the remainder as a distribution in the momentum, and it is stated there that the same proof gives it at fixed momentum. The expansion implies the power counting of~\cite{Weinberg:1959nj} for renormalised graphs. We work in a mass-independent scheme, and in a scheme in which the coefficient $\theta_I$, which describes a mixing of the source of the mass operator with derivatives of the couplings, vanishes. Such a scheme exists by a redefinition of the mass source~\cite{Jack:2013sha}, as we show in appendix~\ref{app:mass}. Note that the result is not restricted to this scheme. The redefinition of the mass source does not change $B^I$, and under a redefinition of the couplings, which relates two mass-independent schemes, \eqref{eq:introG} keeps its form with $B^I$ transforming as a vector. The gradient property of $B^I$ therefore holds in every mass-independent scheme.

Let us summarise the relation to earlier work. For multiscalar theories the gradient property was shown to hold through three loops by Wallace and Zia~\cite{Wallace:1974dx,Wallace:1974dy}. From four loops on it imposes constraints on the coefficients of the beta function, which are satisfied at four loops~\cite{Jack:2018oec}. Pannell and Stergiou showed that they are satisfied through five loops, and that at six loops they are satisfied by $B^I$, but not by the conventional beta function, if and only if four linear relations hold among the five- and six-loop coefficients of the field rotation, which were not known~\cite{Pannell:2024sia}. Our proof implies the four relations. We also explicitly computed the coefficients which enter and indeed find that the relations hold - as required by our proof. This is described in appendix~\ref{app:checks}.

The Ward identity which we use also has precursors, which are also Ward identities for special conformal transformations away from fixed points. For a single coupling and without the anomaly, Osborn obtained in this way a relation between the two-point function of two operators and its response to a gradient of the coupling~\cite{Osborn:1993cr}. An identity of the same structure, with a term proportional to the beta function times an operator insertion weighted by its position, is used in QCD to obtain the off-diagonal entries of the anomalous dimension matrix of twist-two operators at a given order from the special conformal anomaly at one order less~\cite{Braun:2003rp}. The identity for the generating functional with several local couplings away from fixed points is given in~\cite{Jack:2013sha}, eq.~(4.9) there, together with its first derivative, eq.~(4.11). Its second derivative, the identity for the two-point function, is given there at a fixed point, eq.~(4.23), see also~\cite{Fortin:2012hn}. What is new here is the second derivative at constant couplings away from fixed points, with the sources for the currents and the mass operator, and the relation of its contact term to $E_{[IJ]}$.

With \eqref{eq:introE} the antisymmetric part of $T_{IJ}$ is itself a sum of three such terms, see \eqref{eq:Tdecomp}. This is a more specific statement than the gradient property. One can therefore ask whether it implies relations among the coefficients of the beta function beyond those which follow from \eqref{eq:introG}. We investigate this in appendices~\ref{app:equiv} and~\ref{app:cut}. Through five loops we find that it does not. On the graphs which can contribute, the form \eqref{eq:Tdecomp} and the gradient property impose the same linear conditions on the coefficients. We also show that the two statements are equivalent near a zero of $B^I$ at which the matrix $\pr_JB^I$ is invertible, in the absence of field rotations which act on the couplings. Whether they are equivalent at higher orders around the free theory remains open.

The paper is structured as follows. In section~\ref{sec:setup} we introduce the theory, the sources and the anomaly, and we collect the consistency conditions which we use. In section~\ref{sec:ward} we derive the Ward identity. In section~\ref{sec:form} we obtain the form \eqref{eq:introE} of the antisymmetric coefficient. We also discuss the properties of renormalised perturbation theory on which the derivation rests, and the choice of scheme. In section~\ref{sec:metric} we construct the symmetric metric. We conclude in section~\ref{sec:discussion}. The appendices are of three kinds. Appendices~\ref{app:mass} and~\ref{app:ir} complete the proof by two technical steps, which extend known results of perturbative renormalisation to the specific cases required for this work and are not needed to follow the main conceptual argument. In appendix~\ref{app:mass} we show that the redefinition of the mass source which leads to $\theta_I=0$, given in~\cite{Jack:2013sha} for infinitesimal parameters, can be made for finite parameters. In appendix~\ref{app:ir} we define the first moment with a mass as infrared regulator and show that the Ward identity holds for it, using power counting at large momenta. Appendices~\ref{app:equiv} and~\ref{app:cut} are not needed for the proof. We use them to investigate the question whether the form \eqref{eq:Tdecomp} of $T_{[IJ]}$ implies relations beyond the gradient property. In appendix~\ref{app:equiv} we compare the conditions imposed by this form and by the gradient property through six loops. This comparison uses a property of the counterterms of graphs with a cut vertex, which we derive in appendix~\ref{app:cut}. Finally, appendix~\ref{app:checks} contains consistency checks. We verify the conclusion of the theorem at six loops by an explicit computation of the five- and six-loop coefficients of the field rotation which enter, and we test the Ward identity against correlation functions which are known independently.

While this work was being completed we learned of~\cite{Benfatto:toappear}, which also studies the gradient property of perturbative renormalisation group flows and appears simultaneously.

\section{Setup}
\label{sec:setup}

In this section we explain our setup. We first introduce the theory and its sources. We then give the conformal variation of the sources, the anomaly, and the consistency conditions of~\cite{Jack:2013sha}.

\subsection{Theory and sources}

We consider $N$ real scalar fields $\phi_i$ in four Euclidean dimensions with the Lagrangian
\begin{equation}
\cL = \tfrac12\,\pr_\mu\phi_i\,\pr^\mu\phi_i+\tfrac{1}{4!}\,\lambda_{ijkl}\,\phi_i\phi_j\phi_k\phi_l \, ,
\end{equation}
where $\lambda_{ijkl}$ is a symmetric tensor. We write $g^I$ for the couplings, so that an index $I$ stands for a symmetrised set of four field indices. The kinetic term is invariant under rotations $\phi_i\to R_{ij}\phi_j$ of the fields, which form the group $O(N)$. We call it the group of field rotations. It is the group of linear redefinitions of the fields which preserve the kinetic term. It is a symmetry of the free theory, and it is not a symmetry of the interacting theory unless $\lambda_{ijkl}$ is invariant under this group as well. The redundancy of the field redefinitions implies a redundancy of the couplings. The couplings $R_{ii'}R_{jj'}R_{kk'}R_{ll'}\lambda_{i'j'k'l'}$ describe the same theory as $\lambda_{ijkl}$. We can therefore equivalently let the group of field rotations act on the couplings. The Lie algebra of this group consists of the real antisymmetric matrices $\omega_{ij}=-\omega_{ji}$ in the field indices. We refer to it as the Lie algebra in the following, and we say that a quantity takes values in the Lie algebra if it is such a matrix for each value of its other indices.\footnote{We use this language because it carries over to theories with other fields, in which the group of field rotations is a different group.} For an element $\omega$ of the Lie algebra we write $(\omega g)^I$ for the corresponding infinitesimal change of the couplings, which is linear in $g$. It is the sum of the rotations of the four indices of $\lambda_{ijkl}$, $(\omega\lambda)_{ijkl}=\omega_{im}\lambda_{mjkl}+\omega_{jm}\lambda_{imkl}+\omega_{km}\lambda_{ijml} +\omega_{lm}\lambda_{ijkm}$.

We promote the couplings to functions $g^I(x)$, which we call local couplings. The generating functional $W[g]$ of connected correlation functions in flat space is the logarithm of the partition function. The marginal operators $\cO_I$ are defined by its derivatives,
\begin{equation}
W_{I_1\ldots I_n}(x_1,\ldots,x_n) = \frac{\delta^n W}{\delta g^{I_1}(x_1)\cdots\delta g^{I_n}(x_n)}
= \vev{\cO_{I_1}(x_1)\cdots\cO_{I_n}(x_n)} \, ,
\label{eq:Wn}
\end{equation}
so that $\cO_I$ is the renormalised quartic operator, up to a sign and a factor. The couplings on which correlation functions are evaluated are constant unless stated otherwise.

With local couplings two further sources are needed~\cite{Jack:2013sha}. The first is a source for the currents of the field rotations, $J^\mu_{ij}=\phi_i\pr^\mu\phi_j-\phi_j\pr^\mu\phi_i$. When the couplings depend on the position, the theory requires counterterms which are absent for constant couplings. One of them is the product of a derivative of a coupling and a current, $\pr_\mu g^I\,N_I(g)\cdot J^\mu$, where $N_I$ takes values in the Lie algebra. This term has dimension four and vanishes for constant couplings. To renormalise it we need a source for the current. We therefore couple the currents to a background field $a_\mu(x)$ with values in the Lie algebra, which enters as a gauge field for the field rotations. A rotation with a parameter $\omega(x)$ which depends on the position then leaves the generating functional invariant, if the couplings are rotated and $a_\mu$ is transformed as a gauge field. Derivatives of the couplings appear only in the covariant combination $D_\mu g^I=\pr_\mu g^I+(a_\mu g)^I$, and the counterterm is a shift of $a_\mu$ by $N_ID_\mu g^I$.

The second is a source for the operators $\phi_i\phi_j$ of dimension two. With local couplings there are also counterterms in which these operators are multiplied by $\pr^2g^I$ or by $\pr_\mu g^I\pr^\mu g^J$. These terms have dimension four and vanish for constant couplings. To renormalise them we introduce a source $\mathsf{M}(x)$ for the operators $\phi_i\phi_j$. We call $\mathsf{M}$ the mass source and denote by $V_{\mathsf M}$ the space of symmetric matrices in which it takes values. The two sources couple to the operators through the terms
\begin{equation}
a_\mu\cdot J^\mu = \tfrac12\,a_\mu^{ij}J^\mu_{ij} \, , \qquad
\mathsf{M}\cdot\phi^2 = \tfrac12\,\mathsf{M}_{ij}\,\phi_i\phi_j \, ,
\end{equation}
where $a_\mu^{ij}$ is antisymmetric and $\mathsf{M}_{ij}$ is symmetric in the field indices. Here and below a dot denotes this contraction of a pair of field indices. We use it also for coefficient functions which carry such a pair of indices. The two factors are then either both antisymmetric in the field indices, as $a_\mu$, or both symmetric, as $\mathsf{M}$. The generating functional is $W[g,a,\mathsf{M}]$, and its derivatives with respect to $a_\mu$ and $\mathsf{M}$ are the current and the mass operator.

Since rotated couplings describe the same theory, the beta function is defined only up to an infinitesimal rotation of the couplings, $\beta^I\to\beta^I-(\omega g)^I$. The conventional choice $\beta^I$ is the one for which the anomalous dimension matrix of the fields is symmetric. A second choice is
\begin{equation}
B^I = \beta^I-(\xi g)^I \, ,
\label{eq:B}
\end{equation}
where the element $\xi(g)$ of the Lie algebra is the coefficient of $\pr_\mu\sigma$ in the Weyl variation of the source $a_\mu$, given in \eqref{eq:deltav} below.\footnote{This element is denoted by $\upsilon$ in~\cite{Jack:2013sha} and by $S$ in~\cite{Pannell:2024sia}.} It is determined by the counterterm $N_I$ of the current introduced above. With this choice the trace of the energy-momentum tensor is $B^I\cO_I$, without a further term proportional to the divergence of the current. It is $B^I$ which enters the consistency conditions~\cite{Jack:1990eb,Osborn:1991gm,Jack:2013sha}, and which we show to be a gradient. The two choices differ by a motion along the directions which do not change the theory. In particular, $B^I=0$ is equivalent to conformal invariance, whereas $\beta^I$ need not vanish at a conformal point. In multiscalar $\phi^4$ theory $\xi$ can be non-zero from five loops, and at six loops $\beta^I$ is not a gradient while $B^I$ can be~\cite{Pannell:2024sia}.

\subsection{Conformal variation of the sources}

Our aim is a Ward identity for the generating functional in flat space. The generating functional depends on the metric and on the sources. A conformal transformation of flat space can be written as a diffeomorphism followed by a Weyl rescaling which restores the flat metric. Since the metric is unchanged, the whole effect of the transformation is a change of the sources. The generating functional is invariant under diffeomorphisms, and its response to a Weyl rescaling is known. The change of the sources which we compute in this subsection therefore gives an identity for $W[g,a,\mathsf{M}]$ in flat space, which we state in the next subsection. Two derivatives of this identity with respect to the couplings give the Ward identity for the two-point function in section~\ref{sec:ward}.

Let $v^\mu$ be a conformal Killing vector of flat space, with $\pr_\mu v_\nu+\pr_\nu v_\mu=2\sigma\,\delta_{\mu\nu}$. Then $\sigma=\alpha+2\,b\cdot x$ with a constant $\alpha$ and a constant vector $b$, and $\pr_\mu v^\mu=4\sigma$.

We obtain the action of a conformal transformation on the sources by writing it as a diffeomorphism followed by a Weyl rescaling. The functional $W$ is invariant under diffeomorphisms when the sources transform as tensors. We therefore start from the local Weyl variation of the sources. On a curved background with metric $\gamma_{\mu\nu}$ a rescaling $\gamma_{\mu\nu}\to(1+2\sigma)\gamma_{\mu\nu}$ with an arbitrary function $\sigma(x)$ is compensated, up to the anomaly, by the variation~\cite{Jack:2013sha}
\begin{align}
\delta^{\rm W}_\sigma g^I &= \sigma\,\beta^I \, , \notag\\
\delta^{\rm W}_\sigma a_\mu &= \sigma\,\rho_I\,D_\mu g^I-\pr_\mu\sigma\,\xi \, , \notag\\
\delta^{\rm W}_\sigma\mathsf{M} &= -\sigma\left((2-\gamma_{\mathsf M})\mathsf{M}+\delta_I\,D^2g^I
+\epsilon_{IJ}\,D_\mu g^ID^\mu g^J\right)-2\,\pr_\mu\sigma\,\theta_I\,D^\mu g^I+\ldots \, ,
\label{eq:weylvar}
\end{align}
where the superscript stands for Weyl.\footnote{This variation is denoted by $\Delta_\sigma$ in~\cite{Jack:2013sha}.} The dots stand for two terms which are proportional to the curvature scalar $R$ and to $\nabla^2\sigma$. The first line is the definition of the beta function with local couplings. The other two lines contain all terms which are allowed by dimensional analysis and covariance, and they define the coefficient functions which appear in them. Here $\beta^I$ is the conventional choice.\footnote{A different choice $\beta^I-(\omega g)^I$ of the beta function can be used in \eqref{eq:weylvar} if at the same time $\xi$ is replaced by $\xi-\omega$ and $\rho_I$ by $\rho_I+\pr_I\omega$. The two forms of \eqref{eq:weylvar} differ by a background gauge transformation with the parameter $\sigma\,\omega(g)$, under which $W$ is invariant. The combinations $B^I=\beta^I-(\xi g)^I$ and $\hat\rho_I=\rho_I+\pr_I\xi$, which enter our results, are the same for both.}

We use a mass-independent scheme throughout. By this we mean that the renormalisation constants do not depend on the mass source. The beta function and the coefficient functions in \eqref{eq:weylvar} and in the anomaly are then functions of the couplings alone, and the mass source enters only through the terms which are displayed. Minimal subtraction is an example. The massless theory then has no scale apart from the renormalisation scale, on which correlation functions depend through logarithms. In particular one-point functions vanish at constant couplings and $a_\mu=\mathsf{M}=0$.

We now specialise to flat space. The diffeomorphism generated by a conformal Killing vector $v$ changes the flat metric by $2\sigma\,\delta_{\mu\nu}$, and a Weyl rescaling with $-\sigma$ restores it. Under the diffeomorphism $g^I$ and $\mathsf{M}$ transform as scalars and $a_\mu$ as a one-form with Lie derivative $\cL_va_\mu$. The combination of the two leaves the flat metric unchanged and is a symmetry up to the anomaly. Its action on the sources is the change under the diffeomorphism minus the Weyl variation \eqref{eq:weylvar} with the same $\sigma$, $\delta_v=\cL_v-\delta^{\rm W}_\sigma$, where $\cL_v$ acts on scalars as $v^\mu\pr_\mu$. In flat space $R=0$, and $\nabla^2\sigma=0$ for a conformal Killing vector, so that the two terms which we did not display in \eqref{eq:weylvar} vanish. The action on the sources in flat space of the diffeomorphism generated by $v$, followed by the Weyl rescaling with $-\sigma$, where $\sigma$ is the function determined by $v$, is then
\begin{align}
\delta_vg^I &= v^\mu\pr_\mu g^I-\sigma\,\beta^I \, , \notag\\
\delta_va_\mu &= \cL_va_\mu-\sigma\,\rho_I\,D_\mu g^I+\pr_\mu\sigma\,\xi \, , \notag\\
\delta_v\mathsf{M} &= v\cdot\pr\mathsf{M}+\sigma\left((2-\gamma_{\mathsf M})\mathsf{M}+\delta_I\,D^2g^I
+\epsilon_{IJ}\,D_\mu g^ID^\mu g^J\right)+2\,\pr_\mu\sigma\,\theta_I\,D^\mu g^I \, ,
\label{eq:deltav}
\end{align}
where $\cL_va_\mu=v^\nu\pr_\nu a_\mu+\pr_\mu v^\nu a_\nu$ and all coefficients are evaluated on the local couplings. The coefficients $\rho_I$ and $\xi$ take values in the Lie algebra. They describe the mixing of $a_\mu$ with $D_\mu g^I$. The matrix $\gamma_{\mathsf M}$ is the anomalous dimension of the mass operator, and $\delta_I$, $\epsilon_{IJ}=\epsilon_{JI}$ and $\theta_I$ take values in $V_{\mathsf M}$. They describe the mixing of $\mathsf{M}$ with derivatives of the couplings. All are functions of the couplings which are determined by the counterterms with local couplings. At a fixed point with $\beta^I=-(4-\Delta_I)\,g^I$ to linear order, the first line is the transformation of a source for an operator of dimension $\Delta_I$. The action on a functional of the sources is
\begin{equation}
\delta_vW = \int d^4x\left(\delta_vg^I\,\frac{\delta W}{\delta g^I}
+\delta_va_\mu\cdot\frac{\delta W}{\delta a_\mu}
+\delta_v\mathsf{M}\cdot\frac{\delta W}{\delta\mathsf{M}}\right) .
\label{eq:deltavW}
\end{equation}

\subsection{The Weyl anomaly with local couplings}

The Ward identity which we use does not require the theory to be close to a fixed point. It is the analogue for special conformal transformations of the Callan--Symanzik equation, which is the Ward identity for dilatations in a theory which is not scale invariant. Away from a fixed point the trace of the energy-momentum tensor does not vanish. It is $\beta^I\cO_I$, up to terms with the current and the mass operator. The current of a special conformal transformation is then not conserved, and its divergence is the trace multiplied by the function $\sigma(x)$, which is linear in $x$. The breaking of the symmetry is therefore given by the beta function at any value of the couplings. The conformal Ward identity away from fixed points reads~\cite{Jack:2013sha}, eq.~(4.9) there,
\begin{equation}
\delta_vW = \cA[\sigma;g,a,\mathsf{M}] \, ,
\label{eq:ward}
\end{equation}
where $\cA$ is the Weyl anomaly restricted to flat space. The invariance under diffeomorphisms holds for vector fields of compact support, whereas a conformal Killing vector grows at large distances. For a conformal Killing vector the identity \eqref{eq:ward} therefore requires that the integrals in $\delta_vW$ converge at large distances. We show in appendix~\ref{app:ir} that it holds, with a mass as infrared regulator. At $a_\mu=\mathsf{M}=0$ the anomaly takes the form
\begin{equation}
\cA[\sigma;g] = \int d^4x \left( \tfrac12\,\sigma\,S_{IJ}\,\pr^2g^I\,\pr^2g^J
+ \pr_\mu\sigma\,E_{IJ}\,\pr^\mu g^I\,\pr^2g^J \right) + \ldots \, ,
\label{eq:anomaly}
\end{equation}
where the dots denote terms with three or four factors of $\pr g$. These terms vanish after two derivatives with respect to the couplings at constant couplings and do not enter the two-point identity. The two terms which are displayed are the most general terms with at most two factors of $\pr g$, up to total derivatives, as we show below. The coefficients $S_{IJ}(g)$ and $E_{IJ}(g)$ are functions of the couplings, with $S_{IJ}$ symmetric and $E_{IJ}$ a general matrix.\footnote{The notation of~\cite{Jack:2013sha} is different. Our $S_{IJ}$ is proportional to the coefficient called $A_{IJ}$ there, and our $E_{IJ}$ to the coefficient called $S_{IJ}$ there. We apologise that for us the association of $S_{IJ}$ with a symmetric tensor and of $A_{IJ}$ with an antisymmetric one is too strong to follow this notation. Further, \eqref{eq:ward} is the identity of~\cite{Jack:2013sha} for the flat-space functional with the opposite sign convention for $v$.}

The form \eqref{eq:anomaly} is fixed by general properties. The anomaly is a local functional, because it arises from the counterterms which are needed for local sources, and it is linear in $\sigma$. In a mass-independent scheme there is no dimensionful parameter, and the marginal sources are dimensionless. Therefore the integrand must contain four derivatives, distributed over $\sigma$ and the sources. Terms without derivatives or with two derivatives would require a coefficient of dimension four or two and are absent. Finally the integrand is a Lorentz scalar which depends on $g^I$ only through functions of $g^I$ multiplying derivatives of $g^I$.

We can list the terms with four derivatives up to integration by parts. Terms with two or more derivatives on $\sigma$ vanish because $\sigma$ is linear in $x$. Terms with a single factor of $\pr g$, such as $\sigma\,\pr^4g^I$ times a function of the couplings, are reduced by integration by parts to such terms and to terms with two or more factors of $\pr g$, which come from the derivatives of the coefficient function. With two factors of $\pr g$ the possible terms are $\sigma\,\pr^2g^I\pr^2g^J$, $\sigma\,\pr_\mu\pr_\nu g^I\pr^\mu\pr^\nu g^J$, $\pr_\mu\sigma\,\pr^\mu g^I\pr^2g^J$ and $\pr_\mu\sigma\,\pr_\nu g^I\pr^\mu\pr^\nu g^J$. The second and the fourth reduce to the other two by integration by parts, up to terms with three factors of $\pr g$ from the derivatives of the coefficient functions. Thus the two remaining structures are the ones displayed in \eqref{eq:anomaly}. Note that the symmetric part of $E_{IJ}$ multiplies a total derivative up to terms with three factors of $\pr g$, since $\pr^\mu g^{(I}\pr^2g^{J)}=\pr_\nu(\pr^\mu g^{(I}\pr^\nu g^{J)}) -\tfrac12\,\pr^\mu(\pr_\nu g^I\pr^\nu g^J)$.

With the sources $a_\mu$ and $\mathsf{M}$ the derivatives in \eqref{eq:anomaly} become covariant, and the anomaly contains further terms~\cite{Jack:2013sha}. Each of them contains the field strength $f_{\mu\nu}$ of $a_\mu$ or the mass source. However, we need only two of their coefficients below. These are the coefficients $Q_I$ and $L_I$ of~\cite{Jack:2013sha}, which belong to the terms $\pr_\mu\sigma\,f^{\mu\nu}\cdot D_\nu g^I$ and $\pr_\mu\sigma\,\mathsf{M}\cdot D^\mu g^I$. We also need the coefficient $R_{IJK}$ of the term $\pr_\mu\sigma\,\pr^\mu g^I\,\pr^\nu g^J\pr_\nu g^K$ in \eqref{eq:anomaly}. We use $Q_I$ and $L_I$ in the normalisation of~\cite{Jack:2013sha}. Since our anomaly \eqref{eq:anomaly} is normalised differently, these two terms appear in it with the coefficients $\kappa^{-1}Q_I$ and $\kappa^{-1}L_I$, where $\kappa$ is the constant in \eqref{eq:Ehat} below.

\subsection{Weyl consistency conditions}
\label{sec:consistency}

In this subsection we recall the origin of the relation \eqref{eq:introT} of~\cite{Jack:2013sha}, from which we start. The function $A$ and the tensors $G_{IJ}$ and $T_{[IJ]}$ in this relation are built from coefficients of the Weyl anomaly, and the relation follows from the consistency conditions on the anomaly. Some of these coefficients appear only in curved space. On a background with metric $\gamma_{\mu\nu}$ the variation of the generating functional under a local Weyl rescaling contains, up to an overall normalisation, the terms
\begin{equation}
\int d^4x\sqrt{\gamma}\,\left[\sigma\left(\tfrac14\,A_{\rm E}\,\mathcal{E}_4+G_{IJ}\,\mathcal{G}^{\mu\nu}\pr_\mu g^I\pr_\nu g^J\right)
+2\,\pr_\mu\sigma\,W_I\,\mathcal{G}^{\mu\nu}\pr_\nu g^I\right] ,
\label{eq:curved}
\end{equation}
where $\mathcal{E}_4$ is the Euler density and $\mathcal{G}^{\mu\nu}$ the Einstein tensor. This defines a function $A_{\rm E}(g)$, a symmetric tensor $G_{IJ}(g)$ and a one-form $W_I(g)$ on the space of couplings. At a fixed point $A_{\rm E}$ is the $a$-anomaly. The remaining terms of the anomaly reduce in flat space to those of the previous subsection.

Two Weyl variations commute. Imposing this on the variations of the sources and on the anomaly gives relations among the coefficients~\cite{Osborn:1991gm,Jack:2013sha}. We use the following results of section~2 of~\cite{Jack:2013sha}. The first two concern the variations of the sources,
\begin{equation}
\hat\rho_I\,B^I = 0 \, , \qquad
\Psi_I{}^J\delta_J+\epsilon_{IJ}B^J = \mathcal{D}\,\theta_I \, ,
\label{eq:rhoB}
\end{equation}
where
\begin{equation}
\hat\rho_I \equiv \rho_I+\pr_I\xi \, , \qquad
\Psi_J{}^K \equiv \delta_J{}^K+\pr_JB^K+\tfrac12\,(\hat\rho_J\,g)^K \, ,
\label{eq:Psi}
\end{equation}
and $\mathcal{D}$ is a linear differential operator of first order on the space of couplings, which is given in~\cite{Jack:2013sha}. In the proof we use the second relation only in a scheme with $\theta_I=0$, in which its right-hand side vanishes, so that the form of $\mathcal{D}$ is not needed there. It is needed only in the check of section~\ref{app:third}.\footnote{The combination $\hat\rho_I$ is denoted by $\tilde\rho_I$ in~\cite{Jack:2013sha}. We use a hat for $\hat\rho_I$ and for the combinations $\hat E_{IJ}$ and $\hat M$ below, which contain it.} The remaining three relations concern the anomaly. The terms with the Einstein tensor give
\begin{equation}
\pr_IA_{\rm E} = G_{IJ}\,B^J-B^J\pr_JW_I-\pr_IB^J\,W_J-(\hat\rho_I\,g)^JW_J \, .
\label{eq:J3}
\end{equation}
The second and third terms on the right-hand side are the Lie derivative of the one-form $W_I$ along the vector field $B^I$ on the space of couplings, and the last term is a correction from the field rotations. The terms with two derivatives of $\sigma$ relate the curl of the one-form to the flat-space coefficients,
\begin{equation}
\pr_IW_J-\pr_JW_I = -2\kappa\,\hat E_{[IJ]} \, ,
\label{eq:J2}
\end{equation}
where
\begin{equation}
\hat E_{IJ} \equiv \Psi_J{}^KE_{IK}+R_{IJK}B^K+\kappa^{-1}\left(L_J\cdot\theta_I+\tfrac12\,\hat\rho_J\cdot Q_I\right)
\label{eq:Ehat}
\end{equation}
and $\kappa$ is a constant which relates our normalisation of the anomaly to that of~\cite{Jack:2013sha}. In terms of the coefficients $S_{IJ}$ and $T_{IJK}$ of~\cite{Jack:2013sha} we have $E_{IJ}=\kappa^{-1}S_{IJ}$ and $R_{IJK}=\kappa^{-1}T_{IJK}$. The value and the sign of $\kappa$ depend on the normalisation and the sign convention of the generating functional, and they are not needed below. Finally the terms with the field strength of $a_\mu$ give, for every element $\omega$ of the Lie algebra,
\begin{equation}
(\omega g)^IW_I = -\,\omega\cdot Q_I\,B^I \, .
\label{eq:WQ}
\end{equation}

We can now show that \eqref{eq:J3} implies a relation of the form \eqref{eq:introT}. Using \eqref{eq:WQ} with $\omega=\hat\rho_I$ and the identity $B^J\pr_JW_I+\pr_IB^JW_J=B^J(\pr_JW_I-\pr_IW_J)+\pr_I(B^JW_J)$ we find
\begin{equation}
\pr_IA = \left(G_{IJ}+\pr_IW_J-\pr_JW_I+\hat\rho_I\cdot Q_J\right)B^J \, , \qquad
A \equiv A_{\rm E}+W_IB^I \, .
\label{eq:Tpre}
\end{equation}
A relation of this form is preserved if we add to the tensor in brackets a term which vanishes when it is contracted with $B^J$. Since $\hat\rho_JB^J=0$, the term $-\hat\rho_J\cdot Q_I$ is of this kind. Adding it and using \eqref{eq:J2} we obtain
\begin{equation}
\pr_IA = T_{IJ}\,B^J \, , \qquad T_{IJ} \equiv G_{IJ}+T_{[IJ]} \, , \qquad
T_{[IJ]} \equiv -2\kappa\,\hat E_{[IJ]}+2\,\hat\rho_{[I}\cdot Q_{J]} \, .
\label{eq:Tfull}
\end{equation}
This is the tensor of~\cite{Jack:2013sha}. With this choice the symmetric part of $T_{IJ}$ is $G_{IJ}$, and the remainder is antisymmetric. It is given by the antisymmetric part of $\hat E_{IJ}$, and through \eqref{eq:Ehat} by $E_{IJ}$. In section~\ref{sec:metric} we use the same freedom again, together with a change of the function $A$, to replace $T_{IJ}$ by a symmetric tensor.

The relations \eqref{eq:rhoB} to \eqref{eq:WQ} are all that we take from the analysis of the anomaly in curved space in~\cite{Jack:2013sha}. The rest of our argument uses the flat-space Ward identity \eqref{eq:ward} only.

\section{Ward identity}
\label{sec:ward}

In this section we derive from \eqref{eq:ward} a Ward identity for the two-point function of the marginal operators at constant couplings. We proceed in four steps. We first rewrite the left-hand side of \eqref{eq:ward} with the help of the invariance under background gauge transformations, in order to express it through $B^I$ and $\hat\rho_I$, which are the combinations that enter the consistency conditions, instead of $\beta^I$, $\rho_I$ and $\xi$. We then apply two derivatives with respect to the marginal sources to both sides and set the couplings to a constant value and $a_\mu=\mathsf{M}=0$. The constant value is not assumed to be a fixed point. The left-hand side becomes a sum of two- and three-point functions, and the right-hand side becomes a sum of derivatives of a delta function multiplied by the anomaly coefficients $S_{IJ}$ and $E_{[IJ]}$. Next we specialise to a special conformal transformation and separate the terms which depend on the midpoint of the two points from those which do not. The former give the Callan--Symanzik equation. The latter give the identity \eqref{eq:localfull}, which is the main result of this section. It relates $E_{[IJ]}$ to the two-point function, to the first moment of the three-point function and to two-point functions with the current and the mass operator.

\subsection{Background gauge identity}
\label{sec:gaugeid}

The left-hand side \eqref{eq:deltavW} of the Ward identity contains, through the variation \eqref{eq:deltav} of $a_\mu$, the term
\begin{equation}
\int d^4x\;\pr_\mu\sigma\,\xi\cdot\frac{\delta W}{\delta a_\mu} \, ,
\label{eq:xiterm}
\end{equation}
in which $\sigma$ is differentiated and the rotation $\xi$ appears explicitly. We want to express it through terms which are proportional to $\sigma$ itself. For this we use that the functional $W[g,a,\mathsf{M}]$ is invariant under background gauge transformations of the group of field rotations. Under an infinitesimal transformation with a parameter $\omega(x)$ which takes values in the Lie algebra the sources change by $\delta_\omega a_\mu=D_\mu\omega$ and $\delta_\omega g^I=-(\omega g)^I$, and the invariance reads
\begin{equation}
\delta_\omega W \equiv
\int d^4x\left(D_\mu\omega\cdot\frac{\delta}{\delta a_\mu}-(\omega g)^I\frac{\delta}{\delta g^I}\right)W = 0 \, ,
\label{eq:gauge}
\end{equation}
where we have dropped a term with $\delta/\delta\mathsf{M}$ which vanishes at $\mathsf{M}=0$. We use \eqref{eq:gauge} with $\omega=\sigma\,\xi(g)$. At $a_\mu=0$ we have $D_\mu\omega=\pr_\mu\sigma\,\xi+\sigma\,\pr_K\xi\,\pr_\mu g^K$, and therefore
\begin{equation}
\int d^4x\;\pr_\mu\sigma\,\xi\cdot\frac{\delta W}{\delta a_\mu}
= \int d^4x\;\sigma\left((\xi g)^K\frac{\delta W}{\delta g^K}
-\pr_K\xi\,\pr_\mu g^K\cdot\frac{\delta W}{\delta a_\mu}\right) .
\end{equation}
The left-hand side is the term \eqref{eq:xiterm}. We see that it shifts $\beta^K$ to $B^K$ in the variation of the couplings and $\rho_K$ to $\hat\rho_K$ in the variation of $a_\mu$. At $a_\mu=\mathsf{M}=0$ and for general $g^I(x)$ the Ward identity becomes
\begin{align}
&\int d^4x\left[\left(v^\mu\pr_\mu g^K-\sigma B^K\right)\frac{\delta W}{\delta g^K}
-\sigma\,\hat\rho_K\,\pr_\mu g^K\cdot\frac{\delta W}{\delta a_\mu}\right. \notag\\
&\qquad\left.+\left(\sigma\,\delta_K\,\pr^2g^K+\sigma\,\epsilon_{KL}\,\pr_\mu g^K\pr^\mu g^L
+2\,\pr_\mu\sigma\,\theta_K\,\pr^\mu g^K\right)\cdot\frac{\delta W}{\delta\mathsf{M}}\right] = \cA[\sigma;g] \, ,
\label{eq:wardB}
\end{align}
with $B^I$ and $\hat\rho_I$ as in \eqref{eq:B} and \eqref{eq:Psi}. We refer to the first term as the coupling term, to the second as the current term and to the third as the mass term. The rotation $\xi$ no longer appears explicitly.

\subsection{Second derivative at constant couplings}

We now apply $\delta^2/\delta g^I(y)\,\delta g^J(z)$ to \eqref{eq:wardB} and set the couplings constant. We write $u=y-z$ and denote the two-point function by $F_{IJ}(u)=W_{IJ}(y,z)$. Besides $F_{IJ}$ and the three-point function $W_{KIJ}$ the result contains the two-point functions of the current and of the mass operator with a marginal operator,
\begin{equation}
\frac{\delta^2W}{\delta a_\mu(y)\,\delta g^J(z)} = J^\mu_J(u) \, , \qquad
\frac{\delta^2W}{\delta\mathsf{M}(y)\,\delta g^J(z)} = e_J(u) \, ,
\label{eq:je}
\end{equation}
which depend only on $u$ by translation invariance. The function $J^\mu_J(u)$ carries a coupling index and takes values in the Lie algebra. By rotation invariance $e_J$ is even in $u$ and $J^\mu_J$ is odd, since the reflection $u\to-u$ is a rotation in four dimensions. We treat the four parts of \eqref{eq:wardB} in turn.

\paragraph{Couplings.} The first derivative of the first term of \eqref{eq:wardB} is
\begin{equation}
-\left(v\cdot\pr+4\sigma\right)\vev{\cO_I(y)}-\sigma(y)\,\pr_IB^K\,\vev{\cO_K(y)}
+\int d^4x\left(v^\mu\pr_\mu g^K-\sigma B^K\right)(x)\,W_{KI}(x,y) \, .
\end{equation}
The first term comes from the derivative term after integration by parts, and the second from the dependence of $B^K$ on the local coupling. At constant couplings the factor $v^\mu\pr_\mu g^K-\sigma B^K$ in the last term reduces to $-\sigma B^K$. The second derivative at constant couplings is
\begin{align}
&-\left(v(y)\cdot\pr_y+4\sigma(y)+v(z)\cdot\pr_z+4\sigma(z)\right)F_{IJ}
-\sigma(y)\,M^K{}_I\,F_{KJ}-\sigma(z)\,M^K{}_J\,F_{IK} \notag\\
&\qquad-B^K\int d^4x\;\sigma(x)\,W_{KIJ}(x,y,z) \, ,
\label{eq:part1}
\end{align}
where $M^K{}_I=\pr_IB^K$ is the stability matrix. Here we have dropped the term
\begin{equation}
-\sigma(y)\,\pr_I\pr_JB^K\,\delta^4(u)\,\vev{\cO_K(y)} \, ,
\end{equation}
since the one-point function vanishes at constant couplings.

\paragraph{Current.} The second term of \eqref{eq:wardB}, which we called the current term, contains the factor $\pr_\mu g^K$. We set $\pr_\mu g^K=0$ after taking the two derivatives. Hence we obtain a nonzero contribution only if one of the two derivatives acts on this factor. The other derivative then acts either on $\delta W/\delta a_\mu$, which gives the function $J^\mu_J$ of \eqref{eq:je}, or on $\hat\rho_K$. The term in which it acts on $\hat\rho_K$ is proportional to the one-point function of the current and vanishes. After integration by parts we obtain
\begin{equation}
\pr^y_\mu\Big(\sigma(y)\,\hat\rho_I\cdot J^\mu_J(u)\Big)
-\pr^z_\mu\Big(\sigma(z)\,\hat\rho_J\cdot J^\mu_I(u)\Big) \, ,
\label{eq:part2}
\end{equation}
where the second term is the first with the two points and the two indices exchanged, and we used $J^\mu_I(-u)=-J^\mu_I(u)$.

\paragraph{Mass operator.} In the third term of \eqref{eq:wardB} the term with $\epsilon_{KL}$ contains two factors of $\pr g$. Hence both derivatives must act on these factors, and the result is proportional to the one-point function of the mass operator, which vanishes. The terms with $\delta_K$ and $\theta_K$ are linear in derivatives of $g$, and we find as for the current
\begin{equation}
\pr_y^2\Big(\sigma(y)\,\delta_I\cdot e_J\Big)-2\,\pr^y_\mu\Big(\pr^\mu\sigma\;\theta_I\cdot e_J\Big)
+\pr_z^2\Big(\sigma(z)\,\delta_J\cdot e_I\Big)-2\,\pr^z_\mu\Big(\pr^\mu\sigma\;\theta_J\cdot e_I\Big) \, .
\label{eq:part3}
\end{equation}

\paragraph{Anomaly.} The right-hand side receives contributions only from the two terms displayed in \eqref{eq:anomaly}, since all other terms contain at least three factors of $\pr g$, or the field strength of $a_\mu$, or the mass source. With $\pr_\mu\pr_\nu\sigma=0$ the term with $S_{IJ}$ gives $S_{IJ}\,\pr_y^2\big(\sigma(y)\,\pr^2\delta^4(u)\big)$. The term with $E_{IJ}$ gives $-\pr_\mu\sigma\,(E_{IJ}-E_{JI})\,\pr^\mu\pr^2\delta^4(u)$. We see that only the antisymmetric part of $E_{IJ}$ remains.

\subsection{Special conformal transformations}

We specialise to a special conformal transformation, $v^\mu=2(b\cdot x)x^\mu-x^2b^\mu$ and $\sigma=2\,b\cdot x$, and introduce the midpoint $c=(y+z)/2$. The identity is then linear in $b$, and we find that each term is either proportional to $b\cdot c$ or independent of $c$. The function $F_{IJ}(u)$ is rotation invariant, even in $u$ and symmetric in its indices, because it is a second functional derivative of a Poincar\'e invariant functional.

We start with \eqref{eq:part1}. The derivative terms combine into $2\,b\cdot c\;u\cdot\pr_uF_{IJ}$, and $4\sigma(y)+4\sigma(z)=16\,b\cdot c$. With $\sigma(y)=2\,b\cdot c+b\cdot u$ and $\sigma(z)=2\,b\cdot c-b\cdot u$ the terms with the stability matrix become
\begin{equation}
-2\,b\cdot c\left(M^TF+FM\right)_{IJ}-b\cdot u\left(M^TF-FM\right)_{IJ} \, ,
\end{equation}
where $(M^TF)_{IJ}=M^K{}_IF_{KJ}$ and $(FM)_{IJ}=F_{IK}M^K{}_J$. In the three-point term we write $\sigma(x)=2\,b\cdot c+2\,b\cdot(x-c)$. The first part gives $-2\,b\cdot c\,B^K\pr_KF_{IJ}$, because the integral of $W_{KIJ}$ over $x$ is the derivative of the two-point function with respect to the constant coupling. The second part involves the first moment of the three-point function about the midpoint,
\begin{equation}
m^\mu_{K;IJ}(u) = \int d^4x\;(x-c)^\mu\,W_{KIJ}(x,c+\tfrac u2,c-\tfrac u2) \, ,
\label{eq:moment}
\end{equation}
which does not depend on $c$. Indeed, at constant couplings $W_{KIJ}$ depends only on the differences of its arguments, and the shift $x\to x+c$ of the integration variable removes $c$ from the integrand. The integral is defined with a mass as infrared regulator, as described in appendix~\ref{app:ir}. The three-point term contributes $-2\,b_\mu B^Km^\mu_{K;IJ}$. By rotation invariance $m^\mu_{K;IJ}$ is a vector which is odd in $u$. Exchanging the two points and the two indices leaves $W_{KIJ}$ and $c$ unchanged and maps $u\to-u$, so that
\begin{equation}
m^\mu_{K;IJ}(u) = m^\mu_{K;JI}(-u) = -\,m^\mu_{K;JI}(u) \, .
\label{eq:momentanti}
\end{equation}
At separated points rotation invariance gives in addition
\begin{equation}
m^\mu_{K;IJ}(u) = u^\mu\,h_{K;IJ}(u^2) \, , \qquad J^\mu_J(u) = u^\mu\,j_J(u^2) \, , \qquad u\neq0 \, ,
\label{eq:scalarlifts}
\end{equation}
with $h_{K;IJ}=-h_{K;JI}$. We use the scalar functions $h_{K;IJ}$ and $j_J$ only at separated points.

In the current term \eqref{eq:part2} the derivative acts on $\sigma$ or on the two-point function. The second possibility gives the divergence of the two-point function of the current. This follows from the derivative of \eqref{eq:gauge} with respect to $g^J(z)$ at constant couplings,
\begin{equation}
\omega\cdot\pr_\mu J^\mu_J(u) = -(\omega g)^K F_{KJ}(u) \, ,
\end{equation}
for every $\omega$ in the Lie algebra. Thus this part of the current term is equal to $-\sigma(y)\,(\hat\rho_I\,g)^KF_{KJ}-\sigma(z)\,(\hat\rho_J\,g)^KF_{IK}$. It has the same form as the terms with the stability matrix in \eqref{eq:part1}, and we combine the two. The result is the terms with the stability matrix with $M^K{}_I$ replaced by
\begin{equation}
\hat M^K{}_I \equiv \pr_IB^K+(\hat\rho_I\,g)^K \, .
\label{eq:Mhat}
\end{equation}
The part in which the derivative acts on $\sigma$ is $2\,b_\mu\,(\hat\rho_I\cdot J^\mu_J-\hat\rho_J\cdot J^\mu_I)$.

In the mass term \eqref{eq:part3} we use $\pr_y^2(\sigma\,e_J)=\sigma\,\pr^2e_J+4\,b\cdot\pr_ue_J$, where the derivatives act on $e_J$ as a distribution in $u$. The terms independent of $c$ are
\begin{equation}
b_\mu\left[\delta_I\cdot\left(u^\mu\pr^2+4\,\pr^\mu\right)e_J-\delta_J\cdot\left(u^\mu\pr^2+4\,\pr^\mu\right)e_I
-4\,\theta_I\cdot\pr^\mu e_J+4\,\theta_J\cdot\pr^\mu e_I\right] ,
\end{equation}
and the terms proportional to $b\cdot c$ are $2\,b\cdot c\,(\delta_I\cdot\pr^2e_J+\delta_J\cdot\pr^2e_I)$.

For the anomaly we use $-4\,b\cdot\pr_u\pr^2\delta^4(u)=b\cdot u\,\pr^4\delta^4(u)$, which follows from $[\pr^4,u_\mu]=4\,\pr_\mu\pr^2$. We obtain for the right-hand side
\begin{equation}
2\,b\cdot c\;S_{IJ}\,\pr^4\delta^4(u)+E_{[IJ]}\;b\cdot u\;\pr^4\delta^4(u) \, .
\label{eq:rhs}
\end{equation}

The identity holds for all $b$, $c$ and $u$. Therefore the terms proportional to $b\cdot c$ and the terms independent of $c$ hold separately. The terms proportional to $b\cdot c$ are even under the exchange of the two points. They read
\begin{equation}
-\left[\left(u\cdot\pr_u+8+B^K\pr_K\right)F+\hat M^TF+F\hat M\right]_{IJ}
+\delta_I\cdot\pr^2e_J+\delta_J\cdot\pr^2e_I = S_{IJ}\,\pr^4\delta^4(u) \, ,
\end{equation}
which is the Callan--Symanzik equation of the two-point function. It is also obtained from a dilatation. The terms independent of $c$ are odd under the exchange and are not obtained from a dilatation. We consider them in the next subsection.

\subsection{A new Ward identity for \texorpdfstring{$E_{[IJ]}$}{E[IJ]}}

Collecting the terms which are linear in $b$ and independent of $c$ we find an identity of the form $b_\mu V^\mu_{IJ}(u)=E_{[IJ]}\;b\cdot u\;\pr^4\delta^4(u)$, where $V^\mu_{IJ}$ is a vector of distributions in $u$. Since $b$ is arbitrary, we obtain
\begin{align}
&-u^\mu\left(\hat M^TF-F\hat M\right)_{IJ}-2\,B^Km^\mu_{K;IJ}+4\,\hat\rho_{[I}\cdot J^\mu_{J]}
+2\,\delta_{[I}\cdot\left(u^\mu\pr^2+4\,\pr^\mu\right)e_{J]}-8\,\theta_{[I}\cdot\pr^\mu e_{J]} \notag\\
&\qquad = E_{[IJ]}\;u^\mu\,\pr^4\delta^4(u) \, .
\label{eq:localfull}
\end{align}
This is the Ward identity which we use. The terms on the left are distributions which are not local, and \eqref{eq:localfull} states that their sum is local. The three-point function with an integrated current does not appear, because it was absorbed in the replacement of $\beta^I$ by $B^I$.

At separated points we can use \eqref{eq:scalarlifts} and $\pr^\mu e_J=2\,u^\mu\,\pr_{u^2}e_J$, where $\pr_{u^2}$ is the derivative of a function of $u^2$ with respect to $u^2$. The right-hand side of \eqref{eq:localfull} vanishes there, and \eqref{eq:localfull} is $u^\mu$ times the scalar identity
\begin{equation}
-\left(\hat M^TF-F\hat M\right)_{IJ}-2\,B^Kh_{K;IJ}+4\,\hat\rho_{[I}\cdot j_{J]}
+2\,\delta_{[I}\cdot\left(\pr^2+8\,\pr_{u^2}\right)e_{J]}-16\,\theta_{[I}\cdot\pr_{u^2}e_{J]}
= 0 \, , \qquad u\neq0 \, .
\label{eq:localsep}
\end{equation}
Note that the vector identity \eqref{eq:localfull} holds for all $u$, including $u=0$, while the scalar identity \eqref{eq:localsep} holds only at separated points. The coefficient $E_{[IJ]}$ multiplies a distribution supported at the origin and therefore does not appear in \eqref{eq:localsep}. In section~\ref{sec:form} we obtain $E_{[IJ]}$ from the vector identity \eqref{eq:localfull}. We use the scalar identity only in the consistency checks of appendix~\ref{app:checks}, where it is simpler to evaluate.

At a fixed point with a diagonal stability matrix and without the current and mass terms, \eqref{eq:localfull} reduces at separated points to the statement that the two-point function vanishes between operators of different dimension~\cite{Jack:2013sha,Fortin:2012hn}.\footnote{It is shown in~\cite{Jack:2013sha} that the symmetric part of the coefficient which we call $E_{IJ}$ drops out of the identity for the generating functional, since it multiplies a total derivative. The two-point identity is written there at a fixed point, eq.~(4.23), with the remark that the term with this coefficient cancels when the coefficient is antisymmetric. We find instead that the antisymmetric part remains, as in \eqref{eq:localfull}. The relations of~\cite{Jack:2013sha} which we use do not depend on this remark.}

\section{Form of the antisymmetric coefficient}
\label{sec:form}

\subsection{Pairing with a test function}

The functions $F_{IJ}$, $m^\mu_{K;IJ}$, $J^\mu_J$ and $e_J$ are distributions in $u$, including at $u=0$, because they are obtained from functional derivatives of the finite functional $W$. We can therefore pair the vector identity \eqref{eq:localfull} with a vector of test functions. We choose
\begin{equation}
\psi_\mu(u) = \frac{u_\mu\,|u|^2}{192}\;\chi(\mu_0|u|) \, , \qquad
\varphi(u) = u^\mu\psi_\mu(u) = \frac{|u|^4}{192}\;\chi(\mu_0|u|) \, ,
\label{eq:testfn}
\end{equation}
where $\chi$ is a smooth function with compact support which is equal to one in a neighbourhood of the origin, and $\mu_0$ is a fixed scale. We choose $\chi$ and $\mu_0$ once and for all. In four dimensions $\pr^4|u|^4=192$, so that $\varphi(0)=0$ and $\pr^4\varphi(0)=1$. We denote the pairing of a vector of distributions with $\psi_\mu$, summed over $\mu$, by $\vev{\,\cdot^{\,\mu},\psi_\mu}$, and the pairing of a distribution with $\varphi$ by $\vev{\,\cdot\,,\varphi}$. The right-hand side of \eqref{eq:localfull} gives $E_{[IJ]}\,\pr^4\varphi(0)=E_{[IJ]}$, and we obtain the following.

\begin{proposition}
\label{prop:form}
The antisymmetric anomaly coefficient is of the form
\begin{equation}
E_{[IJ]} = \left(\hat M^Tc-c\hat M\right)_{IJ}+B^K\tilde\tau_{K;IJ}+\hat\rho_{[I}\cdot\tilde X_{J]}
+\delta_{[I}\cdot Z_{J]}+\theta_{[I}\cdot Y_{J]} \, ,
\label{eq:formfull}
\end{equation}
with $\hat M$ given in \eqref{eq:Mhat} and
\begin{align}
c_{IJ} &= -\vev{F_{IJ},\varphi} \, , & \tilde\tau_{K;IJ} &= -2\,\vev{m^\mu_{K;IJ},\psi_\mu} \, , &
\tilde X_J &= 4\,\vev{J^\mu_J,\psi_\mu} \, , \notag\\
Z_J &= 2\,\vev{(u^\mu\pr^2+4\,\pr^\mu)\,e_J,\psi_\mu} \, , & Y_J &= -8\,\vev{\pr^\mu e_J,\psi_\mu} \, .
\label{eq:coeffs}
\end{align}
The matrix $c_{IJ}$ is symmetric and $\tilde\tau_{K;IJ}$ is antisymmetric in its last two indices. In perturbation theory all five coefficients are power series in the couplings.
\end{proposition}

The form \eqref{eq:formfull} follows from \eqref{eq:localfull} by the pairing. We write a tilde on two of the coefficients, because we reserve $\tau_{K;IJ}$ and $X_J$ for the combinations which enter the form \eqref{eq:introE} of the introduction. We define them in \eqref{eq:formscheme}. We discuss the statement that the coefficients are power series in the next subsection.

In the first term we used $\vev{u^\mu F_{IJ},\psi_\mu}=\vev{F_{IJ},\varphi}$. In the same way, when the vectors are $u^\mu$ times scalar distributions, as in \eqref{eq:scalarlifts} at separated points, the coefficients are pairings of the scalar functions with $\varphi$, for instance $\tilde\tau_{K;IJ}=-2\,\vev{h_{K;IJ},\varphi}$ and $\tilde X_J=4\,\vev{j_J,\varphi}$.

The individual coefficients depend on the choice of $\chi$ and $\mu_0$ and on the renormalisation scale, while $E_{[IJ]}$ does not. This reflects the freedom to add the local counterterm $\pr^2g^I\pr^2g^J$ with a symmetric coefficient, which shifts $F_{IJ}$ by a contact term. A change of the test function which preserves the normalisation $\pr^4\varphi(0)=1$ changes $c_{IJ}$ by a symmetric matrix, and it changes the other coefficients such that the sum is the same. Thus only the sum in \eqref{eq:formfull} is fixed. At fixed $\mu_0$ and fixed renormalisation scale the coefficients are functions of the couplings alone. Note that at a point where $B^I\neq0$ every antisymmetric matrix can be written as $B^K\tau_{K;IJ}$ with some $\tau$. The content of the proposition is that the coefficients are regular where $B^I$ and its derivative vanish, and in particular as formal power series around the free theory.

\subsection{Inputs from renormalised perturbation theory}

We call a quantity regular if it is a power series in the couplings. The regularity of the coefficients \eqref{eq:coeffs} uses two properties of perturbative renormalisation, which are inputs of our argument.

First, we use that $W$ with local sources is finite order by order. The coefficients of the expansions of $F_{IJ}$, $W_{KIJ}$, $J^\mu_J$ and $e_J$ in the constant coupling are then distributions. This is Hypothesis~\ref{hyp:finite} below. It is the starting point of~\cite{Jack:2013sha}. A construction of perturbation theory in which the couplings are test functions is given in~\cite{Epstein:1973gw}, in a different framework.

Second, we use that the first moment \eqref{eq:moment} exists at each order as a distribution in $u$ on all of $\mathbb{R}^4$, and that the Ward identity \eqref{eq:localfull} holds for it. We show this in appendix~\ref{app:ir}. There we give the scalar fields a mass through a constant value of the mass source. In the massive theory the first moment converges at large distances and the Ward identity is exact, and we show that both have a limit when the mass is removed. The limit exists because the three-point function with an additional mass operator grows at most quadratically in the momentum of the two marginal operators. This follows from power counting at large momenta~\cite{Weinberg:1959nj,Smirnov:1990rz} and from the fact that all vertices have an even number of lines. It corresponds to the statement that the operators of lowest dimension in the product of two marginal operators, apart from the identity, are the bilinears $\phi_i\phi_j$.

\subsection{Scheme for the mass source}
\label{sec:scheme}

The terms with $\delta_I$ and $\theta_I$ in \eqref{eq:formfull} differ from the first three terms. They are not proportional to $B^I$ or its derivative, and there is no relation of the type $\theta_IB^I=0$. However, we can remove them by a choice of scheme. The mass source can be redefined by local terms with $D^2g^I$ and $D_\mu g^ID^\mu g^J$. This preserves the consistency conditions and shifts $\theta_I$ by $\Psi_I{}^Jd_J+e_{IJ}B^J$ with arbitrary regular $d_J$ and $e_{IJ}$~\cite{Jack:2013sha}. The shift is stated in~\cite{Jack:2013sha} for an infinitesimal redefinition. We show in appendix~\ref{app:mass} that it holds without change for a finite one. Since $\Psi_I{}^J$ is the unit matrix at lowest order, it is invertible as a power series, and we can choose a scheme with
\begin{equation}
\theta_I = 0 \, .
\label{eq:theta0}
\end{equation}
This scheme is not minimal subtraction. The redefinition changes the operators $\cO_I$ by $\pr^2$ acting on dimension-two operators. The correlation functions $F_{IJ}$ and $m^\mu_{K;IJ}$ and the anomaly coefficients therefore change as well. However, $\beta^I$, $\xi$ and $\rho_I$ do not change, because the variations of $g^I$ and $a_\mu$ in \eqref{eq:deltav} do not contain the mass source. Hence $B^I$ and $\hat\rho_I$ are the same in both schemes. The consistency conditions of section~\ref{sec:setup} and the Ward identity \eqref{eq:localfull} hold in the new scheme in the same form. From here on all quantities are those of the scheme \eqref{eq:theta0}.

\subsection{Three kinds of terms}
\label{sec:three}

In the scheme \eqref{eq:theta0} the second relation in \eqref{eq:rhoB} gives $\delta_I=-(\Psi^{-1})_I{}^J\epsilon_{JK}B^K$. The term with $\delta_I$ in \eqref{eq:formfull} is then proportional to $B^K$, and the term with $\theta_I$ vanishes. We also separate the part of $\hat M$ which contains $\hat\rho$. For a Lie algebra valued $\rho$ and a matrix $n_{IJ}$ we write $\rho\cdot(gn)_J=(\rho g)^Kn_{KJ}$. We then find that \eqref{eq:formfull} becomes
\begin{align}
E_{[IJ]} &= B^K\tau_{K;IJ}+\hat\rho_{[I}\cdot X_{J]}+\left(M^Tc-cM\right)_{IJ} \, , \notag\\
\tau_{K;IJ} &= \tilde\tau_{K;IJ}-(\Psi^{-1})_{[I}{}^L\epsilon_{|LK|}\cdot Z_{J]} \, , \qquad
X_J = \tilde X_J+2\,(gc)_J \, ,
\label{eq:formscheme}
\end{align}
where $M^K{}_I=\pr_IB^K$ as before and the indices between bars are not antisymmetrised. This is the form \eqref{eq:introE} of the introduction, which consists of three kinds of terms. In section~\ref{sec:removable} we refer to them as the first, the second and the third type, in this order.

\section{Symmetric metric}
\label{sec:metric}

We now show that \eqref{eq:formscheme} implies a gradient formula with a symmetric tensor. We work in the scheme \eqref{eq:theta0}.

\subsection{Removable terms}
\label{sec:removable}

We call an antisymmetric tensor $\Omega_{IJ}$ removable if $\Omega_{IJ}B^J$ is the sum of a regular symmetric tensor contracted with $B^J$ and the gradient of a regular function. Three types of terms are removable. In this subsection $\tau_{IJK}$, $X_J$ and $C_{IJ}$ stand for arbitrary regular coefficients. In the next subsection we apply the result to the coefficients of section~\ref{sec:form}, which are regular by Proposition~\ref{prop:form}. The symmetric tensors and the functions which we obtain below are built from these coefficients, from $B^I$ and from $\hat\rho_I$ by products and derivatives, and are therefore regular as well.

The first type is $\Omega_{IJ}=\tau_{IJK}B^K$, where $\tau_{IJK}$ is antisymmetric in its first two indices. Then
\begin{equation}
\Omega_{IJ}B^J = \tau_{IJK}B^K\,B^J = \left(\tau_{IKJ}+\tau_{JKI}\right)B^K\,B^J \, .
\label{eq:type1}
\end{equation}
In the last step the second term vanishes, because $\tau_{JKI}$ is antisymmetric in $J$ and $K$, and the first term is the previous expression with the names of the summation indices exchanged. The right-hand side is the tensor $(\tau_{IKJ}+\tau_{JKI})B^K$, which is symmetric in $I$ and $J$, contracted with $B^J$. No gradient is needed.

The second type is $\Omega_{IJ}=\hat\rho_{[I}\cdot X_{J]}$. Here $\hat\rho_JB^J=0$ implies
\begin{equation}
\Omega_{IJ}B^J = \hat\rho_{[I}\cdot X_{J]}\,B^J = \tfrac12\,\hat\rho_I\cdot X_JB^J = \hat\rho_{(I}\cdot X_{J)}\,B^J \, .
\label{eq:type2}
\end{equation}
The right-hand side is a symmetric tensor contracted with $B^J$, and again no gradient is needed.

The third type is $\Omega_{IJ}=(M^TC-CM)_{IJ}$ with a symmetric $C_{IJ}$. For any symmetric $C$ we have
\begin{equation}
\pr_I\left(C_{JK}B^K\right)-\pr_J\left(C_{IK}B^K\right)
= \left(M^TC-CM\right)_{IJ}+\left(\pr_IC_{JK}-\pr_JC_{IK}\right)B^K \, ,
\label{eq:curlC}
\end{equation}
so that $M^TC-CM$ differs from the curl of the one-form $C_{IK}B^K$ by a term of the first type. The contraction of the curl with $B^J$ is
\begin{equation}
B^J\left[\pr_I\left(C_{JK}B^K\right)-\pr_J\left(C_{IK}B^K\right)\right]
= \pr_I\left(B^JC_{JK}B^K\right)-\left(\cL_BC\right)_{IK}B^K \, ,
\end{equation}
where $(\cL_BC)_{IK}=B^J\pr_JC_{IK}+M^J{}_IC_{JK}+M^J{}_KC_{IJ}$ is the Lie derivative of $C$, which is symmetric. Combining the two steps gives
\begin{equation}
\Omega_{IJ}B^J = \pr_I\left(B^JC_{JK}B^K\right)
-\left[\pr_I\!\left(C_{JK}B^K\right)+\pr_J\!\left(C_{IK}B^K\right)-B^K\pr_KC_{IJ}\right]B^J \, ,
\label{eq:type3}
\end{equation}
where the tensor in the bracket is symmetric. This is the gradient of a function plus a symmetric tensor contracted with $B^J$.

\subsection{Antisymmetric part of the tensor}

We now go through the terms of $T_{[IJ]}$ in \eqref{eq:Tfull} and show that each is of one of the three types. We write $E_{IJ}=s_{IJ}+a_{IJ}$ with $s_{IJ}=E_{(IJ)}$ and $a_{IJ}=E_{[IJ]}$ given by \eqref{eq:formscheme}. For $\theta_I=0$ the antisymmetric part of $\hat E_{IJ}$ in \eqref{eq:Ehat} is
\begin{equation}
\hat E_{[IJ]} = a_{IJ}+\tfrac12\left(sM-M^Ts\right)_{IJ}+\tfrac12\left(aM+M^Ta\right)_{IJ}
-\tfrac12\,\hat\rho_{[I}\cdot(gE^T)_{J]}+R_{[IJ]K}B^K-\tfrac{1}{2\kappa}\,\hat\rho_{[I}\cdot Q_{J]} \, ,
\label{eq:Ehatanti}
\end{equation}
where $(gE^T)_J$ is defined by $\rho\cdot(gE^T)_J=(\rho g)^KE_{JK}$. The second and third term come from $\pr_JB^KE_{IK}$ and the fourth from $\tfrac12(\hat\rho_J\,g)^KE_{IK}$ in $\Psi_J{}^KE_{IK}$.

The first term is a sum of the three types by \eqref{eq:formscheme}. The second term is of the third type with $C=-\tfrac12\,s$. The fourth and the last term are of the second type, and the fifth is of the first type. For the third term we need the form of $a$. The part $M^Tc-cM$ of $a$ contributes
\begin{equation}
\tfrac12\left((M^T)^2c-cM^2\right) = M^Tc''-c''M \, , \qquad c''=\tfrac12\left(M^Tc+cM\right) ,
\end{equation}
since the terms $M^TcM$ cancel. Hence this contribution is of the third type. The part $B^K\tau_{K;IJ}$ of $a$ contributes a term proportional to $B^K$, which is of the first type. The part $\hat\rho_{[I}\cdot X_{J]}$ of $a$ contributes terms with $\hat\rho_I\cdot X_K\,\pr_JB^K$, which are of the second type, and terms with $\hat\rho_K\,\pr_JB^K$. By $\hat\rho_KB^K=0$ the latter equal $-\pr_J\hat\rho_K\,B^K$ and are of the first type.

Each term of $\hat E_{[IJ]}$ is thus a sum of terms of the three types. The same holds for $T_{[IJ]}=-2\kappa\,\hat E_{[IJ]}+2\,\hat\rho_{[I}\cdot Q_{J]}$, since the additional term is of the second type. We now add up all contributions of the same type. In this way we write $T_{[IJ]}$ as the sum of a single term of each type,
\begin{equation}
T_{[IJ]} = -2\kappa\,B^K\tau'_{K;IJ}+\hat\rho_{[I}\cdot\bar X_{J]}-2\kappa\left(M^TC-CM\right)_{IJ} \, .
\label{eq:Tdecomp}
\end{equation}
The three terms are of the first, the second and the third type, in this order. Their coefficients are
\begin{align}
\tau'_{K;IJ} &= \tau_{K;IJ}+\tfrac12\left(\tau_KM+M^T\tau_K\right)_{IJ}+R_{[IJ]K}
-\tfrac12\,\pr_{[I}\hat\rho_{|K|}\cdot X_{J]} \, , \notag\\
\bar X_J &= -2\kappa\left(X_J+\tfrac12\,X_K\,\pr_JB^K-\tfrac12\,(gE^T)_J\right)+3\,Q_J \, , \notag\\
C &= c-\tfrac12\,s+\tfrac12\left(M^Tc+cM\right) ,
\label{eq:CoX}
\end{align}
where $\tau_K$ denotes the matrix $\tau_{K;IJ}$. The matrix $C$ is symmetric and $\tau'_{K;IJ}$ is antisymmetric in its last two indices. We checked the coefficients \eqref{eq:CoX} by computer algebra. For three couplings and a single generator of field rotations we inserted \eqref{eq:Ehat} and \eqref{eq:formscheme} into \eqref{eq:Tfull}, with arbitrary functions of the couplings for all coefficients and with $\hat\rho_I=n_{IJ}B^J$ for an arbitrary antisymmetric $n_{IJ}$, so that $\hat\rho_IB^I=0$. The difference of the two sides of \eqref{eq:Tdecomp} then vanishes identically. Note that \eqref{eq:Tdecomp} is the tensor $T_{[IJ]}$ of \eqref{eq:Tfull} in a different form, and we have not changed it yet. In the next subsection we replace each of the three terms by a symmetric tensor and a gradient.

\subsection{Metric and potential}

We can now state the main result. In addition to the steps above, its proof uses the consistency conditions \eqref{eq:rhoB} to \eqref{eq:WQ} of~\cite{Jack:2013sha} and the following two hypotheses.

\begin{hypothesis}
\label{hyp:finite}
In minimal subtraction the generating functional $W[\gamma,g,a,\mathsf{M}]$ of multiscalar $\phi^4$ theory with local couplings, background field and mass source is finite at each order of perturbation theory, with the bare sources and the field independent counterterms of~\cite{Jack:2013sha}. It is invariant under diffeomorphisms and background gauge transformations, and its Weyl variation is given by \eqref{eq:weylvar} and a local anomaly.
\end{hypothesis}

This hypothesis is also assumed in~\cite{Osborn:1991gm,Jack:2013sha}. The consistency conditions are derived from it there, and they are therefore consequences of it and not further assumptions. Related results are proven in other frameworks. In the causal approach perturbation theory is constructed with couplings which are test functions~\cite{Epstein:1973gw}. For massless $\phi^4$ theory the correlation functions with insertions of composite operators are constructed with flow equations in~\cite{Holland:2014pna}. The operators are coupled to sources which are smooth functions, and the additional divergences are removed by counterterms which are local and multilinear in the sources, as in the framework which we use. For an arbitrary massless renormalisable theory and an arbitrary number of insertions it is shown in~\cite{Frob:2015uqy}, Theorems~1 and~3 there, that these correlation functions are tempered distributions at each order. In both works the renormalisation conditions are defined by the flow equations. These results do not prove the hypothesis in the form which we use, which refers to minimal subtraction and to the form of the anomaly.

\begin{hypothesis}
\label{hyp:expansion}
The asymptotic expansion \eqref{eq:irsmirnov} holds for the renormalised graphs of Lemma~\ref{lem:irW} at Euclidean momenta $p\neq0$, with a remainder which is smaller than any given power of $\rho$ if the orders of the Taylor expansions are chosen large enough.
\end{hypothesis}

The expansion is derived in~\cite{Smirnov:1990rz}. The status of the estimate of the remainder is described in appendix~\ref{app:ir}.

\begin{theorem}
\label{thm:main}
Assume Hypotheses~\ref{hyp:finite} and~\ref{hyp:expansion}. Consider multiscalar $\phi^4$ theory in four dimensions to all orders in perturbation theory, in minimal subtraction with the mass source redefined as in appendix~\ref{app:mass}, so that $\theta_I=0$. Then
\begin{equation}
\pr_IA' = G'_{IJ}\,B^J
\label{eq:main}
\end{equation}
holds with
\begin{align}
A' &= A+2\kappa\,B^IC_{IJ}\,B^J \, , \notag\\
G'_{IJ} &= G_{IJ}+2\kappa\left[\pr_I\!\left(C_{JK}B^K\right)+\pr_J\!\left(C_{IK}B^K\right)
-B^K\pr_KC_{IJ}-\left(\tau'_{I;JK}+\tau'_{J;IK}\right)B^K\right] \notag\\
&\quad+\hat\rho_{(I}\cdot \bar X_{J)} \, ,
\label{eq:explicitfull}
\end{align}
where $C$, $\tau'$ and $\bar X$ are given in \eqref{eq:CoX}, \eqref{eq:formscheme} and \eqref{eq:coeffs}. Here $A$ is the function in \eqref{eq:Tfull}, $G_{IJ}$ is the symmetric tensor in \eqref{eq:curved}, $\kappa$ is the constant in \eqref{eq:Ehat} and $\hat\rho_I$ is defined in \eqref{eq:Psi}. The coefficients $C$, $\tau'$ and $\bar X$ depend on the test function $\chi$, on the scale $\mu_0$ and on the renormalisation scale. The tensor $G'_{IJ}$ is symmetric. The function $A'$ is invariant under field rotations, and it equals $A$ at zeros of $B^I$. Both $A'$ and $G'_{IJ}$ are power series in the couplings. The term of order zero of $G'_{IJ}$ is that of $G_{IJ}$, which is a positive definite matrix, so that $G'_{IJ}$ is positive definite to all orders in perturbation theory.
\end{theorem}

\begin{proof}
We contract \eqref{eq:Tdecomp} with $B^J$. The first term gives $-2\kappa\,(\tau'_{I;JK}+\tau'_{J;IK})B^KB^J$ by \eqref{eq:type1}, applied to $\tau_{IJK}=\tau'_{K;IJ}$. The second term gives $\hat\rho_{(I}\cdot\bar X_{J)}B^J$ by \eqref{eq:type2}. The third term gives \eqref{eq:type3}, multiplied by $-2\kappa$. Hence
\begin{equation}
T_{[IJ]}B^J = \left(G'_{IJ}-G_{IJ}\right)B^J-2\kappa\,\pr_I\left(B^JC_{JK}B^K\right) ,
\end{equation}
and \eqref{eq:main} follows from \eqref{eq:Tfull} with $A'=A+2\kappa\,B^IC_{IJ}B^J$. The coefficients $c$, $\tau$, $X$ and $Z$ are regular by Proposition~\ref{prop:form}, and the anomaly coefficients and the coefficients in \eqref{eq:deltav} are regular because they are given by counterterms. The statement on the lowest order is shown in the third remark below.\footnote{We also checked $\pr_IA'=G'_{IJ}B^J$ with \eqref{eq:explicitfull} by computer algebra, starting from \eqref{eq:Tfull} and \eqref{eq:formscheme}, in the same way as for the coefficients \eqref{eq:CoX}.}
\end{proof}

The first two terms in the square bracket of \eqref{eq:explicitfull} are the symmetrised derivative of the one-form $C_{IK}B^K$. The last term is the symmetrised product of $\hat\rho_I$ with a Lie algebra valued one-form.

\subsection{Remarks}

We add four remarks. First, the theorem gives the potential $A'$ and the metric $G'_{IJ}$ explicitly, in \eqref{eq:explicitfull}, and no equation remains to be solved. In particular the proof does not proceed order by order. In the explicit tests of the gradient property one makes an ansatz for the potential and the metric at a given loop order and solves $\pr_IA'=G'_{IJ}B^J$ for their coefficients~\cite{Wallace:1974dy,Jack:2018oec,Pannell:2024sia}. A solution exists only if the coefficients of the beta function satisfy relations, which have to be checked at each order. Here the potential is supplied by the anomaly through \eqref{eq:J3}, and $A'$ and $G'_{IJ}$ are obtained from the anomaly coefficients and the coefficients of Proposition~\ref{prop:form} by the algebraic steps above. The relations therefore hold at every order. To evaluate \eqref{eq:explicitfull} at a given order one has to compute these coefficients, which requires more than the beta function. In practice, if one wants to compute the metric and potential at a given order it might be simpler to solve for an ansatz, which by the theorem has a solution at every order.

Second, we do not show that there is a scheme in which the tensor $T_{IJ}$ of the consistency conditions is itself symmetric. It was conjectured in~\cite{Poole:2019kcm} that such a scheme can always be reached by a choice of local counterterms. Our result is a weaker statement. We obtain $G'_{IJ}$ from $T_{IJ}$ by a shift of the potential and by using that only the contraction $T_{IJ}B^J$ enters, as described in section~\ref{sec:removable}. Only a part of the difference $G'_{IJ}-G_{IJ}$, the Lie derivative of $2\kappa\,C_{IJ}$ along $B^I$, can be attributed to a change of scheme by a local term in the generating functional~\cite{Jack:2013sha}. In the companion paper~\cite{Reehorst:sphere} we show, under a hypothesis on finiteness of the same kind as Hypothesis~\ref{hyp:finite}, that in four dimensions there is a scheme in which the one-form $W_I$ of \eqref{eq:curved} vanishes. In this scheme $T_{[IJ]}=2\,\hat\rho_{[I}\cdot Q_{J]}$, so that $T_{IJ}$ is symmetric in the absence of field rotations.

Third, the new metric is positive definite to all orders in perturbation theory, in the sense stated in the introduction. Every term of $G'_{IJ}-G_{IJ}$ in \eqref{eq:explicitfull} contains a factor of $B^K$, of its derivative $\pr_JB^K$, or of $\hat\rho_I$. In multiscalar $\phi^4$ theory the beta function starts at second order in the couplings and its derivative at first order. Moreover, the coefficient $\hat\rho_I$ has no term of order zero, because the sources do not mix in the free theory, and its expansion starts with the two-loop term of~\cite{Jack:2013sha}. The tensor $G_{IJ}$ starts at order zero with a positive matrix~\cite{Jack:1990eb,Jack:2013sha}. Therefore the difference $G'_{IJ}-G_{IJ}$ is at least of first order in the couplings, and the term of order zero of $G'_{IJ}$ is this positive definite matrix. This has two consequences. The metric has an inverse as a power series, so that the gradient formula can be solved for the beta function, $B^I=(G'^{-1})^{IJ}\pr_JA'$. Moreover, the metric truncated at any finite order is a positive definite matrix for sufficiently small couplings. We do not make a statement about the sum of the series. Note that the symmetry of $G'_{IJ}$ is shown in the proof of the theorem without the use of positivity.

Fourth, the gradient property holds in every mass-independent scheme, although the theorem is stated in a particular one. Besides finite local terms in the generating functional, which change $A$, $G_{IJ}$ and the other coefficients of the anomaly but not $B^I$, there are three changes of scheme to consider. The first is the redefinition of the mass source by which we reached the scheme with $\theta_I=0$. It does not change $B^I$, as shown in section~\ref{sec:scheme}, so that \eqref{eq:main} holds for the same $B^I$ before the redefinition. The second is a redefinition of the couplings. Under it $B^I$ transforms as a vector. Hence \eqref{eq:main} holds in the new couplings with the same function $A'$ and with $G'_{IJ}$ transformed as a tensor, which is again symmetric.

The third is a redefinition of the background field by a finite local term, $a_\mu\to a_\mu+r_I\,D_\mu g^I$ with a regular $r_I$ in the Lie algebra~\cite{Jack:2013sha}. It changes the rotation $\xi$ by a term which is linear in the vector field, so that the vector field $B'^I$ of the new scheme is related to $B^I$ by
\begin{equation}
B^I = B'^I+(w\,g)^I \, , \qquad w = \tilde r_L\,B'^L \, ,
\end{equation}
with a regular $\tilde r_L$ in the Lie algebra. In this case $B^I$ does not transform as a vector, and a short argument is needed. For $\omega$ in the Lie algebra let $X_I(\omega)=(\omega g)^JG'_{JI}$. Inserting the relation above into \eqref{eq:main} we find $\pr_IA'=[G'_{IL}+X_I(\tilde r_L)]B'^L$, where the matrix in brackets is not symmetric. However, the function $A'$ is invariant under field rotations, because $A$ is invariant and $C_{IJ}$ is covariant. Hence $(\omega g)^I\pr_IA'=0$ for every $\omega$, and with \eqref{eq:main} this gives $X_K(\omega)B^K=0$. For $\omega=\tilde r_I$ this reads
\begin{equation}
X_L(\tilde r_I)\,B'^L+Y_{IL}\,B'^L = 0 \, , \qquad
Y_{IL} = (\tilde r_Ig)^J\,G'_{JK}\,(\tilde r_Lg)^K \, ,
\end{equation}
where $Y_{IL}$ is symmetric. Adding this to the previous relation we obtain
\begin{equation}
\pr_IA' = \left[G'_{IL}+X_I(\tilde r_L)+X_L(\tilde r_I)+Y_{IL}\right]B'^L \, .
\end{equation}
The matrix in brackets is symmetric and regular, and its term of order zero is that of $G'_{IJ}$. The gradient property therefore holds for $B'^I$ with the same potential. Note that the explicit expressions \eqref{eq:explicitfull} for the metric and the potential are those of the scheme of the theorem.

\section{Discussion}
\label{sec:discussion}

We have shown that the renormalisation group flow of multiscalar $\phi^4$ theory in four dimensions is a gradient flow to all orders in perturbation theory. More precisely, we showed that $\pr_IA'=G'_{IJ}B^J$ holds with a symmetric metric $G'_{IJ}$ which is positive definite to all orders, and we gave the potential and the metric explicitly in \eqref{eq:explicitfull}. The gradient property holds in every mass-independent scheme.

The main new ingredient was the Ward identity \eqref{eq:localfull} for special conformal transformations away from fixed points. The gradient formula of~\cite{Osborn:1991gm,Jack:2013sha} contains a tensor which is not known to be symmetric. Its antisymmetric part is determined by the anomaly coefficient $E_{[IJ]}$. The Ward identity expresses this coefficient through the two-point functions and the first moment of the three-point function of the marginal operators. From it we found that $E_{[IJ]}$ is a sum of three kinds of terms, \eqref{eq:formscheme}. We then showed that each kind, contracted with $B^J$, is equal to a symmetric tensor contracted with $B^J$ and a gradient. This allowed us to replace the tensor of~\cite{Osborn:1991gm,Jack:2013sha} by a symmetric one.

In addition to these steps the proof used the consistency conditions of~\cite{Osborn:1991gm,Jack:2013sha} and two hypotheses. The first is the finiteness of the generating functional with local sources, which is also assumed in~\cite{Osborn:1991gm,Jack:2013sha} and from which the consistency conditions are derived. The second is that the expansion of renormalised Feynman integrals at momenta which are large compared to the mass holds with a bound on its remainder at each non-zero momentum. From it we obtained the massless limit of the first moment \eqref{eq:moment} in appendix~\ref{app:ir}. A proof of the two hypotheses in the form in which we use them would make the result unconditional.

We checked consistency of our proof with perturbative loop data in two ways. At six loops the conclusion of the theorem is equivalent to the four relations among coefficients of the field rotation which were derived in~\cite{Pannell:2024sia} as the condition for the gradient property. These coefficients were not known. These relations now follow from the theorem. Additionally we checked the relations indeed hold through explicit computation of the missing coefficients. We also tested the Ward identity that plays a central role in our proof against correlation functions at first order near a conformal fixed point, in a resonant model in three dimensions and at third order in $\phi^4$ theory. These checks are described in appendix~\ref{app:checks}. We should point out their limits. The checks of the Ward identity are at separated points, where the anomaly term does not contribute, and they use the first moment computed directly in the massless theory. The check at third order is not independent of the relations of~\cite{Jack:2013sha}. The only test which involves the contact term $E_{[IJ]}$ and the field rotation is the one at six loops.

The proof gives more than the gradient property. It shows that the antisymmetric part of the tensor of~\cite{Osborn:1991gm,Jack:2013sha} has the specific form \eqref{eq:formscheme}, and the gradient property follows from this form. It is therefore natural to ask whether the form has further consequences, in particular relations among the coefficients of the beta function which do not follow from the gradient property alone. We investigated this in appendix~\ref{app:equiv}. Through five loops we found no such relation. On the graphs which can contribute, the form and the gradient property impose the same linear conditions, and the two statements are equivalent near a zero of $B^I$ with an invertible stability matrix. Whether this holds to all orders in the expansion around the free theory is an open question.

The methods of this paper extend in three directions. The first and the third are the subject of work in preparation, and the second is the subject of the companion paper~\cite{Reehorst:sphere}. The first is the gradient property for general renormalisable theories in four dimensions, with gauge fields and fermions~\cite{Reehorst:toappear}. The consistency conditions and the Ward identity have the same structure there, and the steps of the proof appear to generalise.

The second direction is $d=4-\vep$, where the beta function contains the term $-\vep\,g^I$ and the theory has fixed points within perturbation theory. For fixed $\vep$ a gradient formula always exists if coefficients with poles at $\vep=0$ are allowed, as noted in~\cite{Pannell:2024sia}, so that the statement of interest is the gradient property with a metric and a potential which are regular at $\vep=0$. This is a stronger statement than the theorem of this paper. In~\cite{Pannell:2024sia} it was found to require relations among the coefficients of the beta function at five and six loops which are independent of the relations in four dimensions, and which hold in modified minimal subtraction. In~\cite{Pannell:2025sph} the gradient equation was solved order by order with a symmetric metric, and the free coefficients of the solution were chosen such that at fixed points the potential agrees with the quantity $\tilde F$ through order $\vep^5$, and such that the metric agrees at leading order with the two-point function of the nearly marginal operators on the sphere, up to a normalisation of the operators. In~\cite{Reehorst:sphere} we replace the Ward identity for the two-point function in flat space by the Ward identity of a conformal Killing vector of the sphere. Under a hypothesis on the finiteness of the generating functional with local couplings in flat space, of the same kind as Hypothesis~\ref{hyp:finite}, we show there that the vector field $-\vep\,g^I+B^I$ is a gradient to all orders, with a symmetric metric and a potential which are regular at $\vep=0$. The metric contains the two-point function on the sphere and is positive definite to all orders in the couplings and in $\vep$. At fixed points the potential is proportional to $\tilde F$ minus its value for free scalars, where $\tilde F$ is defined with the couplings to the curvature at the zeros of their beta functions. This explains the observations of~\cite{Pannell:2025sph}, and it implies that $\tilde F$ decreases towards the infrared along a flow between two fixed points within perturbation theory, and that the matrix of derivatives of the vector field at a fixed point has real eigenvalues. The companion paper also states properties of the generating functional on the sphere from which the gradient formula follows for other theories near their critical dimensions. These are sextic scalars near three dimensions, scalars with quartic and Yukawa couplings and gauge fields with fermions near four, cubic scalars near six and fermions near two dimensions, and for them the results are conditional on these properties, with some points open.

The construction on the sphere also has a consequence at $\vep=0$. Under the hypothesis of~\cite{Reehorst:sphere} it gives a second gradient formula in four dimensions for the same vector field $B^I$, without our second hypothesis, and with a metric and a potential which differ from \eqref{eq:explicitfull}. The metric \eqref{eq:explicitfull} contains the coefficient $c_{IJ}$, which is the two-point function of the marginal operators in flat space paired with a test function at the scale $\mu_0$, while the metric from the sphere contains the two-point function on the sphere, whose scale is the radius. It would be interesting to understand whether the two are related by the freedom in the choice of the metric and the potential. A related question is whether there is a preferred choice within this freedom, for which the metric and the potential have a meaning along the flow and not only at fixed points.

The third direction is two dimensions. The general gradient formulas there contain an antisymmetric tensor~\cite{Osborn:1991gm,Friedan:2009ik}, as in four dimensions, and in section~\ref{app:twod} we related this tensor to the quantities of this paper. In~\cite{Reehorst:twod} we consider conformal field theories perturbed by marginal operators. We believe that the Ward identity of a conformal Killing vector of the sphere gives a gradient formula with a symmetric metric to all orders in conformal perturbation theory, under a hypothesis of finiteness with local couplings of the same kind as Hypothesis~\ref{hyp:finite}. The potential would agree with the $c$-function up to terms of second order in the beta function. It would be interesting to extend this to nearly marginal perturbations, which describe flows between different fixed points.

There are further questions which we have not studied. We considered only the quartic couplings, with the mass operator as a source. The framework of~\cite{Osborn:1991gm} also treats the masses and the cubic couplings as couplings, and it would be interesting to obtain the gradient property on this larger space. In supersymmetric theories the metric on the space of couplings was compared with exact results for the central charge $a$ in~\cite{Freedman:1998rd}, and $a$ is determined at fixed points by a maximisation principle~\cite{Intriligator:2003jj}. It would be interesting to see how the symmetric metric and the potential of this paper are related to these results. For boundaries in two dimensions a gradient formula was proven in~\cite{Friedan:2003yc}, and the flow on line defects was shown to be irreversible in~\cite{Cuomo:2021rkm}, while the gradient property of defects is known only at low orders of perturbation theory~\cite{Pannell:2025sph}. A Ward identity on a sphere which contains the defect might extend our method to this case. Finally, it would be interesting to prove Hypotheses~\ref{hyp:finite} and~\ref{hyp:expansion}. The flow equations with which closely related statements were proven in~\cite{Holland:2014pna,Frob:2015uqy} may be a suitable framework.

\section*{Acknowledgements}
I thank Slava Rychkov for motivating me to prove statements about gradient flow and for useful conversations. I thank Slava Rychkov and Benoit Sirois for an initial collaboration in this direction, which used a different approach from the one in this paper. I thank Franz Herzog and Ben Ruijl for their replies to my questions about the evaluation of the counterterms, and especially Ben Ruijl for sharing the code with which I checked the six-loop relations that the gradient property implies. Claude and ChatGPT were used extensively and played an instrumental role in every stage of this research. At King's College London I was funded by UK Research and Innovation (UKRI) under the UK government's Horizon Europe funding guarantee [grant number EP/X042618/1]. Currently, I am supported by a Royal Society University Research Fellowship URF\textbackslash R1\textbackslash 241371.

\appendix

\section{Redefinition of the mass source}
\label{app:mass}

In section~\ref{sec:scheme} we use a scheme with $\theta_I=0$, which is reached by a finite redefinition of the mass source. In this appendix we show three properties of such a redefinition. First, the variation of the new mass source has the form \eqref{eq:deltav} with new coefficients $\delta_I$, $\epsilon_{IJ}$ and $\theta_I$, and the change of these coefficients is exactly linear in the parameters of the redefinition. In~\cite{Jack:2013sha} the change is given to first order in the parameters. Since there are no terms of higher order, this formula is exact, and $\theta_I$ can be set to zero. Second, the consistency conditions hold for the coefficients of the new scheme. Third, $B^I$ and $\hat\rho_I$ do not change, while the marginal operators change by a total derivative.

We start with the first property. Let
\begin{equation}
\mathsf{M}' = \mathsf{M}+f \, , \qquad f = d_I\,D^2g^I+e_{IJ}\,D_\mu g^ID^\mu g^J \, ,
\end{equation}
with arbitrary regular $d_I(g)$ and $e_{IJ}(g)$ which take values in $V_{\mathsf M}$, and let $W'[g,a,\mathsf{M}']=W[g,a,\mathsf{M}'-f]$. The variations $\delta_vg^I$ and $\delta_va_\mu$ do not contain $\mathsf{M}$, which has dimension two, and $f$ does not depend on $\mathsf{M}$. Therefore the variation of $f$ is determined by $\delta_vg^I$ and $\delta_va_\mu$ alone, and it is linear in $d_I$ and $e_{IJ}$. From \eqref{eq:deltav} we find
\begin{align}
\delta_v\mathsf{M}' &= v\cdot\pr\mathsf{M}'+\sigma\left((2-\gamma_{\mathsf M})\mathsf{M}'+\delta_I\,D^2g^I
+\epsilon_{IJ}\,D_\mu g^ID^\mu g^J\right)+2\,\pr_\mu\sigma\,\theta_I\,D^\mu g^I+\Phi \, , \notag\\
\Phi &= \delta_vf-v\cdot\pr f-\sigma\,(2-\gamma_{\mathsf M})\,f \, .
\end{align}
The remainder $\Phi$ is a local expression of dimension two in the sources $g^I$ and $a_\mu$. It is covariant, linear in $\sigma$ and linear in $d_I$ and $e_{IJ}$, and it contains no second derivatives of $\sigma$ for a conformal Killing vector of flat space. It is therefore of the form
\begin{equation}
\Phi = \sigma\left(\Delta\delta_I\,D^2g^I+\Delta\epsilon_{IJ}\,D_\mu g^ID^\mu g^J\right)
+2\,\pr_\mu\sigma\,\Delta\theta_I\,D^\mu g^I \, ,
\end{equation}
where $\Delta\delta_I$, $\Delta\epsilon_{IJ}$ and $\Delta\theta_I$ are linear in $d_I$ and $e_{IJ}$. The variation of $\mathsf{M}'$ has the form \eqref{eq:deltav} with $\delta_I+\Delta\delta_I$, $\epsilon_{IJ}+\Delta\epsilon_{IJ}$ and $\theta_I+\Delta\theta_I$. Since no terms of higher order in $d_I$ and $e_{IJ}$ occur, the shifts are given exactly by their terms of first order. For $\theta_I$ we find from the terms with $\pr_\mu\sigma$ in $\delta_vf$
\begin{equation}
\Delta\theta_I = -\Psi_I{}^Jd_J-e_{IJ}B^J \, .
\end{equation}
This is the relation $\delta\theta_I=\Psi_I{}^Jd_J+e_{IJ}B^J$ of~\cite{Jack:2013sha}.\footnote{The parameters $d_I$ and $e_{IJ}$ of~\cite{Jack:2013sha} have the opposite sign, since the redefinition considered there corresponds to $W'[g,a,\mathsf{M}]=W[g,a,\mathsf{M}+f]$.}

We now choose $e_{IJ}=0$ and $d_I=(\Psi^{-1})_I{}^J\theta_J$. This $d_I$ is regular, because $\Psi_I{}^J$ is the unit matrix at lowest order. With this choice $\theta_I+\Delta\theta_I=0$.

For the second property, note that the functional $W'$ differs from $W$ only by a change of variables. The consistency conditions follow from the fact that two Weyl variations commute. This does not depend on the choice of variables, and the derivation uses only the form \eqref{eq:deltav} of the variations. The conditions therefore hold for the coefficients of the new scheme.

For the third property, the variations $\delta_vg^I$ and $\delta_va_\mu$ are unchanged, and with them $B^I$ and $\hat\rho_I$. The marginal operators are the derivatives of the generating functional with respect to $g^I$ at fixed mass source, and they depend on whether $\mathsf{M}$ or $\mathsf{M}'$ is held fixed,
\begin{equation}
\cO'_I(x) = \cO_I(x)-\int d^4y\,\frac{\delta f(y)}{\delta g^I(x)}\cdot\frac{\delta W}{\delta\mathsf{M}(y)} \, .
\end{equation}
At constant couplings and $a_\mu=0$ the second term is $-d_I\cdot\pr^2\,\delta W/\delta\mathsf{M}(x)$, which is a total derivative of the operators of dimension two. The correlation functions which enter the Ward identity in the scheme with $\theta_I=0$ are those of the operators $\cO'_I$.

\section{Infrared regularisation by a mass}
\label{app:ir}

The Ward identity \eqref{eq:localfull} contains the first moment \eqref{eq:moment} of the three-point function, which is an integral over all positions of the third operator. In the massless theory the correlation functions decay only as powers of the distance, so that the convergence of this integral at large distances is not obvious. Moreover, the identity \eqref{eq:ward}, from which \eqref{eq:localfull} was derived, rests on the invariance of the generating functional under the diffeomorphism generated by a conformal Killing vector. This invariance holds for vector fields of compact support, whereas a conformal Killing vector grows at large distances. The identity for a conformal Killing vector therefore also requires that the correlation functions decay sufficiently fast, which has to be shown. In this appendix we have to show three statements. First, the first moment can be defined in the massless theory as a tempered distribution in $u$. Second, with this definition it has the properties which we used in section~\ref{sec:ward}, namely the antisymmetry \eqref{eq:momentanti} and the form $u^\mu h_{K;IJ}(u^2)$ at $u\neq0$. Third, the Ward identity \eqref{eq:localfull} holds for it.

We proceed in two steps. In the first step we give the scalar fields a mass $m$ by setting the mass source of section~\ref{sec:setup} to a constant. In the massive theory the correlation functions decay exponentially, the first moment converges, and the derivation of section~\ref{sec:ward} gives a Ward identity which is exact. It contains additional terms which are due to the mass source. In the second step we show that every term of this identity has a limit for $m\to0$, and that the additional terms vanish in the limit. The tool is Lemma~\ref{lem:irW}, a bound on the growth of correlation functions at momenta which are large compared to $m$. It follows from the asymptotic expansion of renormalised Feynman integrals at large momenta~\cite{Smirnov:1990rz} and from power counting~\cite{Weinberg:1959nj}. The bound implies that the derivative of each term with respect to $\log m$ is of order $m$ up to logarithms. We then define the first moment of the massless theory as the limit of the massive one. The inputs of the argument and its relation to known results are stated after the proof of the lemma.

\subsection{Ward identity in the massive theory}

We set $\mathsf{M}_{ij}=m^2\delta_{ij}$ and keep the couplings constant and $a_\mu=0$. Since the scheme is mass independent, the beta function and all coefficient functions of sections~\ref{sec:setup} and~\ref{sec:ward} are the same as for $m=0$. For $m>0$ the Fourier transforms of the connected correlation functions are analytic in the momenta in a strip around the real Euclidean momenta, and the correlation functions decay exponentially at large distances. In particular the first moment
\begin{equation}
m^\mu_{K;IJ}(u;m) = \int d^4x\;(x-c)^\mu\,\vev{\cO_K(x)\,\cO_I(c+\tfrac u2)\,\cO_J(c-\tfrac u2)}_m
\label{eq:irmm}
\end{equation}
converges at large $x$. The integrand is a distribution, which is defined at coinciding points by renormalisation, and the moment is equal to $-i$ times the derivative of the Fourier transform of the three-point function with respect to the momentum of $\cO_K$ at zero momentum. It is a distribution in $u$ on all of $\mathbb{R}^4$.

The derivation of section~\ref{sec:ward} used the identity \eqref{eq:ward} for a conformal Killing vector $v_{\rm c}$, which grows at large distances. The invariance under diffeomorphisms, from which \eqref{eq:ward} follows, holds for vector fields of compact support. In the massless theory the correlation functions decay only as powers of the distance, and the step from compact support to a conformal Killing vector is not obvious. For $m>0$ the step can be justified, as we now show. Apart from this step the derivation of section~\ref{sec:ward} goes through in the same way for $m>0$, with additional terms from the mass source, which we collect afterwards.

We replace $v_{\rm c}$ by the vector field
\begin{equation}
v^\mu(x) = \chi_R(x)\,v^\mu_{\rm c}(x) \, ,
\end{equation}
where $\chi_R$ is a smooth function which is equal to one inside a ball of radius $R$ and which vanishes outside the ball of radius $2R$. This $v$ has compact support. It is a conformal Killing vector inside the ball of radius $R$, but not in the shell between $R$ and $2R$. Under the diffeomorphism generated by $v$ the flat metric changes by its Lie derivative, $\pr_\mu v_\nu+\pr_\nu v_\mu$. We split this symmetric tensor into its trace and its traceless part,
\begin{equation}
\pr_\mu v_\nu+\pr_\nu v_\mu = 2\sigma\,\delta_{\mu\nu}+2\,\tau_{\mu\nu} \, , \qquad \sigma=\tfrac14\,\pr_\mu v^\mu \, ,
\qquad \tau_{\mu\nu}=\tfrac12\left(\pr_\mu v_\nu+\pr_\nu v_\mu\right)-\tfrac14\,\delta_{\mu\nu}\,\pr_\rho v^\rho \, .
\end{equation}
The traceless part $\tau_{\mu\nu}$ vanishes inside the ball of radius $R$, where $v=v_{\rm c}$, and outside the ball of radius $2R$, where $v=0$. It is therefore different from zero only in the shell between $R$ and $2R$. The invariance of $W$ under the diffeomorphism states that the change of the sources as tensors, together with this change of the metric, leaves $W$ unchanged. The Weyl rescaling with $-\sigma$ and the Weyl variation of the sources remove the term $2\sigma\,\delta_{\mu\nu}$ up to the anomaly, as in section~\ref{sec:setup}. The term $2\,\tau_{\mu\nu}$ remains as a change of the metric, and it contributes the derivative of $W$ with respect to the metric. The identity \eqref{eq:ward} then holds with an additional term,
\begin{equation}
\delta_vW+\int d^4x\;\tau_{\mu\nu}\,T^{\mu\nu} = \cA[\sigma;g,a,\mathsf{M}] \, , \qquad
T^{\mu\nu} = 2\,\frac{\delta W}{\delta\gamma_{\mu\nu}} \, ,
\end{equation}
where $T^{\mu\nu}$ is the stress tensor, defined as the response of the generating functional to a change of the metric at fixed sources. After two derivatives with respect to the couplings the additional term is the integral of $\tau_{\mu\nu}(x)$ with the three-point function of the stress tensor at $x$ and the two marginal operators. For $m>0$ this function decays exponentially when $x$ is far from the two operators, while $\tau_{\mu\nu}$ in the shell grows at most as a power of $R$. The additional term therefore vanishes for $R\to\infty$, and the other terms converge to their values for the conformal Killing vector. In the shell $\sigma$ is not linear in $x$. The terms with $\pr^2\sigma$ which we did not display in \eqref{eq:weylvar} and the terms of the anomaly with two or more derivatives of $\sigma$ therefore do not vanish there. After two derivatives with respect to the couplings at fixed $y$ and $z$ the terms of the anomaly are supported where $x$ coincides with $y$ and $z$, which lie outside the shell for large $R$. The term with $\pr^2\sigma$ and the mass operator is a correlation function with an operator in the shell, which decays exponentially. Hence these terms vanish for $R\to\infty$ as well.

We now turn to the additional terms. The resulting identity differs from \eqref{eq:localfull} by the terms which are due to the mass source. The variation \eqref{eq:deltav} contains $\sigma\,(2-\gamma_{\mathsf M})\mathsf{M}$, which contributes the first moment of the three-point function with the mass operator at the third point and terms with $\pr_I\gamma_{\mathsf M}$ and the two-point function $e_J$. The anomaly contains terms with one or two factors of $\mathsf{M}$, whose coefficients do not depend on $m$, and the one-point functions of $\cO_K$ and of the mass operator do not vanish. These contribute derivatives of $\delta^4(u)$ with coefficients which are positive powers of $m$ times powers of $\log(m/\mu)$. The term which was dropped in \eqref{eq:gauge} vanished there because $\mathsf{M}=0$. Here $\mathsf{M}$ does not vanish, but the term still vanishes, because the change of $\mathsf{M}$ under a field rotation is its commutator with the parameter of the rotation and $\mathsf{M}$ is proportional to the unit matrix. We obtain
\begin{equation}
\big[\text{left-hand side of \eqref{eq:localfull}}\big]_m+r^\mu_{IJ}(u;m) = E_{[IJ]}\;u^\mu\,\pr^4\delta^4(u) \, ,
\label{eq:irmassive}
\end{equation}
where the bracket is evaluated in the massive theory with the first moment \eqref{eq:irmm}, and
\begin{equation}
r^\mu_{IJ} = 2\,m^2\,(2-\gamma_{\mathsf M})\cdot m^\mu_{\phi^2;IJ}(u;m)
-2\,m^2\,u^\mu\,\pr_{[I}\gamma_{\mathsf M}\cdot e_{J]}(u;m)+m^2\,\ell_{IJ}(m)\,\pr^\mu\delta^4(u) \, .
\label{eq:irrem}
\end{equation}
Here $m^\mu_{\phi^2;IJ}$ is the first moment with the mass operator at the third point, and $\ell_{IJ}$ is a polynomial in $\log(m/\mu)$.

\subsection{Bound at large momentum}

In this subsection we prove Lemma~\ref{lem:irW}, a bound on correlation functions at momenta which are large compared to $m$. We use it in the next subsection to remove the mass.

The lemma concerns the following functions. Let $A_1$ and $A_2$ be two local operators and let $B_1,\ldots,B_k$ with $k\geq1$ be further local operators. All of them have an even number of fields. This is sufficient for our purpose, since multiscalar $\phi^4$ theory is invariant under $\phi\to-\phi$, and the operators which enter the Ward identity, namely the marginal operators, the mass operator and the currents, are even. The restriction is essential for the bound, because it excludes that the two parts of a graph are connected by a single line. We consider the Fourier transform of their connected correlation function in the massive theory, in which $A_1$ and $A_2$ carry the momenta $p$ and $-p$ and the further operators carry zero momentum,
\begin{equation}
Q(p;m) = \prod_{a=1}^k\pr_{q_a}^{\,j_a}\int d^4x\,\prod_{a=1}^kd^4y_a\;
e^{ip\cdot x+i\sum_aq_a\cdot y_a}\,\vev{A_1(x)\,A_2(0)\,B_1(y_1)\cdots B_k(y_k)}_m\,\Big|_{q_a=0} \, .
\label{eq:irQ}
\end{equation}
Here $j_a\geq0$ derivatives with respect to the momentum $q_a$ of $B_a$ are taken before the momentum is set to zero. By the dimension of an operator we mean its canonical dimension, which is the number of fields plus the number of derivatives. Let $\Delta_1$ and $\Delta_2$ be the dimensions of $A_1$ and $A_2$, and let $d_a\leq4$ be the dimension of $B_a$. We denote by $D$ the mass dimension of $Q$,
\begin{equation}
D = \Delta_1+\Delta_2-4+\sum_{a=1}^k\left(d_a-4-j_a\right) .
\end{equation}
Each integration over a position lowers the dimension by four, and each derivative with respect to a momentum lowers it by one. Besides $p$ and $m$ the function $Q$ depends on the couplings and on the renormalisation scale $\mu$. As for all correlation functions in this paper we do not write these two arguments. The semicolon in $Q(p;m)$ separates the momentum from the parameter $m$. By dimensional analysis $Q$ is $m^D$ times a function of the ratios $p/m$ and $\mu/m$, and the dependence on $\mu$ enters the bound below through $\log(m/\mu)$.

\begin{lemma}
\label{lem:irW}
Let $Q(p;m)$ be a function of the form \eqref{eq:irQ}, with $\Delta_1$, $\Delta_2$ and $D$ as defined above. Then for $|p|\geq m$
\begin{equation}
|Q(p;m)| \leq C\,|p|^{\alpha}\,m^{D-\alpha}\,\big(1+\log(|p|/m)\big)^N\big(1+|\log(m/\mu)|\big)^N \, , \qquad
\alpha = \max(D,\,\Delta_1+\Delta_2-6) \, ,
\label{eq:irbound}
\end{equation}
with constants $C$ and $N$ which depend on the order.
\end{lemma}

\begin{figure}[t]
\centering
\includegraphics[width=0.8\textwidth]{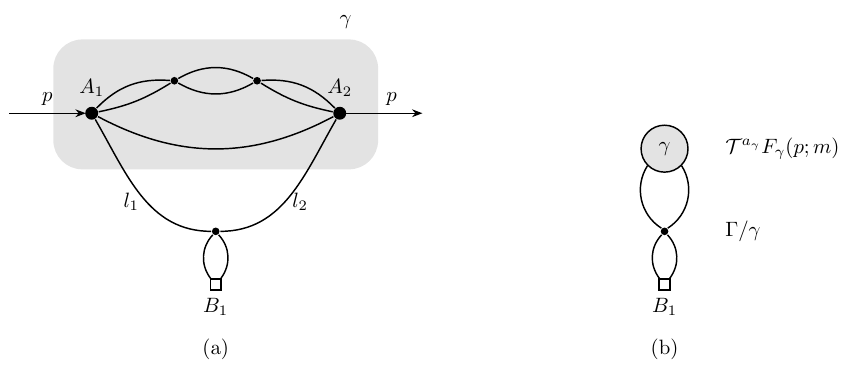}
\caption{(a) A graph which contributes to $Q$. The operators $A_1$ and $A_2$ carry the momentum $p$, the small dots are quartic vertices, and the square is an operator $B_1$ at zero momentum, here the mass operator. The shaded region is one of the subgraphs $\gamma$ in the sum \eqref{eq:irsmirnov}. It contains the two external vertices, and $n=2$ lines of the graph, $l_1$ and $l_2$, leave it. Other subgraphs in the sum include the whole graph, with $n=0$, and the single line between $A_1$ and $A_2$, with $n=6$. (b) The corresponding term of the expansion \eqref{eq:irsmirnov}. The subgraph $\gamma$ is expanded in the momenta $l_1$, $l_2$ and in the mass, and it is inserted as a vertex into the reduced graph $\Gamma/\gamma$, which depends only on $m$. Since all vertices have an even number of lines, a subgraph $\gamma\neq\Gamma$ cannot be connected to the rest by a single line.}
\label{fig:lemma}
\end{figure}

\begin{proof}
The proof has three steps. We first write $Q$ as a sum of renormalised two-point graphs, then we use the asymptotic expansion of such graphs at large momentum, and finally we count the powers.

\emph{Reduction to two-point graphs.} We expand the correlation function in \eqref{eq:irQ} in the couplings. Each order is a sum of Feynman graphs. Their vertices are the operators $A_1$, $A_2$ and $B_a$ and the quartic vertices of the interaction, and their lines are massive propagators. The Fourier transform in \eqref{eq:irQ} assigns the momentum $p$ to $A_1$, $-p$ to $A_2$ and $q_a$ to $B_a$. An operator at zero momentum is a vertex without external momentum. A graph of $Q$ is therefore a graph with two external vertices, which carry the momenta $p$ and $-p$, and with internal vertices of several kinds, as in figure~\ref{fig:lemma}(a). These are the quartic interaction, the operators at zero momentum, and the vertices which arise from the derivatives with respect to the momenta of these operators. To see the last kind, consider a line with momentum $l$ through which the momentum $q$ of an operator flows. Its derivative with respect to $q$ at $q=0$ is
\begin{equation}
\frac{\pr}{\pr q_\mu}\,\frac{1}{(l+q)^2+m^2}\,\bigg|_{q=0}
= \frac{1}{l^2+m^2}\,\big(-2\,l^\mu\big)\,\frac{1}{l^2+m^2} \, .
\end{equation}
The right-hand side has the form of two lines with the same momentum $l$ which are joined by a vertex with the factor $-2\,l^\mu$. A derivative therefore replaces a line by two lines and a new vertex, which has two lines and a factor linear in the momentum. Each vertex $v$ has an even number of lines $n_v$ and a factor which is a homogeneous polynomial of degree $a_v$ in the momenta, and we call $d_v=n_v+a_v$ its dimension. For the two external vertices $d_v$ is $\Delta_1$ and $\Delta_2$, and for all internal vertices $d_v\leq4$. We denote by $F_\Gamma(p;m)$ the unrenormalised Feynman amplitude of such a graph $\Gamma$ in dimensional regularisation, and by $R$ the $R$-operation of minimal subtraction, which subtracts the divergences of all subgraphs and of the graph itself, so that $RF_\Gamma(p;m)$ is the renormalised amplitude. The derivative with respect to an external momentum commutes with the $R$-operation, because the counterterms are polynomials in the external momenta of the subgraphs. Thus $Q$ is a finite sum of renormalised amplitudes $RF_\Gamma(p;m)$ of such graphs. The correlation functions also contain contact terms, in which two or more of the operators are at the same point. These arise from the counterterms with local couplings. In terms of graphs, a contact term is a graph in which a divergent subgraph that contains several operator vertices is replaced by a single local vertex. This local vertex is the counterterm which the $R$-operation subtracts for that subgraph. The contact terms are therefore contained in $RF_\Gamma$, and no further graphs have to be added.

\emph{Expansion at large momentum.} We first reduce the problem to a single ratio. The function $Q$ depends on the two ratios $p/m$ and $\mu/m$. At a fixed order in the couplings its dependence on $\mu$ is a polynomial in $\log\mu$, of a degree $N$ which is bounded by the order. The reason is that in minimal subtraction $\mu$ enters only through the factors $\mu^\vep$ of the bare couplings, which are multiplied by poles in $\vep$ of bounded order. A polynomial of degree $N$ is determined by its values at $N+1$ points. We choose $N+1$ different constants $c_i$ and obtain, by interpolation in the variable $\log(\mu/m)$,
\begin{equation}
Q(p;m)\big|_{\mu} = \sum_{i=0}^{N}P_i\big(\log(\mu/m)\big)\;Q(p;m)\big|_{\mu=c_im} \, ,
\end{equation}
where the $P_i$ are polynomials of degree $N$ which depend only on the $c_i$. It is therefore sufficient to consider the scales $\mu=c_im$. For these the second ratio is a constant, and dimensional analysis gives $Q=m^Dq_i(p/m)$. We need the growth of $q_i$ for large $p/m$. This is the limit of a large external momentum at fixed masses. For this limit we use the asymptotic expansion of~\cite{Smirnov:1990rz}, eqs.~(3.4) and~(4.3) there. It replaces a renormalised graph by a sum of terms, in each of which a subgraph that carries the large momentum is replaced by a local vertex. This is useful because each term is a product of a factor which depends on $p$ but not on the rest of the graph, and a factor which does not depend on $p$. The dependence on $p$ of each term is then a power which can be read off from the subgraph, and the growth of the whole graph is given by the largest of these powers. The expansion reads
\begin{equation}
RF_\Gamma(p/\rho;m) = \sum_\gamma RF_{\Gamma/\gamma}(m)\circ R^*\,\mathcal{T}^{a_\gamma}F_\gamma(p/\rho;m)
+\text{remainder} \qquad (\rho\to0) \, .
\label{eq:irsmirnov}
\end{equation}
The sum is over the subgraphs $\gamma$ which contain the two external vertices and which are connected and one-particle irreducible after these two vertices are identified. These are the parts of the graph through which the large momentum can flow from $A_1$ to $A_2$. The sum includes $\gamma=\Gamma$, for which the reduced graph is trivial. One example is the shaded subgraph in figure~\ref{fig:lemma}(a), and the corresponding term is shown in figure~\ref{fig:lemma}(b). In general some lines of $\Gamma$ which are attached to vertices of $\gamma$ do not belong to $\gamma$. They become the lines of the reduced graph at the vertex to which $\gamma$ is contracted, and their number $n$ enters the power counting below. For each $\gamma$ we denote by $\Gamma/\gamma$ the reduced graph, which is obtained from $\Gamma$ by contracting $\gamma$ to a single vertex, and by $F_{\Gamma/\gamma}$ its amplitude. The symbol $\circ$ stands for the insertion of the factor to its right into this vertex. The operator $\mathcal{T}^{a}$ acts on the amplitude $F_\gamma$ of the subgraph. This amplitude depends on the large momentum $p$, on the masses of the lines of $\gamma$, and on the momenta which enter $\gamma$ through the lines that connect it to the rest of the graph, such as $l_1$ and $l_2$ in figure~\ref{fig:lemma}(a). The operator $\mathcal{T}^{a}$ replaces $F_\gamma$ by its Taylor expansion to order $a$ in these masses and momenta, which are small compared to $p$. The result is a polynomial in them, with coefficients which depend on $p$. It is the factor which is inserted into the vertex of the reduced graph. The reduced graph has no external momentum and depends only on $m$. The Taylor expansion sets the masses and the small momenta in $\gamma$ to zero, which produces infrared divergences in $\gamma$, and it inserts polynomials into the reduced graph, which produces further ultraviolet divergences there. The two cancel in the sum. The operation $R^*$ of~\cite{Smirnov:1990rz} is the extension of the $R$-operation which subtracts the infrared divergences of $\gamma$ as well as the ultraviolet ones, so that each term is finite separately. The term of Taylor order $j$ is homogeneous of degree $\omega(\gamma)-j$ in $p$ up to powers of $\log|p|$, where $\omega(\gamma)$ is the superficial degree of divergence of $\gamma$. This is the power with which the amplitude of $\gamma$ scales when all its momenta are scaled and the masses are neglected, namely four for each loop, minus two for each line, plus the degree $a_v$ of the polynomial at each vertex. The remainder is smaller than any given power of $\rho$ if the orders $a_\gamma$ are chosen large enough. Therefore $RF_\Gamma$ grows at most as $|p|^{\alpha}$ times a power of $\log|p|$, where $\alpha$ is the maximum of $\omega(\gamma)$ over the subgraphs in the sum.

\emph{Power counting.} Let $\gamma$ be a connected subgraph which contains the two external vertices, with $L$ internal lines and $h$ loops, and let $n$ be the number of lines which connect it to the rest of $\Gamma$. In figure~\ref{fig:lemma}(a) we have $n=2$. Then $\sum_{v\in\gamma}n_v=2L+n$ and
\begin{equation}
\omega(\gamma) = 4h-2L+\sum_{v\in\gamma}a_v = 4-n-\sum_{v\in\gamma}(4-d_v)
\leq \Delta_1+\Delta_2-4-n \, ,
\end{equation}
where we used $d_v\leq4$ for the internal vertices. For $\gamma=\Gamma$ we have $n=0$ and $\omega(\Gamma)=D$. Indeed, in this case the right-hand side is $\Delta_1+\Delta_2-4$ plus $d_v-4$ for each internal vertex. The quartic vertices have $d_v=4$ and do not contribute, and an operator $B_a$ with $j_a$ derivatives contributes $d_a-4-j_a$, because each derivative adds a vertex with $d_v=3$. This is the dimension $D$ of $Q$, which is the sum of the dimensions of the operators minus four for each operator but one, minus the number of derivatives. For $\gamma\neq\Gamma$ the number $n$ is even, because all $n_v$ are even, and $n\neq0$ because $\Gamma$ is connected. Thus $n\geq2$, which gives \eqref{eq:irbound}.
\end{proof}

\paragraph{Inputs.} Besides the framework of~\cite{Jack:2013sha} we used classical properties of massive perturbation theory in Euclidean space, namely analyticity in the momenta and the polynomial dependence on $\log\mu$ at fixed order, and the asymptotic expansion \eqref{eq:irsmirnov}. This expansion is derived in~\cite{Smirnov:1990rz} for graphs with massive lines and polynomial vertex factors, in dimensional renormalisation with minimal subtraction. The estimate of the remainder is proved in detail there for the remainder regarded as a tempered distribution in the momentum, with a renormalisation of analytic type. For the remainder as a function at non-exceptional momenta, and for dimensionally renormalised integrals, it is stated there that the same proof gives the estimate. We use the expansion in this second form, for functions at $p\neq0$. This is Hypothesis~\ref{hyp:expansion}.

\paragraph{Relation to Weinberg's theorem.} The exponent $\alpha$ in Lemma~\ref{lem:irW} is the one which Weinberg's power counting theorem~\cite{Weinberg:1959nj} predicts. According to this theorem an amplitude grows with a large momentum as a power which is the largest degree of divergence among the subgraphs through which this momentum flows. The lemma is this statement for the functions \eqref{eq:irQ}, with the degrees of divergence evaluated. We state and prove it separately because the theorem does not cover our case. The theorem of~\cite{Weinberg:1959nj} is proved for absolutely convergent integrals, whereas our graphs are renormalised. The rule is stated there for renormalised graphs in eq.~(20) of section~V there, with the remark that the proof that the subtractions do not change it is omitted. Our proof obtains it for renormalised graphs from the expansion \eqref{eq:irsmirnov}. It is also noted there that such a subgraph is connected to the rest by at least one or at least two lines, depending on the class of graphs. In our case the even number of fields gives the condition $n\geq2$.\footnote{For renormalised graphs the existence of an expansion in powers and logarithms is proven in~\cite{Bergere:1977ft}, and the leading power is obtained for convergent graphs in~\cite{Bergere:1974zh}. Bounds of this type for the correlation functions of the field in massive $\phi^4$ theory are proven with flow equations in~\cite{Kopper:2001to}. The approach to the massless theory through the dependence on the mass and power counting at large momenta goes back to~\cite{Symanzik:1973vg}.}

\subsection{Massless limit}

In this subsection we remove the mass. We apply Lemma~\ref{lem:irW} three times. First, we show that the first moment and the two-point functions in \eqref{eq:irmassive} have a limit for $m\to0$, and we define the first moment of the massless theory as this limit. Second, we obtain the properties of this moment which were used in section~\ref{sec:ward}. Third, we show that the additional terms \eqref{eq:irrem} vanish in the limit, so that the Ward identity \eqref{eq:localfull} holds in the massless theory. These are the three statements listed at the beginning of the appendix.

To show that a function $f(m)$ has a limit for $m\to0$ we bound its derivative with respect to $\log m$. If this derivative vanishes as a positive power of $m$, up to logarithms, then it is integrable with respect to $dm/m$ at $m=0$, and the limit exists. The lemma applies to this derivative for the following reason. The derivative of a correlation function with respect to $m^2$, at fixed $\mu$ and fixed couplings, is the correlation function with an additional mass operator at zero momentum, because $m^2$ is the constant value of the source $\mathsf{M}$. It is therefore again a function of the form \eqref{eq:irQ}, and its dimension is lower by two.

We apply this to the Fourier transform of \eqref{eq:irmm}, which we denote by $m^\mu_{K;IJ}(p;m)$ and which has $D=3$. Note that the moment \eqref{eq:irmm} is taken about the midpoint, while in \eqref{eq:irQ} the operator $A_2$ sits at the origin. The two differ in the routing of the momentum $q$ of the third operator, half of which flows to $A_1$ for the midpoint. The proof of the lemma does not depend on this routing, because a derivative with respect to $q$ lowers the dimension of a vertex by one in either case. Its derivative with respect to $\log m$ is $2m^2$ times an amplitude with $D=1$, and the lemma with $\Delta_1=\Delta_2=4$ gives $\alpha=2$. Therefore
\begin{equation}
\left|\,m\,\frac{\pr}{\pr m}\,m^\mu_{K;IJ}(p;m)\right|
\leq C\,m\,|p|^2\,\big(1+\log(|p|/m)\big)^N\big(1+|\log(m/\mu)|\big)^N \, .
\label{eq:irder}
\end{equation}
The right-hand side is integrable with respect to $dm/m$ at $m=0$. Hence the limit $m\to0$ exists for $p\neq0$. For $|p|<m$ the moment and its derivative with respect to $\log m$ are bounded by $m^3$ up to logarithms, by dimensional analysis and continuity. For a Schwartz function $f$ we split the integral over $p$ at $|p|=m$ and find
\begin{equation}
\left|\int d^4p\;f(p)\;m\,\frac{\pr}{\pr m}\,m^\mu_{K;IJ}(p;m)\right|
\leq C_f\left(m+m^7\right)\big(1+|\log(m/\mu)|\big)^{2N} \, ,
\end{equation}
where the first term comes from $|p|\geq m$ and \eqref{eq:irder}, and the second from $|p|<m$. This is integrable with respect to $dm/m$ at $m=0$, and so the limit also exists in the sense of tempered distributions. In the same way the two-point functions $F_{IJ}$, $J^\mu_J$ and $e_J$ have limits. Their derivatives with respect to $\log m$ are bounded by $m^2|p|^2$, $m^2|p|$ and $m^2$ up to logarithms. Note that the number of lines $n\geq2$ is what makes \eqref{eq:irder} vanish with $m$. A subgraph with $n=1$ would give a logarithmic divergence of the limit, and it would correspond to an operator of dimension one in the product of the two marginal operators.

We define the first moment of the massless theory as the limit,
\begin{equation}
m^\mu_{K;IJ}(u) = \lim_{m\to0}\,m^\mu_{K;IJ}(u;m) \, .
\label{eq:irdef}
\end{equation}
It is antisymmetric in $I$ and $J$, because \eqref{eq:momentanti} holds for each $m$. Its Fourier transform is a rotation-covariant vector of dimension three which depends only on $p$ and is a polynomial in $\log\mu$. It is therefore $p^\mu p^2$ times a polynomial in $\log(p^2/\mu^2)$, and $m^\mu_{K;IJ}(u)$ is a tempered distribution which is equal to $u^\mu h_{K;IJ}(u^2)$ for $u\neq0$.

Finally the remainder \eqref{eq:irrem} vanishes in the limit. The moment $m^\mu_{\phi^2;IJ}$ has $D=1$ and $\alpha=2$, so that the first term is of order $m\,|p|^2$ up to logarithms. The second term is $m^2$ times a function which has a limit, and the third is of order $m^2$ up to logarithms. The limit of \eqref{eq:irmassive} is the Ward identity \eqref{eq:localfull} with the first moment \eqref{eq:irdef}, as an identity of tempered distributions in $u$.

The same argument applies in the scheme with $\theta_I=0$ of section~\ref{sec:scheme}. In this scheme the generating functional is $W'[g,a,\mathsf{M}']=W[g,a,\mathsf{M}'-f]$ with the function $f$ of appendix~\ref{app:mass}, and Lemma~\ref{lem:irW} was proved for the correlation functions of minimal subtraction. The derivatives of $W'$ with respect to the couplings at fixed $\mathsf{M}'$ differ from those of $W$ by terms which contain derivatives of $f$. In these terms one or two of the marginal operators are replaced by the mass operator, which is multiplied by $d_I$ or its derivatives and by a polynomial of second order in the momentum, or two operators are replaced by a contact term which multiplies a correlation function with one operator less. Each of these terms is a regular function of the couplings times a polynomial in the momenta times a function of the form \eqref{eq:irQ} in minimal subtraction, with even operators of dimension at most four. The lemma therefore applies to each of them, and the limits exist as before. The one-point function of the mass operator, which also appears, is of order $m^2$ up to logarithms and vanishes in the limit.

Note that Theorem~\ref{thm:main} uses the first moment only through the definition \eqref{eq:irdef} as a limit. It does not require that this limit is equal to the integral \eqref{eq:moment} evaluated directly in the massless theory with a cutoff at large distance, and we have not shown this. The checks of the Ward identity in appendix~\ref{app:checks} do assume it, because there we evaluate the moment directly in the massless theory, which is simpler.

\section{Comparison with the gradient property}
\label{app:equiv}

The proof of Theorem~\ref{thm:main} establishes two statements about the antisymmetric part $T_{[IJ]}$. The first is that it has the form \eqref{eq:Tdecomp}. The second is the gradient property, which follows from the first. In this appendix we ask whether the second statement also implies the first. If it does not, the form \eqref{eq:Tdecomp} contains information beyond the gradient property.

To compare the two statements we regard them as conditions on an arbitrary antisymmetric tensor $\Omega_{IJ}(g)$, of which $T_{[IJ]}$ is a particular case. The starting point is the relation \eqref{eq:introT}, which with $\Omega_{IJ}$ in place of $T_{[IJ]}$ reads $\pr_IA=(G_{IJ}+\Omega_{IJ})B^J$. The gradient property states that there are a regular symmetric tensor $G'_{IJ}$ and a regular function $A'$ with $\pr_IA'=G'_{IJ}B^J$. Subtracting the two relations we find
\begin{equation}
\Omega_{IJ}B^J = \tilde S_{IJ}B^J+\pr_Ia \, , \qquad \tilde S_{IJ}=G'_{IJ}-G_{IJ} \, , \qquad a=A-A' \, .
\label{eq:grem}
\end{equation}
Conversely, if this equation holds for some regular symmetric $\tilde S_{IJ}$ and some regular function $a$, then $G'_{IJ}=G_{IJ}+\tilde S_{IJ}$ and $A'=A-a$ satisfy the gradient property. The gradient property is therefore equivalent to the condition that the one-form $\Omega_{IJ}B^J$ is the sum of a symmetric tensor contracted with $B^J$ and a gradient. This is the form in which we use it in this appendix, because it is a condition on $\Omega_{IJ}$ alone. The form \eqref{eq:Tdecomp} is a sum of the three removable types of section~\ref{sec:metric},
\begin{equation}
\Omega_{IJ} = \tau_{IJK}B^K+\hat\rho_{[I}\cdot X_{J]}+\left(M^TC-CM\right)_{IJ} \, ,
\label{eq:rform}
\end{equation}
with regular $C_{IJ}=C_{JI}$, $\tau_{IJK}=-\tau_{JIK}$ and $X_J$. The form \eqref{eq:rform} implies \eqref{eq:grem} by the identities of section~\ref{sec:metric}.

We compare the two conditions in two ways. First, we compare them order by order in the expansion around the free theory, through six loops. At a given degree in the couplings the antisymmetric tensors $\Omega_{IJ}$ form a vector space of finite dimension. Each of the two conditions states that $\Omega_{IJ}$ lies in a linear subspace, namely the subspace of tensors for which suitable $\tilde S_{IJ}$ and $a$ exist, or suitable $C_{IJ}$, $\tau_{IJK}$ and $X_J$. The number of independent linear conditions which define a subspace is its codimension. We compute this number for both conditions. Since \eqref{eq:rform} implies \eqref{eq:grem}, the two subspaces are equal if and only if the two numbers agree, and a difference would count the additional relations which the form imposes. Second, we show that the two conditions are equivalent near a zero of $B^I$ at which the stability matrix is invertible. We restrict the comparison to the tensors which can occur in $T_{[IJ]}$, as described below. Note that for an arbitrary antisymmetric tensor the two conditions are not equivalent in the expansion around the free theory. A proof of equivalence to all orders would therefore have to use further properties of $T_{[IJ]}$, and we do not have one.

\paragraph{Tensors and graphs.} To count conditions we need an explicit basis of the space of antisymmetric tensors $\Omega_{IJ}$ of a given degree. We obtain it from graphs. The theory has $N$ scalar fields $\phi_i$, and the couplings are the components of the symmetric tensor $\lambda_{ijkl}$, so that an index $I$ of a coupling stands for four field indices. In perturbation theory every quantity on the space of couplings, such as the beta function or $T_{IJ}$, is a sum of Feynman graphs. It is therefore a polynomial in $\lambda_{ijkl}$ in which the field indices are contracted in pairs with Kronecker deltas, with numerical coefficients which do not depend on $N$. The one-loop beta function $\lambda_{ijmn}\lambda_{mnkl}$ plus permutations is an example. We consider the tensors $\Omega_{IJ}$ of this kind, because $T_{[IJ]}$ is one of them. A term of degree $n$ of $\Omega_{IJ}$ therefore consists of $n$ factors of $\lambda$ and of the two sets of four indices $I$ and $J$, with a rule which says which field index is contracted with which. Such a rule is a graph. Each factor of $\lambda$ is a vertex with four lines, the index sets $I$ and $J$ are two marked vertices with four lines, and each contraction is a line. Since $\lambda_{ijkl}$ is totally symmetric, two rules give the same tensor if and only if the graphs are the same. The vacuum graphs with $n$ vertices and two marked vertices, antisymmetrised under the exchange of the marked vertices, are therefore a basis of the antisymmetric tensors of degree $n$, for a general number of fields. The enumeration of the basis is then the enumeration of graphs, which we do by computer.

The graphs are also the natural language for the restriction which we need. We call a vertex of a connected graph a cut vertex if its removal disconnects the graph. We use the following property, which is Proposition~\ref{prop:cut} and is proved in appendix~\ref{app:cut}. In minimal subtraction the tensor $T_{[IJ]}$ has no component along graphs which are disconnected or have a cut vertex. The reason is that the coefficients of $T_{[IJ]}$ are counterterms of Feynman graphs, and the same graphs label the tensors.

\paragraph{Counts.} We restrict $\Omega_{IJ}$ to the graphs which are connected and have no cut vertex. The spaces of such antisymmetric two-forms of degree 3, 4 and 5 have dimension 8, 45 and 336. Both \eqref{eq:grem} and \eqref{eq:rform} are sets of linear conditions on the coefficients of $\Omega_{IJ}$ up to a given degree, and we determine the number of independent conditions by linear algebra. For \eqref{eq:grem} we impose the equation on the one-forms through degree $n+2$, with $\tilde S_{IJ}$ of degree at most $n$. For \eqref{eq:rform} we take generic $C_{IJ}$, $\tau_{IJK}$ and $X_J$ and a generic $\hat\rho_I$ which satisfies $\hat\rho_IB^I=0$. We used the beta function of minimal subtraction through six loops~\cite{Bednyakov:2021ojn,Pannell:2024sia} and $B^I=\beta^I-(\xi g)^I$ with the five- and six-loop terms of $\xi$. The rotation $\xi$ has four coefficients at five loops and 19 at six loops. For the four five-loop coefficients and for eight of the six-loop coefficients we used the values of section~\ref{app:sixloop}. The remaining eleven six-loop coefficients are not known. They enter the condition \eqref{eq:grem} at degree 5 through the term of highest degree of $B^I$, and they do not enter the other entries of the table. The counts are ranks of linear systems whose entries depend polynomially on these coefficients. Such a rank has the same value for all values of the coefficients outside a set of measure zero. For special values it could be different, which would create or remove a linear dependence among the conditions. We computed the counts with random numbers for the eleven unknown coefficients, which gives the value for generic coefficients. We obtain the same counts when the eleven coefficients are set to zero, and when each of the 19 six-loop coefficients is changed separately. We cannot exclude that the true values are special, but we see no sign of it. The entries which depend on this assumption are marked in the table. The results are given in table~\ref{tab:counts}. The rows without the current term refer to the span of the first two terms of \eqref{eq:rform}, with $M=\pr B$.

\begin{table}[t]
\centering
\begin{tabular}{lccc}
 & degree $\leq3$ & degree $\leq4$ & degree $\leq5$ \\
\hline
dimension of the space & 8 & 53 & 389 \\
gradient property \eqref{eq:grem} & 0 & 9 & 95$^*$ \\
form \eqref{eq:rform} & 0 & 9 & 95 \\
form \eqref{eq:rform} without the current term & 0 & 13 & 108 \\
\end{tabular}
\caption{Number of independent linear conditions on an antisymmetric tensor $\Omega_{IJ}$ in multiscalar $\phi^4$ theory, with $B^I=\beta^I-(\xi g)^I$ through six loops. The tensor is restricted to the subspace of graphs which are connected and have no cut vertex, to which $T_{[IJ]}$ belongs. The entry with an asterisk depends on eleven coefficients of $\xi$ at six loops which are not known. We give the value for generic coefficients, as explained in the text.}
\label{tab:counts}
\end{table}

To read the table, compare the rows for the gradient property and for the form. They agree, with 9 conditions through degree 4 and 95 through degree 5. Since the form implies the gradient property, equal numbers mean that the two conditions select the same tensors on this subspace.\footnote{We also computed the counts on the space of all antisymmetric tensors, including graphs which are disconnected or have a cut vertex. There the form imposes more conditions than the gradient property, namely 13 instead of 12 through degree 4 and 121 instead of 118 through degree 5, and the additional conditions concern only these graphs. For an arbitrary tensor the two conditions are therefore not equivalent. In another mass-independent scheme $T_{[IJ]}$ may contain such graphs. This does not lead to new relations. The tensor of the new scheme is obtained from the tensor of minimal subtraction by a change of coordinates on the space of couplings, and a tensor of the form \eqref{eq:rform} keeps this form under a change of coordinates. It therefore satisfies the additional conditions automatically.}

Hence through degree 4, which corresponds to five loops, the form \eqref{eq:rform} implies no relation beyond the gradient property. At degree 5, which corresponds to six loops, the same holds for generic values of the eleven unknown coefficients. If their true values were special, the number of conditions of the gradient property on this subspace could differ from 95, and the form would then imply additional relations. Without the current term there are 4 and 13 further conditions on this subspace.

The field rotation does not change the counts. We find the same numbers with $\xi=0$, with the values of section~\ref{app:sixloop}, and with random coefficients for all four five-loop and all 19 six-loop graphs. The counts without the current term are also the same if $B^I$ is replaced by $\beta^I$ in $M$. The rotation enters \eqref{eq:grem} through the products of $(\xi g)^I$ with the terms of degree 0 and 1 of $\tilde S_{IJ}$, and it enters \eqref{eq:rform} at degree 5 through the term of degree 0 of $C_{IJ}$. With $\xi=0$ or with random $\xi$ the vector field is not a gradient at six loops, while with the values of section~\ref{app:sixloop} it is. Thus the counts do not depend on this property.

We obtained the counts with modular arithmetic. The values $\zeta_3$, $\zeta_5$, $\zeta_7$, $\zeta_9$, $\zeta_{3,5}$ and $\pi$ in the coefficients were replaced by independent random residues, and we used the primes $2^{31}-1$, $10^9+7$ and $998244353$ with two random points each. All six evaluations agree. We checked the data for the beta function by evaluating the graphs for the $O(N)$ symmetric coupling, which reproduces the beta function of~\cite{Kompaniets:2017yct} as a polynomial in $N$ at each order through six loops. This tests linear combinations of the coefficients and not each coefficient separately. We checked the implementation of $\xi$ in the same linear system. The equation $\tilde S_{IJ}B^J=\pr_Ia$ with $\tilde S_{IJ}=\delta_{IJ}$ at lowest order has a solution through six loops for the values of section~\ref{app:sixloop}. It has no solution if one of the four five-loop coefficients or one of the eight six-loop coefficients which enter \eqref{eq:R3} and \eqref{eq:R4} is shifted, or if the overall normalisation of $\xi$ is changed. It is not affected by the remaining eleven coefficients. This agrees with~\cite{Pannell:2024sia}. The counts for \eqref{eq:rform} through degree 4, with and without the current term, were also confirmed by numerical evaluation with explicit tensors at $N=3$. The counts with the current term do not depend on the higher orders of $\hat\rho_I$, as long as $\hat\rho_IB^I=0$ holds.

\paragraph{Fixed points with invertible stability matrix.} The counts above concern the expansion around the free theory. The free theory is a zero of the vector field $B^I$ at which the matrix of its first derivatives also vanishes, since $B^I$ starts at second order in the couplings. In a different situation we can prove that the gradient property \eqref{eq:grem} implies the form \eqref{eq:rform}, so that the two conditions are equivalent. This is the expansion around a point $g_*$ where $B^I$ vanishes and the stability matrix $M^I{}_J=\pr_JB^I$ is invertible, in the absence of field rotations which act on the couplings. All functions are power series around $g_*$, and \eqref{eq:rform} is taken without the last term. The components $B^I$ are coordinates near $g_*$, because $M$ is invertible. By \eqref{eq:grem} the derivatives of $a$ vanish at $g_*$. A function whose first derivatives vanish at the origin of a coordinate system is a constant plus a quadratic form in the coordinates with coefficients which are power series, so that
\begin{equation}
a = a(g_*)+B^IC_{IJ}B^J
\end{equation}
with a symmetric $C_{IJ}(g)$. Then $\pr_Ia=2\,(M^TCB)_I+B^J\pr_IC_{JK}B^K$, and \eqref{eq:grem} becomes
\begin{equation}
\Xi_{IJ}B^J = 0 \, , \qquad \Xi_{IJ} = \Omega_{IJ}-\tilde S_{IJ}-2\,(M^TC)_{IJ}-B^K\pr_IC_{KJ} \, .
\end{equation}
For fixed $I$ this states that a vector of functions is orthogonal to the vector of coordinates. Such a vector is of the form $\Xi_{IJ}=\Xi_{IJK}B^K$ with $\Xi_{IJK}=-\Xi_{IKJ}$. Taking the antisymmetric part in $I$ and $J$ we find
\begin{equation}
\Omega_{IJ} = \left(M^TC-CM\right)_{IJ}
+\left(\tfrac12\,\pr_IC_{KJ}-\tfrac12\,\pr_JC_{KI}+\Xi_{[IJ]K}\right)B^K \, ,
\end{equation}
which is \eqref{eq:rform}.

This proof does not extend to the expansion around the free theory. There the stability matrix vanishes and a function whose gradient vanishes with $B^I$ need not be quadratic in $B^I$, as the example $B=g^2$, $a=g^3$ with one coupling shows. With field rotations the components $B^I$ also satisfy the linear relations $\hat\rho_IB^I=0$, and a matrix which annihilates $B^I$ can contain terms of the form $X_I\cdot\hat\rho_J$, which give the second removable type. Whether there are further matrices of this kind at higher degree is not known to us.

\section{Counterterms of graphs with a cut vertex}
\label{app:cut}

The result of this appendix is not needed for Theorem~\ref{thm:main}. It is used only in appendix~\ref{app:equiv}, where we investigate whether the form \eqref{eq:Tdecomp} of $T_{[IJ]}$ implies relations beyond the gradient property.

We show that in minimal subtraction the antisymmetric part of the tensor $T_{IJ}$ in \eqref{eq:Tfull} has no component along graphs which are disconnected or have a cut vertex. We call a vertex of a connected graph a cut vertex if its removal disconnects the graph, and we call a graph 2-connected if it is connected and has no cut vertex. The argument has three steps, which correspond to the three parts of this appendix.

The tensor $T_{[IJ]}$ is built from the coefficients of the Weyl variation and of the anomaly, such as $E_{IJ}$, $\hat\rho_I$ and $\theta_I$. In minimal subtraction each of these coefficients is a combination of residues of simple poles of counterterms. The first step is a product formula for the counterterm of a graph with a cut vertex. It shows that such a counterterm is a product of two series of poles, so that it starts with a double pole and has no simple pole. For scalar graphs in momentum space this formula is known~\cite{Herzog:2017bjx}, and we give the proof for the graphs with local sources which occur here. In the second step we go through the coefficients one by one and conclude that each of them contains only 2-connected graphs, with one exception. The exception is a part of the coefficient $\theta_I$ whose graphs have a cut vertex. A third step is needed because $T_{[IJ]}$ is not linear in the coefficients. It contains products of two coefficients, which correspond to gluing two graphs. We show that gluing two 2-connected graphs gives a 2-connected graph. Finally, the exceptional part of $\theta_I$ enters $T_{IJ}$ only through a term which is symmetric in $I$ and $J$, so that it does not contribute to the antisymmetric part $T_{[IJ]}$.

\paragraph{Product formula.} We work in position space, with sources which differ from constants by test functions. Let $\Gamma$ be a connected vacuum graph with source vertices which contributes to the field independent counterterms, and let $v$ be a cut vertex of $\Gamma$. Then $\Gamma$ consists of two parts $\Gamma_1$ and $\Gamma_2$ which share only $v$. Every other vertex has an even number of lines, four for a vertex with a coupling and two for a vertex with $a_\mu$ or $\mathsf{M}$. Therefore each part has an even number of lines at $v$, and hence two. We denote by $\Gamma_a(x)$ the part $\Gamma_a$ with the vertex $v$ at the point $x$ and with all other vertices integrated. This is a graph of the one-point function of a bilinear $\phi_i\phi_j(x)$ in the presence of the sources, and
\begin{equation}
\Gamma = \int d^dx\;\Gamma_1(x)\,\Gamma_2(x) \, ,
\label{eq:cutfactor}
\end{equation}
where the indices of the two bilinears are contracted with the coupling at $v$.

Let $K$ denote the pole part in $\vep$ and $\bar R$ the subtraction of all proper subdivergences. The counterterm of a graph $\gamma$ is $\Delta(\gamma)=-K\bar R\,\gamma$, and $R\,\gamma=\bar R\,\gamma+\Delta(\gamma)$ is finite. The counterterm is a local functional of the sources. For $\Gamma_a(x)$ it is a scalar of dimension two at the point $x$, which is a combination of $D^2g^I$, $D_\mu g^ID^\mu g^J$ and $\mathsf{M}$ with coefficients which contain only poles in $\vep$. Since the sources are test functions, no external momentum is set to zero and no infrared divergence occurs. This is the infrared prescription which we use. In momentum space the same counterterms are obtained with infrared rearrangement if the infrared divergences are subtracted as well, as in the $R^*$ operation~\cite{Herzog:2017bjx}. In particular $\Delta(\Gamma_a)$ is not the pole of the integral $\Gamma_a$ at vanishing external momenta. A scaleless integral vanishes in dimensional regularisation, but its ultraviolet counterterm need not vanish.

\begin{lemma}
\label{lem:cutproduct}
In minimal subtraction the counterterm of a vacuum graph with a cut vertex is
\begin{equation}
\Delta(\Gamma) = \int d^dx\;\Delta(\Gamma_1)(x)\,\Delta(\Gamma_2)(x) \, .
\label{eq:cutproduct}
\end{equation}
\end{lemma}

\begin{proof}
We argue by induction on the number of lines. A connected one-particle irreducible subgraph of $\Gamma$ lies in $\Gamma_1$ or in $\Gamma_2$, or it is the union of a subgraph $\gamma_1$ of $\Gamma_1$ and a subgraph $\gamma_2$ of $\Gamma_2$ which share only $v$. For a proper subgraph of the last kind the counterterm is $\Delta(\gamma_1)\Delta(\gamma_2)$ by the induction hypothesis. The subtractions in $\bar R\,\Gamma$ are labelled by the sets of disjoint divergent proper subgraphs of $\Gamma$. Each such set is a pair of a set for $\Gamma_1$ and a set for $\Gamma_2$, where the two elements which contain $v$, if both are present, are joined to one subgraph. Here we allow the sets for $\Gamma_a$ to contain $\Gamma_a$ itself. The only pair which does not correspond to a proper subgraph of $\Gamma$ is the pair of $\Gamma_1$ and $\Gamma_2$. Summing over the pairs we therefore obtain
\begin{equation}
\bar R\,\Gamma = \int d^dx\,\left[R\,\Gamma_1(x)\,R\,\Gamma_2(x)-\Delta(\Gamma_1)(x)\,\Delta(\Gamma_2)(x)\right] .
\label{eq:cutRbar}
\end{equation}
The functions $R\,\Gamma_a(x)$ are finite and smooth, so that the first term has no pole. The second term is a product of two series which contain only poles, and in minimal subtraction $K$ acts on such a product as the identity. Hence $\Delta(\Gamma)=-K\bar R\,\Gamma$ is given by \eqref{eq:cutproduct}.
\end{proof}

For scalar graphs the product formula is eq.~(4.11) of~\cite{Herzog:2017bjx}, which is proven there by the same induction. The algebraic fact behind it is that the counterterm map of minimal subtraction is multiplicative on products of graphs, because the pole part $K$ satisfies $K(K(A)\,K(B))=K(A)\,K(B)$. For disjoint unions of graphs this is shown in~\cite{Connes:1999yr}, section~3, from the property (23) there. It is pointed out in~\cite{Herzog:2017bjx} that the formula fails for tensor graphs, in which indices are contracted between the two factors. Here both factors are Lorentz scalars, and the counterterms depend on the momenta of the sources only through the derivatives in $D^2g^I$ and $D_\mu g^ID^\mu g^J$.

We use three consequences of Lemma~\ref{lem:cutproduct}. First, a product of two series which contain only poles starts at order $\vep^{-2}$, so that $\Delta(\Gamma)$ has no simple pole. Second, if $\Gamma_2$ contains no source vertex, then $\Delta(\Gamma_2)(x)$ is a constant of dimension two. There is no such constant in the massless theory, and thus $\Delta(\Gamma_2)$ and $\Delta(\Gamma)$ vanish. Note that this follows from locality and dimensional analysis and not from the vanishing of the integral. Third, the same factorisation holds for a one-particle irreducible graph with two external lines of $\phi$, if the cut vertex $v$ separates a part $\Gamma_2$ without external lines. The proof is unchanged. If $\Gamma_1$ contains a loop, the counterterm has no simple pole. If $\Gamma_1$ is the vertex $v$ alone, with a coupling and the two external lines, we find instead
\begin{equation}
\Delta(\Gamma)_{ij} = \lambda_{ijkl}\,\Delta(\Gamma_2)_{kl} \, ,
\label{eq:cutvertexleg}
\end{equation}
which can have a simple pole. Here $\Gamma_2(x)$ is a vacuum graph in which the vertex at $x$ is a vertex of the mass source.

\paragraph{Residues.} In minimal subtraction the coefficients of the consistency conditions are linear in the residues of the simple poles of the counterterms~\cite{Jack:2013sha}. The coefficients $E_{IJ}$, $R_{IJK}$, $L_I$ and $Q_I$ are obtained from the field independent counterterms with four derivatives, with $\mathsf{M}$ and two derivatives, and with $f_{\mu\nu}$ and two derivatives. The beta function, $\xi$, $\hat\rho_I$ and $\theta_I$ are obtained from the counterterms with four and with two fields. The linear relations contain the residues, their derivatives with respect to the couplings and their contractions with $g^I$. These operations mark a vertex or remove a mark and do not change the graph. Hence each coefficient is a combination of the graphs whose counterterms have a simple pole.

For the coefficients with field indices we need the following notion. A coefficient with values in the Lie algebra or in $V_{\mathsf M}$ corresponds to graphs with two legs, and the beta function to graphs with four legs. We define the closed graph by joining the legs at one additional vertex. For $L_I$ and $Q_I$ this vertex is the vertex of $\mathsf{M}$ or $a_\mu$ of the vacuum graph. We call a graph 2-connected if it is connected and has no cut vertex. The consequences of \eqref{eq:cutproduct} then give the following. The graphs of $E_{IJ}$ and $R_{IJK}$ and the closed graphs of $L_I$ and $Q_I$ are 2-connected by the first consequence. The closed graphs of $\beta^I$ are 2-connected by the second consequence, since a part without legs contains no source at constant couplings. The closed graphs of $\xi$ and $\hat\rho_I$ are 2-connected, because a part without legs gives a counterterm with two derivatives of the couplings and no derivative of the fields, which does not contribute to the counterterm $N_I$ of section~\ref{app:sixloop}. For $\theta_I$ the third consequence leaves the graphs \eqref{eq:cutvertexleg}. Their residues are those of the vacuum graphs with one vertex of $\mathsf{M}$, which give $L_I$ by the same linear relation that gives $\theta_I$ from the counterterms with two fields~\cite{Jack:2013sha}. Thus
\begin{equation}
\theta_I = \bar\theta_I+n\,\lambda\,L_I \, , \qquad (\lambda\,L_I)_{ij} = \lambda_{ijkl}\,L_{I,kl} \, ,
\label{eq:thetacut}
\end{equation}
where the closed graphs of $\bar\theta_I$ are 2-connected and $n$ is a number which depends on the normalisation of $L_I$. The second term has both legs at one vertex, which is a cut vertex of the closed graph.

\paragraph{Products.} The tensor $T_{[IJ]}$ is not linear in the residues. It contains products of two coefficients, which correspond to two operations on graphs. In the first a vertex $K$ of a graph is replaced by a graph with four legs, as in $\pr_JB^K E_{IK}$ and $R_{IJK}B^K$. In the second two graphs with two legs are joined leg to leg, as in $\hat\rho_J\cdot Q_I$ and $L_J\cdot\theta_I$.

\begin{lemma}
\label{lem:cutglue}
Let $G$ be a 2-connected graph and $H$ a connected graph with four legs whose closed graph is 2-connected. Then the graph obtained by replacing a vertex of $G$ by $H$ is 2-connected. Let $H_1$ and $H_2$ be graphs with two legs whose closed graphs are 2-connected. Then the graph obtained by joining them leg to leg is 2-connected.
\end{lemma}

\begin{proof}
For the first statement let $K$ be the vertex which is replaced. We remove a vertex $w$ from the new graph. If $w$ belongs to $G$, each component of $G$ without $K$ and $w$ is joined to $H$ by a line which ended at $K$, because $G$ without $w$ is connected, and $H$ is connected. If $w$ belongs to $H$, each component of $H$ without $w$ contains a leg, because the closed graph of $H$ without $w$ is connected. It is therefore joined to $G$ without $K$, which is connected. For the second statement we remove a vertex $w$ of $H_1$. Each component of $H_1$ without $w$ contains a leg and is joined to $H_2$, which is connected.
\end{proof}

The term $\tfrac12(\hat\rho_J\,g)^KE_{IK}$ in \eqref{eq:Ehat} and the term $(\xi g)^K$ of $B^K$ are of the first kind, with $H$ a graph of $\hat\rho_J$ or $\xi$ attached to one leg of a vertex.

\begin{proposition}
\label{prop:cut}
In minimal subtraction the antisymmetric part $T_{[IJ]}$ in \eqref{eq:Tfull} is a combination of 2-connected graphs. The same holds in the scheme \eqref{eq:theta0}.
\end{proposition}

\begin{proof}
By \eqref{eq:Tfull} and \eqref{eq:Ehat} the tensor $T_{[IJ]}$ is the antisymmetric part of a sum of products of the coefficients listed above. By Lemma~\ref{lem:cutglue} each product is a combination of 2-connected graphs, with the exception of the second term of \eqref{eq:thetacut} in $L_J\cdot\theta_I$. This term contributes $n\,L_{J,ij}\,\lambda_{ijkl}\,L_{I,kl}$ to $\hat E_{IJ}$, which is symmetric in $I$ and $J$ and drops out of $\hat E_{[IJ]}$.

The scheme \eqref{eq:theta0} is reached by the redefinition of appendix~\ref{app:mass}. The anomaly of the new scheme is the anomaly of minimal subtraction with $\mathsf{M}$ replaced by $\mathsf{M}'-f$. In the normalisation of \eqref{eq:Ehat} this shifts $E_{IJ}$ by $-\kappa^{-1}L_I\cdot d_J$ and $R_{IJK}$ by $-\kappa^{-1}L_I\cdot e_{JK}$, and it does not change $L_I$, $Q_I$, $B^I$ and $\hat\rho_I$. With $\Delta\theta_I=-\Psi_I{}^Kd_K-e_{IK}B^K$ we find that $\hat E_{IJ}$ changes by
\begin{equation}
-\kappa^{-1}\left[L_I\cdot\left(\Psi_J{}^Kd_K+e_{JK}B^K\right)+L_J\cdot\left(\Psi_I{}^Kd_K+e_{IK}B^K\right)\right] ,
\end{equation}
which is symmetric. Hence $T_{[IJ]}$ is the same tensor in the two schemes.
\end{proof}

Note that the statement does not hold for the individual coefficients. In minimal subtraction $\theta_I$ contains the second term of \eqref{eq:thetacut}, and the symmetric part of $\hat E_{IJ}$ has components along graphs with a cut vertex. In the scheme \eqref{eq:theta0} the same is true of $E_{IJ}$, since $d_J$ contains $\lambda\,L_J$. Only the combination $T_{[IJ]}$ is free of these graphs. We tested Lemma~\ref{lem:cutglue} and the symmetry of $L_J\cdot(\lambda\,L_I)$ on the graphs of low degree with the code which we used for appendix~\ref{app:equiv}.

\section{Consistency checks}
\label{app:checks}

In this appendix we describe the consistency checks which we performed. There are two separate questions. The first is whether the conclusion of Theorem~\ref{thm:main} agrees with explicit results for the beta function. Through five loops the gradient property was known to hold. At six loops it was shown in~\cite{Pannell:2024sia} to be equivalent to four relations among coefficients of the field rotation $\xi$ which had not been computed. Theorem~\ref{thm:main} implies these relations. In section~\ref{app:sixloop} we compute the coefficients and find that the relations hold. This is a test of the theorem as a whole at the lowest order at which the field rotation enters. The second is whether the Ward identity \eqref{eq:localfull}, which is the new ingredient, holds for correlation functions which are known independently. This is the subject of sections~\ref{app:firstorder} to~\ref{app:third}. Finally, in section~\ref{app:twod} we compare our form of the antisymmetric coefficient with the gradient formula which is known in two dimensions. In sections~\ref{app:firstorder} to~\ref{app:third} we consider \eqref{eq:localfull} at separated points, $u\neq0$, where the right-hand side vanishes. In most of the cases which we consider the terms with the currents and with the mass operator are absent or do not contribute. The exception are three structures in section~\ref{app:third}, for which we keep all terms of \eqref{eq:localsep}. In the other cases the identity which we test is
\begin{equation}
\left(M^TF-FM\right)_{IJ}(u)+2\,\beta^Kh_{K;IJ}(u) = 0 \, ,
\label{eq:sep}
\end{equation}
where $F_{IJ}$ is the two-point function of the marginal operators, $M$ is the stability matrix and $h_{K;IJ}$ is the first moment of the three-point function, as in section~\ref{sec:ward}. It relates the two-point function to the three-point function, and both sides can be computed independently.\footnote{In these checks we compute the first moment directly in the massless theory, because this is simpler. Theorem~\ref{thm:main} uses the first moment which is defined as the massless limit \eqref{eq:irdef}. We have not shown that the two agree, and the checks assume it.} Two of these checks are not in four dimensions. They consider a conformal field theory in $d$ dimensions which is perturbed by scalar operators, and a model in three dimensions, because there the correlation functions are known in closed form. This is possible because the derivation of \eqref{eq:sep} in section~\ref{sec:ward} does not use that the dimension is four. The dimension enters that section only through the numbers $4\sigma$ and $8$ in the Callan--Symanzik equation, which become $d\,\sigma$ and $2d$, and it does not appear in \eqref{eq:sep}. In section~\ref{sec:ward} we used that the reflection $u\to-u$ is a rotation. In odd dimensions this is not the case, and there we use in addition that the theory is invariant under parity. These checks therefore test the derivation of the identity at separated points, and not the anomaly term, which is specific to four dimensions.

\subsection{Six-loop relations}
\label{app:sixloop}

In this section we verify the four relations of~\cite{Pannell:2024sia} which are equivalent to the gradient property of $B^I$ at six loops. We first state the relations. They involve four five-loop and eight six-loop coefficients of the field rotation. We then express these coefficients through simple poles of counterterms of propagator graphs, fix the normalisation and the sign, and describe the computation of the counterterms and its validation. The values of the coefficients are given at the end, in table~\ref{tab:fiveloop} and in \eqref{eq:s6}, where we also check the four relations.

\paragraph{Relations.} In the conventions of~\cite{Pannell:2024sia} the rotation in \eqref{eq:B} is an antisymmetric matrix $S_{ij}=\xi_{ij}$, and $B_{ijkl}=\beta_{ijkl}-(S_{im}\lambda_{mjkl}+\text{permutations})$. It has an expansion in antisymmetrised propagator graphs, by which we mean the one-particle irreducible graphs of the two-point function of $\phi$,
\begin{equation}
S_{ij} = \sum_{n\geq5}\sum_m s^{(n)}_m\,\mathcal{S}^{(n)}_{m,ij} \, ,
\end{equation}
where $n$ is the number of loops. A propagator graph contributes only if it is not symmetric under the exchange of its two external legs. There are four such graphs at five loops and 19 at six loops. The gradient property of $\beta^I$ and of $B^I$ holds through five loops. At six loops $B^I$ is a gradient if and only if~\cite{Pannell:2024sia}
\begin{align}
s^{(5)}_2 &= \frac{259}{4608} \, , \label{eq:R1}\\
192\,s^{(5)}_3+384\,s^{(5)}_4 &= 31-36\,\zeta_3 \, , \label{eq:R2}\\
s^{(5)}_1-12\,s^{(5)}_2+s^{(5)}_4 &= 12\,s^{(6)}_8-6\,s^{(6)}_9-6\,s^{(6)}_{16} \, , \label{eq:R3}\\
4\,s^{(5)}_1-2\,s^{(5)}_3 &= -2\,s^{(6)}_3+s^{(6)}_6+4\,s^{(6)}_{10}-2\,s^{(6)}_{15}+s^{(6)}_{17} \, .
\label{eq:R4}
\end{align}
The labels of the graphs are those of~\cite{Pannell:2024sia}. The coefficients were not known, and the relations were stated there as predictions. Theorem~\ref{thm:main} implies that they hold. We computed the twelve coefficients which enter.

\paragraph{Definition of the coefficients.} We first recall where the rotation $\xi$ comes from, in order to identify the diagrams which determine it. A counterterm which is bilinear in the field $\phi$ is determined by the divergences of the one-particle irreducible diagrams with two external lines of $\phi$, as for the wave function renormalisation at constant couplings. These are the propagator graphs introduced above. When the couplings depend on the position, these diagrams have additional divergences which are proportional to $\pr_\mu g^I$. By dimensional analysis and Lorentz invariance the corresponding counterterms have one derivative on a field,
\begin{equation}
K_{I,ij}\;\pr_\mu g^I\,\phi_i\,\pr^\mu\phi_j \, ,
\end{equation}
with a matrix $K_{I,ij}$ of poles in $\vep$. We split this matrix into its symmetric and its antisymmetric part in the field indices. For the symmetric part $\phi_i\,\pr^\mu\phi_j$ can be replaced by $\tfrac12\pr^\mu(\phi_i\phi_j)$. This part is produced by the wave function renormalisation $Z_{ij}(g)$ of the kinetic term when it is evaluated on the local couplings, together with terms without derivatives of the fields, which belong to the bare mass source. It therefore requires no counterterm of a new kind. For the antisymmetric part this is not possible, because $\phi_i\,\pr^\mu\phi_j-\phi_j\,\pr^\mu\phi_i$ is not a total derivative. This combination is the current of the field rotations. A term in which it multiplies $\pr_\mu g^I$ has the form of the coupling of the current to the background field $a_\mu$, and it is absorbed into the bare background field, $a_{0\mu}=a_\mu+N_I\,D_\mu g^I$. Here $N_I$ is the antisymmetric part of $K_I$, a series of poles in $\vep$ with values in the Lie algebra~\cite{Jack:2013sha}. The rotation $\xi$ is the residue of the simple pole of $N_I$ contracted with $g^I$, up to a sign~\cite{Jack:2013sha}.

The diagrams which determine $N_I$ are thus the propagator graphs in which one of the vertices, which we denote by $v$, carries a coupling which depends on the position and injects a momentum $q$. For each propagator graph and each choice of $v$ we need the part of the counterterm which is linear in $q$ and antisymmetric in the field indices of the two legs. The contraction of $N_I$ with $g^I$ in $\xi$ puts the coupling back at $v$ and sums over the choices of $v$. Hence each propagator graph contributes to $\xi_{ij}$ its tensor, antisymmetrised in the two legs, with a coefficient which is a sum over its vertices. This is the expansion in the graphs $\mathcal{S}^{(n)}_{m,ij}$ above, and the coefficient is $s^{(n)}_m$. A graph which is symmetric under the exchange of its legs drops out, which is why the first contributions appear at five loops.

\begin{figure}[t]
\centering
\includegraphics[width=0.75\textwidth]{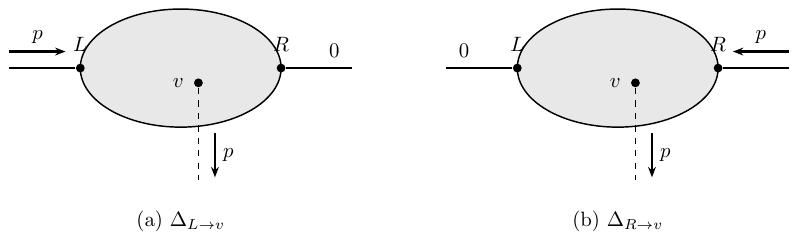}
\caption{The two routings of the external momentum through a propagator graph. The shaded region stands for the graph, $L$ and $R$ are the vertices of the left and the right leg, and $v$ is the internal vertex with the local coupling, which is indicated by the dashed line. (a) The momentum enters at $L$ and leaves at $v$. (b) The momentum enters at $R$ and leaves at $v$.}
\label{fig:routing}
\end{figure}

We now determine the coefficient for a given graph and a given vertex $v$. Let $p_1$ and $p_2$ be the momenta which enter at the two legs and $q$ the momentum which enters at $v$, with $p_1+p_2+q=0$. The counterterm which we need is linear in $q$ and antisymmetric under the exchange of the two legs. By Lorentz invariance it is proportional to $(p_1-p_2)\cdot q$, and with momentum conservation
\begin{equation}
c_v\,(p_1-p_2)\cdot q = -c_v\left(p_1^2-p_2^2\right) ,
\end{equation}
with a coefficient $c_v$ which is a series of poles in $\vep$. We can therefore obtain $c_v$ from two special configurations of the momenta, which are shown in figure~\ref{fig:routing}. For $p_2=0$ the momentum $p=p_1$ enters at the vertex $L$ of the left leg and leaves at $v$, and the counterterm is $-c_v\,p^2$. For $p_1=0$ the momentum $p=p_2$ enters at the vertex $R$ of the right leg and leaves at $v$, and the counterterm is $+c_v\,p^2$. In both configurations no momentum flows through the other leg, so that the diagram is a scalar propagator integral without numerators, with the momentum $p$ routed from $L$ to $v$ or from $R$ to $v$.

We denote by $\Delta_{L\to v}(\Gamma)$ the residue of the simple pole of the counterterm of the propagator graph $\Gamma$ in the configuration of figure~\ref{fig:routing}(a), divided by $p^2$. The counterterm is taken in minimal subtraction, after the subtraction of subdivergences. We define $\Delta_{R\to v}(\Gamma)$ in the same way for the configuration of figure~\ref{fig:routing}(b). The residue of $c_v$ is then proportional to the difference of the two. We fix the factor and the sign in the next paragraph. Summing over the position of the vertex $v$ we obtain the coefficient of the graph $\Gamma_m$ in $S_{ij}$ as
\begin{equation}
s^{(n)}_m = \frac{\eta_n}{\mathrm{sym}_m}\sum_v\left[\Delta_{L\to v}(\Gamma_m)-\Delta_{R\to v}(\Gamma_m)\right] ,
\label{eq:sdef}
\end{equation}
where the sum runs over the internal vertices, $\mathrm{sym}_m$ is the symmetry factor of the graph and $\eta_n=\pm1$.

\paragraph{Normalisation and sign.} The sign $\eta_n$ in \eqref{eq:sdef} follows from the conventions of~\cite{Jack:2013sha}, in which $D_\mu\phi_i=\pr_\mu\phi_i+a_{\mu\,ij}\phi_j$, the bare kinetic term is $\tfrac12(D_{0\mu}\phi_0)^2$ with $a_{0\mu}=a_\mu+N_I\,D_\mu g^I$, and
\begin{equation}
\xi=-N_{I,1}\,g^I \, , \qquad
B_{ijkl}=\beta_{ijkl}-\left(\xi_{im}\lambda_{mjkl}+\text{permutations}\right) ,
\label{eq:upsconv}
\end{equation}
where $N_{I,1}$ is the residue of the simple pole of $N_I$ in $\vep=4-d$. Let $\Pi_{ij}$ be the tensor of a propagator graph with $h$ loops, with the index $i$ on the leg at the vertex $X$ and $j$ on the leg at $Y$. The terms of the bare action which are linear in $\pr_\mu g^I$ multiply $\pr_\mu\phi_i\,\phi_j$ by $\tfrac12\pr_IZ_{ij}+N_{I,ij}$, up to terms without a simple pole. This combination is fixed by the counterterm with the momentum routed from $X$ to the vertex $v$ of the local coupling, and $\tfrac12\pr_IZ_{ij}-N_{I,ij}$ by the routing from $Y$ to $v$. The graph contributes $c\,\Pi_{ij}$ to the anomalous dimension, which equals $-\tfrac{h}{2}$ times the simple pole of $Z_{ij}$~\cite{Jack:2013sha}, and we find $c=(-1)^{h+1}h\,\Delta_{X\to Y}/\mathrm{sym}$ for the 21 propagator structures with two to five loops and for the eight six-loop graphs which we use. The simple pole of $Z_{ij}$ is therefore $2\,(-1)^h\Delta_{X\to Y}/\mathrm{sym}$ times $\Pi_{ij}$. In the same normalisation the difference of the two routings to $v$ gives $2N_{I,ij}$, so that the two factors of 2 cancel. Thus the contribution of the graph to $\xi_{ij}$ is
\begin{equation}
-r\,\Pi_{ij} \, , \qquad
r=\frac{(-1)^{h}}{\mathrm{sym}}\sum_v\left[\Delta_{X\to v}-\Delta_{Y\to v}\right] ,
\label{eq:udef}
\end{equation}
and by \eqref{eq:upsconv} the coefficient of $\Pi_{im}\lambda_{mjkl}$ in $B_{ijkl}$ is $c+r$, while the coefficient of the mirror image is $c-r$. In the ancillary files of~\cite{Pannell:2024sia} the coefficient $s^{(n)}_m$ is added to the coefficient of the graph whose free leg is the left leg of the figure, and subtracted from that of its mirror image. Hence $s^{(n)}_m=r$ with $X=L$, that is $\eta_5=-1$ and $\eta_6=+1$. Note that the text of~\cite{Pannell:2024sia} does not specify the sign of the antisymmetrisation in $\mathcal{S}^{(n)}_{m,ij}$, and we rely on the ancillary files, from which the relations \eqref{eq:R1} to \eqref{eq:R4} were derived.

We checked \eqref{eq:udef} in two ways. First, for the two-loop graph with three parallel lines it reproduces the two-loop results for $a_{0\mu}$ and $\rho_I$ of~\cite{Jack:2013sha}, including the sign. Second, the consistency condition $\hat\rho_IB^I=0$ of~\cite{Jack:2013sha}, with $\hat\rho_I=g^J\left(\pr_JN_{I,1}-\pr_IN_{J,1}\right)$, relates the terms $N_I$ with two to five loops at the order of six loops. We evaluated it for random couplings with the counterterms of all propagator graphs up to five loops and with the beta function up to four loops~\cite{Bednyakov:2021ojn}, and it holds to the numerical precision. If we keep only the two-loop term fixed, the condition determines the overall factors of the terms with three, four and five loops and the four coefficients $s^{(5)}_m$, with the values of table~\ref{tab:fiveloop}. With this the relation \eqref{eq:R1} tests the sign and the magnitude of $s^{(5)}_2$, and none of the four relations contains a free parameter. We did not repeat the second check at seven loops, which would require the remaining six-loop graphs, and the sign $\eta_6$ rests on \eqref{eq:udef} alone.

\paragraph{Computation.} For each graph and each internal vertex $v$ we need the two counterterms $\Delta_{L\to v}$ and $\Delta_{R\to v}$ of figure~\ref{fig:routing}. We computed them with the implementation of the $R^*$ operation of~\cite{Herzog:2017bjx}, which is written in FORM and was provided to us by B.~Ruijl. The method rests on infrared rearrangement. The propagator graphs are quadratically divergent, and their counterterm is proportional to $p^2$. Two derivatives with respect to the external momentum reduce it to counterterms of logarithmically divergent integrals. The counterterm of a logarithmically divergent integral does not depend on the external momenta and masses, so that one may set the external momentum to zero and give a mass to a single line instead. The $n$-loop integral then factorises into a one-loop massive integral and a massless propagator integral with $n-1$ loops, and the $R^*$ operation subtracts the infrared divergences which this introduces. At five loops the remaining integrals have four loops, and the code evaluates them with the program Forcer~\cite{Ruijl:2017cxj}. The four five-loop graphs are listed in table~\ref{tab:fiveloop}.

At six loops the remaining propagator integrals have five loops, which is beyond the reach of Forcer. We avoided them by the choice of the line which is given a mass. Each of the eight graphs which we need contains two vertices which are connected by two or three parallel lines. We chose the massive line among these. The other parallel lines then form a loop between the two vertices which contains no further vertex. The rest of the graph, seen from these two vertices, is a massless propagator integral with at most four loops, which Forcer evaluates and which is a power of its momentum. The remaining integration over the parallel lines is then a one-loop integral with a non-integer power of one propagator, which is a product of Gamma functions. For this we modified the code so that the massive line can be chosen and so that it returns the five-loop terms which it cannot evaluate and which the original code drops, and we evaluated these terms with a separate script. In this way we obtained 80 counterterms, ten for each of the eight graphs. These are the two routings for each of the four internal vertices, and the standard routing between the two legs in both directions, which we use for the validation below. From them we obtained the eight six-loop coefficients which enter \eqref{eq:R3} and \eqref{eq:R4}. We did not compute the eleven six-loop coefficients which do not enter the relations.

\begin{table}[t]
\centering
\begin{tabular}{clcc}
$m$ & lines & $\mathrm{sym}_m$ & $s^{(5)}_m$ \\
\hline
1 & $LT^2,\,LR,\,Ta,\,aR,\,Tb,\,bR,\,ab^2$ & 8 & $-\tfrac{83}{768}+\tfrac{7}{32}\zeta_3$ \\[3pt]
2 & $La,\,ab^3,\,bT,\,TR^2,\,LT,\,LR$ & 12 & $\tfrac{259}{4608}$ \\[3pt]
3 & $LT^2,\,La,\,aR,\,RT,\,Tb,\,bR,\,ab^2$ & 4 & $\tfrac{97}{384}-\tfrac18\zeta_3$ \\[3pt]
4 & $La^2,\,aT^2,\,TR,\,Tb,\,bR^2,\,Lb$ & 8 & $-\tfrac{35}{768}-\tfrac{1}{32}\zeta_3$ \\
\end{tabular}
\caption{The four five-loop propagator graphs which are not symmetric under the exchange of the legs, in the order of~\cite{Pannell:2024sia}, with our results for their coefficients in $S_{ij}$. The vertices $L$ and $R$ carry the external legs, $T$, $a$ and $b$ are internal vertices, and an exponent denotes the number of lines between two vertices.}
\label{tab:fiveloop}
\end{table}

\paragraph{Validation.} We validated the computation on quantities which are known. For each graph we computed, with the same code, the counterterm with the standard routing of the momentum from one leg to the other, which is the counterterm of the ordinary propagator integral. These counterterms are tabulated graph by graph in the ancillary file \texttt{Z.m} of~\cite{Kompaniets:2017yct}, as the entries \texttt{Z2[$\ell$,$j$]} which are described in appendix~A there. At five loops our counterterms of the four graphs agree with these entries pole by pole. From them we also obtain the coefficients with which the four graphs enter the five-loop beta function through the anomalous dimension of the field. They agree with the coefficients of the corresponding terms of the beta function of~\cite{Bednyakov:2021ojn}, which we took from the ancillary files of~\cite{Pannell:2024sia}. At six loops the counterterms of the standard routings of all eight graphs agree with the entries of \texttt{Z.m} pole by pole, including the terms with $\pi^4$.

We also tested the two steps which are specific to our computation. The counterterm does not depend on the choice of the massive line, which we verified at five loops for five choices. The evaluation of the terms beyond the reach of Forcer, described above, reproduces a five-loop term which was computed directly. We should point out that all 80 counterterms come from a single code. The standard routings are validated against the literature. The routings to the internal vertices are validated only by the tests of this paragraph and, at five loops, by the condition $\hat\rho_IB^I=0$ described above.

\paragraph{Results.} The five-loop coefficients are given in table~\ref{tab:fiveloop}. The six-loop coefficients are
\begin{align}
s^{(6)}_3 &= \tfrac{233}{480}-\tfrac{111}{160}\zeta_3+\tfrac{33}{160}\zeta_4 \, , &
s^{(6)}_6 &= -\tfrac{37}{64}+\tfrac{17}{80}\zeta_3+\tfrac{9}{80}\zeta_4 \, , \notag\\
s^{(6)}_8 &= -\tfrac{23}{512}+\tfrac{1}{96}\zeta_3 \, , &
s^{(6)}_9 &= -\tfrac{37}{1024}-\tfrac{1}{96}\zeta_3 \, , \notag\\
s^{(6)}_{10} &= \tfrac{2629}{7680}-\tfrac{89}{320}\zeta_3+\tfrac{9}{320}\zeta_4 \, , &
s^{(6)}_{15} &= \tfrac{23}{256}-\tfrac{9}{80}\zeta_4 \, , \notag\\
s^{(6)}_{16} &= \tfrac{259}{3072} \, , &
s^{(6)}_{17} &= -\tfrac{37}{64}+\tfrac{51}{80}\zeta_3-\tfrac{3}{80}\zeta_4 \, .
\label{eq:s6}
\end{align}
The relation \eqref{eq:R1} holds by table~\ref{tab:fiveloop}. For \eqref{eq:R2} we find
\begin{equation}
192\left(\tfrac{97}{384}-\tfrac18\zeta_3\right)+384\left(-\tfrac{35}{768}-\tfrac{1}{32}\zeta_3\right)
= 31-36\,\zeta_3 \, .
\end{equation}
Both sides of \eqref{eq:R3} are equal to $-\tfrac{53}{64}+\tfrac{3}{16}\zeta_3$, and both sides of \eqref{eq:R4} are equal to $-\tfrac{15}{16}+\tfrac98\zeta_3$. In \eqref{eq:R4} the terms with $\zeta_4$ cancel among the five six-loop coefficients. Thus all four relations hold. Together with the analysis of~\cite{Pannell:2024sia} this shows that $B^I$ is a gradient at six loops in minimal subtraction, as Theorem~\ref{thm:main} requires. This is a test of the conclusion at the first order where it restricts $B^I$.

\subsection{First order near a conformal fixed point}
\label{app:firstorder}

\paragraph{Setup.} We consider a conformal field theory in $d$ dimensions perturbed by scalar operators $\cO_i$ of dimension $\Delta_i=d-\epsilon_i$ with generic $\epsilon_i$, with weight $\exp(-S_{\rm CFT}+g^i\!\int\!\cO_i)$. We test \eqref{eq:sep} at first order in the couplings. In the scheme defined by analytic continuation in the $\epsilon_i$ the beta function is $\beta^i=-\epsilon_ig^i$ at this order, and the stability matrix is diagonal. The ingredients are the two-point function at first order and the first moment of the three-point function at order zero. Both follow from the conformal three-point function
\begin{equation}
\vev{\cO_k(x)\,\cO_i(y)\,\cO_j(z)} = \frac{C_{ijk}}{|u|^{\Delta_i+\Delta_j-\Delta_k}\,|x-y|^{2a}\,|x-z|^{2b}} \, ,
\end{equation}
with $2a=\Delta_k+\Delta_i-\Delta_j$ and $2b=\Delta_k+\Delta_j-\Delta_i$. Hence $a-b=\epsilon_j-\epsilon_i$ and $a+b=d-\epsilon_k$.

\paragraph{Closed formulas.} The integral over $x$ and its first moment about the midpoint can be evaluated with Feynman parameters. We find
\begin{align}
I_0 &= \int d^dx\,\frac{1}{|x-y|^{2a}|x-z|^{2b}}
= \pi^{d/2}\,\frac{\Gamma(\tfrac d2-a)\,\Gamma(\tfrac d2-b)\,\Gamma(a+b-\tfrac d2)}
{\Gamma(a)\,\Gamma(b)\,\Gamma(d-a-b)}\;|u|^{d-2a-2b} \, , \notag\\
I_1^\mu &= \int d^dx\,\frac{(x-c)^\mu}{|x-y|^{2a}|x-z|^{2b}} = \frac{a-b}{2\,(d-a-b)}\;u^\mu\,I_0 \, .
\label{eq:I0I1}
\end{align}
We obtain the ratio as the mean of the Feynman parameter in a Beta distribution. The integrals converge for $2a<d$, $2b<d$ and $2(a+b)>d+1$, and they are defined by analytic continuation otherwise. In terms of the $\epsilon_i$ the first-order two-point function and the moment function are
\begin{equation}
F_{ij} = g^k\int d^dx\,\vev{\cO_k(x)\,\cO_i(y)\,\cO_j(z)} \, , \qquad
h_{k;ij} = \frac{\epsilon_j-\epsilon_i}{2\,\epsilon_k}\int d^dx\,\vev{\cO_k(x)\,\cO_i(y)\,\cO_j(z)} \, .
\label{eq:hfirst}
\end{equation}

\paragraph{Outcome.} With $M=-{\rm diag}(\epsilon_i)$ the first term of \eqref{eq:sep} is $(\epsilon_j-\epsilon_i)F_{ij}$, and with \eqref{eq:hfirst} the second term is $-2\,\epsilon_kg^k\,h_{k;ij}=-(\epsilon_j-\epsilon_i)F_{ij}$. Thus the two terms cancel with the relative coefficient 2 which was derived in section~\ref{sec:ward}, and they do not vanish separately for $\epsilon_i\neq \epsilon_j$. If the coefficient of the three-point term is left free, the identity fixes it to 2. We also verified the Callan--Symanzik equation at this order.

We confirmed \eqref{eq:I0I1} by numerical integration in $d=4$ and $d=5$. For $d=4$ and $(\epsilon_i,\epsilon_j,\epsilon_k)=(0.3,0.7,0.6)$ we find, in units $|u|=1$, $I_0=116.3315$ and $I_1=38.7769$ numerically and $116.3330$ and $38.7777$ from \eqref{eq:I0I1}. For $(0.9,0.35,0.8)$ the values are $92.15206$ and $-31.67726$ against $92.15214$ and $-31.67730$. For $d=5$ and $(0.2,0.9,0.9)$ they are $143.5327$ and $55.8181$ against $143.5348$ and $55.8191$. The relative differences are between $10^{-6}$ and $2\times10^{-5}$, which we attribute to the quadrature. These numbers only confirm the closed formulas \eqref{eq:I0I1} by a direct numerical integration, and the cancellation above does not depend on them. In $d=3$ the numerical integration is less accurate, because the integrable singularities at the positions of the two operators are harder to resolve. There we find agreement with \eqref{eq:I0I1} at the level of $4\times10^{-4}$ to $3\times10^{-3}$ for three choices of the $\epsilon_i$.

\paragraph{Change of scheme.} The test above does not involve a beta function with terms of second order. Such terms arise from a redefinition $g^l=g'^l+A^l{}_{mn}g'^mg'^n$, which gives $\beta'^l=-\epsilon_lg'^l+A^l{}_{mn}(\epsilon_m+\epsilon_n-\epsilon_l)g'^mg'^n$. The operators with local couplings are then $\cO'_i=(\pr g^l/\pr g'^i)\,\cO_l$, and the three-point function acquires the contact terms $2A^l{}_{ki}\,\delta(x-y)F_{lj}+2A^l{}_{kj}\,\delta(x-z)F_{il}$. They sit at $x-c=\pm u/2$ and shift the moment function by $A^l{}_{ki}F_{lj}-A^l{}_{kj}F_{il}$. We find that \eqref{eq:sep} holds in the new scheme, and that the terms which are linear in $A$ cancel only for the coefficient 2 of the three-point term. The contact terms are necessary for this. Finally, at a resonance $\epsilon_j=\epsilon_i+\epsilon_k$ the integral $I_0$ has a pole, which is removed by a redefinition of this type, and the beta function has a term of second order in every scheme. We turn to an example.

\subsection{Resonant case in three dimensions}
\label{app:d3}

\paragraph{Model.} We consider the free scalar field in three dimensions perturbed by a mass term and a quartic interaction, with weight $\exp(-S_0+\int(t\,\cO_t+\lambda\,\cO_\lambda))$, $\cO_t=-\tfrac12\phi^2$ and $\cO_\lambda=-\tfrac{1}{4!}\phi^4$. The couplings have $\epsilon_t=2$ and $\epsilon_\lambda=1$, so that $\epsilon_t=2\,\epsilon_\lambda$ and the dimensions are resonant, as they are for marginal operators. The propagator is $G(x)=1/(4\pi|x|)$. We test the component of \eqref{eq:sep} with $\cO_t$ at $y$ and $\cO_\lambda$ at $z$, at first order in $\lambda$ and at $t=0$.

\paragraph{Beta function.} We use dimensional regularisation with $d=3-\vep$ and minimal subtraction with local sources. We find that the only counterterm which contributes at separated points is
\begin{equation}
t_0(x) = t(x)+\frac{k}{\vep}\,\lambda(x)^2 \, , \qquad k=\frac{1}{192\pi^2} \, ,
\end{equation}
together with $\lambda_0=\mu^\vep\lambda$. It arises from the product of two operators $\phi^4$ with three contractions, since $G^3$ has the pole $\delta^3(x)/(32\pi^2\vep)$. The beta functions are
\begin{equation}
\beta^\lambda = -(1+\vep)\,\lambda \, , \qquad \beta^t = -2\,t+\frac{\lambda^2}{96\pi^2} \, ,
\end{equation}
where the second term is the standard two-loop value. The stability matrix has the components $M^t{}_t=-2$, $M^\lambda{}_\lambda=-1$, $M^t{}_\lambda=\lambda/(48\pi^2)$ and $M^\lambda{}_t=0$ at $\vep=0$. The component of \eqref{eq:sep} which we test is
\begin{equation}
-F_{t\lambda}-M^t{}_\lambda\,F_{tt}+2\,\beta^\lambda\,h_{\lambda;t\lambda} = 0 \, ,
\label{eq:d3id}
\end{equation}
since the term $\beta^th_{t;t\lambda}$ is of second order in $\lambda$ at $t=0$.

\paragraph{Correlation functions.} The local counterterm induces a contact term in the three-point function. With $\cO_\lambda$ at $x$ and $z$ and $\cO_t$ at $y$ we find
\begin{equation}
W_{\lambda t\lambda}(x,y,z) = -\tfrac16\,\mu^{2\vep}\,G(x-y)\,G(y-z)\,G(x-z)^3
+\frac{2k}{\vep}\,\delta^d(x-z)\,F_{tt}(y-z) \, ,
\end{equation}
where the pole of $G^3$ cancels against the contact term. With $\ell=\log(\mu|u|)$ and the constant $K=\tfrac12+\log2+\tfrac12(\log\pi+\gamma_E)$ the three functions which enter \eqref{eq:d3id} are, at $\vep=0$,
\begin{equation}
F_{tt} = \frac{1}{32\pi^2u^2} \, , \qquad
F_{t\lambda} = -\frac{\lambda\,(\ell+1+K)}{1536\,\pi^4u^2} \, , \qquad
h_{\lambda;t\lambda} = \frac{L+K}{3072\,\pi^4u^2} \, .
\end{equation}
The two-point function of $\cO_t$ and $\cO_\lambda$ contains a logarithm at first order in $\lambda$.

\paragraph{Outcome.} The three terms of \eqref{eq:d3id} are
\begin{equation}
\frac{\lambda\,(\ell+1+K)}{1536\,\pi^4u^2}-\frac{\lambda}{1536\,\pi^4u^2}-\frac{\lambda\,(\ell+K)}{1536\,\pi^4u^2} = 0 \, .
\end{equation}
Thus the identity holds, with the logarithm and the constant both cancelling. Note that the term of second order in the beta function, which gives the off-diagonal element $M^t{}_\lambda$, is required for this. The identity also holds for all $\vep$, where the free parts of the correlation functions cancel among themselves by \eqref{eq:I0I1} and the contact term cancels $M^t{}_\lambda F_{tt}$. Note that the moment function is not $-1/(2\lambda)$ times the two-point function. It contains the additional term $-2k\,F_{tt}$, which is the product of the pole of the contact term with the term of order $\vep$ in the ratio of the moments. Without the contact term the moment function has a pole in $\vep$ and the identity fails.

\paragraph{Cutoff scheme.} We repeated the computation in three dimensions with a cutoff $|x-z|>a_0$ in position space and $t_0=t-2k\log(\mu a_0)\lambda^2$, which gives the same beta function. In units $|u|=1$ the cut-off integral and its first moment are
\begin{equation}
J_0 = \int_{|x-z|>a_0}\frac{d^3x}{|x-y|\,|x-z|^3} = 4\pi\left(\log\frac{1}{a_0}+1\right) , \qquad
J_1 = -2\pi\log\frac{1}{a_0} \, ,
\end{equation}
up to terms of order $a_0^2$, and $J_0+2J_1=4\pi$. We confirmed these by numerical integration, with $J_0=50.21185$ and $J_1=-18.82798$ at $a_0=0.05$. In this scheme $F_{t\lambda}=-\lambda(\ell+1)/(1536\pi^4u^2)$ and $h_{\lambda;t\lambda}=\ell/(3072\pi^4u^2)$, and \eqref{eq:d3id} holds again. Here the contact term cancels between $F_{t\lambda}$ and $2\,\beta^\lambda h_{\lambda;t\lambda}$, and the term which balances $M^t{}_\lambda F_{tt}$ is the finite part $J_0+2J_1$ of the cut-off integrals. Therefore the term of second order in the beta function is needed in both schemes. Whether it is matched by the contact term or by the finite part of the moment depends on the scheme.

This check shows that the identity holds with a resonant beta function at the first order where the resonance enters. It does not test the terms of second order in $\lambda$. At $t=0$ in three dimensions the moments of the free three-point function of two operators $\cO_t$ and one $\cO_\lambda$ are infrared divergent, and a test at second order requires $t\neq0$.

\subsection{Third order in \texorpdfstring{$\phi^4$}{phi4} theory}
\label{app:third}

\paragraph{Structures.} We use the correspondence between tensors and graphs of appendix~\ref{app:equiv}. There are no antisymmetric two-forms of degree 1 and 2, so that the identity \eqref{eq:localfull} first becomes non-trivial at third order. At this order we find ten structures. One is disconnected and one has a cut vertex. We consider the eight remaining structures. The rotation $\xi$ vanishes at this order, so that $B^I=\beta^I$.

\paragraph{Independent content.} In momentum space each term of the identity is $p^4$ times a polynomial in $\ell=\log(p^2/\mu^2)$, and the right-hand side contributes only to the constant term. We denote by $x_k$ the coefficient of $\ell^k$ in the left-hand side for a given structure. The left-hand side satisfies a homogeneous Callan--Symanzik equation at separated points, because the commutator of a dilatation with a special conformal transformation is a special conformal transformation. Since the identity is empty at degree 1 and 2, the coefficients $x_2$ and $x_3$ at degree 3 vanish as a consequence of the Callan--Symanzik equations of the individual correlation functions. Hence the independent content at separated points is the single equation $x_1=0$ for each structure. In position space $x_1$ is the coefficient of $u^{-8}$.

\paragraph{Ingredients.} We work in $d=4-\vep$ with minimal subtraction. The renormalised operators with local couplings are the derivatives of the bare action, so that
\begin{equation}
F_{IJ} = Z^K{}_IZ^L{}_J\,F^0_{KL} \, , \qquad
h_{K;IJ} = ZZZ\,h^0_{K;IJ}+\tfrac12\left(Z^M{}_{KI}Z^L{}_J\,F^0_{ML}-Z^M{}_{KJ}Z^L{}_I\,F^0_{LM}\right) ,
\label{eq:Fh0}
\end{equation}
where $F^0$ and $h^0$ are the bare functions, $Z^K{}_I=\pr g_0^K/\pr g^I$ and $Z^K{}_{IJ}$ is the second derivative. The second term in $h_{K;IJ}$ comes from the contact terms which the counterterms induce at $x=y$ and $x=z$, at distance $\pm u/2$ from the midpoint. The relations \eqref{eq:Fh0} hold for the structures in which no vertex with a local coupling sits inside a self-energy. In the remaining structures the counterterms of the current and of the mass source contribute. With $a_{0\mu}=N_I\,\pr_\mu g^I$ and $\mathsf{M}_0=-d_I\,\pr^2g^I-e_{IJ}\,\pr_\mu g^I\pr^\mu g^J$ at $a_\mu=\mathsf{M}=0$, the derivative of the bare action with respect to the local coupling is
\begin{equation}
\cO_I = Z^K{}_I\,\cO^0_K-\pr_\mu\!\left(N_I\cdot J_0^\mu\right)-\pr^2\!\left(d_I\cdot\phi_0^2\right) ,
\label{eq:Ofull}
\end{equation}
where $\cO^0_K$, $J_0^\mu$ and $\phi_0^2$ are the bare operators. At separated points the divergence of the bare current, contracted with an element $\omega$ of the Lie algebra, is $-(\omega g_0)^K\cO^0_K$, so that the second term of \eqref{eq:Ofull} replaces $Z^K{}_I$ by
\begin{equation}
\hat Z^K{}_I = Z^K{}_I+(N_I\,g_0)^K \, ,
\label{eq:Zhat}
\end{equation}
which is not the derivative of a function of the couplings. The second derivatives of $a_{0\mu}$ and $\mathsf{M}_0$ with respect to the couplings give further contact terms in $h_{K;IJ}$, which multiply the two-point functions of the current and of the mass operator with $\cO_J$. We need the following coefficients. For a variation $v^I$ of the couplings, with tensor $v_{ijkl}$, the two-loop counterterms are~\cite{Jack:2013sha}
\begin{equation}
(N_Iv^I)_{ij} = -\frac{1}{24\,\vep}\left(\lambda_{iklm}v_{jklm}-v_{iklm}\lambda_{jklm}\right) , \qquad
(e_{IJ}v^Iv^J)_{ij} = \frac{1}{12\,\vep}\,v_{iklm}v_{jklm} \, ,
\label{eq:Ntwo}
\end{equation}
and $d_I$ vanishes at this order. We translated these from the complex fields of~\cite{Jack:2013sha} and derived them again from the pole of the two-loop self-energy. They give $\hat\rho_I=2N_{I,1}$ and $(\theta_Iv^I)_{ij}=\tfrac{1}{24}(\lambda_{iklm}v_{jklm}+v_{iklm}\lambda_{jklm})$ at lowest order. At the next order we computed the counterterms of the three-loop self-energy with local couplings. With $\Pi[x,y,z]_{ij}=x_{ikab}\,y_{abcd}\,z_{cdkj}$ we find
\begin{align}
(\hat\rho_Iv^I)_{ij} &\supset \tfrac{1}{16}\left(\Pi[\lambda,\lambda,v]-\Pi[v,\lambda,\lambda]\right)_{ij} , \notag\\
(\theta_Iv^I)_{ij} &\supset -\tfrac{1}{32}\left(\Pi[\lambda,\lambda,v]+\Pi[v,\lambda,\lambda]\right)_{ij}
+\tfrac{5}{48}\,\Pi[\lambda,v,\lambda]_{ij} \, , \notag\\
(d_Iv^I)_{ij} &\supset \frac{1}{12\,\vep}\,\Pi[\lambda,v,\lambda]_{ij} \, ,
\label{eq:threeloopcoeff}
\end{align}
up to terms in which both field indices sit on the same vertex, which do not enter below. The coefficients $\tfrac{1}{16}$ and $-\tfrac{1}{32}$ follow from the three-loop anomalous dimension of the field. The coefficient $\tfrac{5}{48}$ and $d_I$ require one further pole, for which we find the residue $\tfrac{5}{96}$. The resulting $\delta_I=3d_{I,1}$ agrees with the value which the relation \eqref{eq:rhoB} gives from the two-loop coefficients.

The bare functions at this order are propagator integrals with three, four and five loops. We evaluated them by repeated one-loop integrals with non-integer powers, together with one three-loop propagator integral which we took from~\cite{Kompaniets:2017yct}. We compared the poles of the three five-loop integrals with an evaluation by the $R^*$ operation~\cite{Herzog:2017bjx}, and they agree. The computation also uses the beta function through three loops~\cite{Pannell:2024sia}.

\paragraph{Structures without current and mass operator.} We first evaluated $x_1$ for two of the eight structures to which the current and the mass operator do not contribute. We describe them by their lines, with $I$ and $J$ the marked vertices and $a$, $b$, $c$ the vertices which carry $\lambda$, and with an exponent for the number of lines. The first structure has the lines $IJ^2$, $Ic^2$, $Jb$, $Jc$, $ab^3$, $ac$. It receives the two-point function with a self-energy insertion on one line. Only the terms with the stability matrix contribute, and the three contributions from the three-, four- and five-loop integrals give
\begin{equation}
x_1 = \frac{1}{432}-\frac{49}{1728}+\frac{5}{192} = 0 \, , \qquad x_2=x_3=0 \, .
\end{equation}

The second structure has the lines $IJ^2$, $Ic^2$, $Ja$, $Jb$, $ab^2$, $ac$, $bc$. Here the first moment of the three-point function contributes as well. To test its coefficient we replace the factor 2 in \eqref{eq:sep} by a parameter $\nu$. The terms with the stability matrix give $(x_1,x_2,x_3)=(-\tfrac{127}{1728},\tfrac{7}{288},-\tfrac{1}{288})$, and the three-point term gives $\nu$ times $(\tfrac{127}{3456},-\tfrac{7}{576},\tfrac{1}{576})$, so that
\begin{equation}
(x_1,x_2,x_3) = (\nu-2)\left(\frac{127}{3456},\,-\frac{7}{576},\,\frac{1}{576}\right) .
\label{eq:nu}
\end{equation}
Thus the identity holds with the coefficient 2 which was derived in section~\ref{sec:ward}, and $\nu=2$ is the only value for which $x_1$ vanishes. The three-point term consists of the bare moment, which contributes $\tfrac{111}{256}$ to $x_1$, and of the contact terms in \eqref{eq:Fh0}, which contribute $\tfrac{991}{6912}-\tfrac{1867}{3456}$. Without the contact terms the coefficients of $\ell^k$ have poles in $\vep$ and the identity fails. In both structures neither the current nor the mass operator contributes. The term with the three-loop beta function contributes $-\tfrac{1}{128}$ to $x_1$ in the first structure and $\tfrac{1}{96}$ in the second.

\paragraph{Current and mass operator.} Three structures receive contributions from the current or from the mass operator. They have the lines $Ib$, $Ic^3$, $Ja^2$, $Jb$, $Jc$, $ab^2$ for the first, $IJ$, $Ic^3$, $Ja$, $Jb^2$, $ab^2$, $ac$ for the second and $Ib$, $Ic^3$, $Ja^2$, $Jb^2$, $ab$, $ac$ for the third. In all three the vertex $I$ is joined to a vertex $c$ by three lines. The commutator involves the two structures of $F_{IJ}$ at second order in which a marked vertex sits inside a two-loop self-energy. With \eqref{eq:Fh0} alone these two functions are not finite. The coefficient of $\ell$ retains the poles $-\tfrac{1}{864\,\vep}$ and $+\tfrac{1}{864\,\vep}$. With $\hat Z$ of \eqref{eq:Zhat} both poles cancel, and they cancel only for the normalisation of $N_I$ in \eqref{eq:Ntwo}. The mass source does not contribute to $F_{IJ}$ at this order.

We then evaluated the full identity \eqref{eq:localsep} for the three structures. To test the coefficients we replace the factors 2, 4 and 16 of the three-point term, the current term and the term with $\theta_I$ by $\nu$, $\nu_J$ and $\nu_{\mathsf M}$. The term with $\delta_I$ does not contribute to the coefficients of $\ell^k$ at this order. The current enters only the difference of the first two structures and the mass operator only their sum and the third structure. We find
\begin{align}
\text{difference:}\quad & x_1 = -\frac{74\,\nu-31\,\nu_J-24}{13824} \, , &
x_2 &= \frac{2\,\nu-\nu_J}{2304} \, , \notag\\
\text{sum:}\quad & x_1 = \frac{296\,\nu-39\,\nu_{\mathsf M}+32}{18432} \, , &
x_2 &= -\frac{8\,\nu-\nu_{\mathsf M}}{3072} \, , \notag\\
\text{third structure:}\quad & x_1 = -\frac{\nu_{\mathsf M}-16}{1536} \, , &
x_2 &= 0 \, ,
\label{eq:xcurrentmass}
\end{align}
and $x_3=0$ in all three cases. All poles in the coefficients of $\ell^k$ cancel. Thus the identity holds for $\nu=2$, $\nu_J=4$ and $\nu_{\mathsf M}=16$, which are the coefficients of \eqref{eq:localsep}. The equation $x_1=0$ is sensitive to the three-loop coefficients \eqref{eq:threeloopcoeff}. If we leave them free, it requires the values $\tfrac{1}{16}$, $-\tfrac{1}{32}$ and $\tfrac{5}{48}$ in the three cases, in agreement with the values which we computed from the counterterms. It also requires the sign of the term $(\hat\rho_I\,g)^K$ in \eqref{eq:Mhat}, the contact terms from the second derivatives of $a_{0\mu}$ and $\mathsf{M}_0$, and the last term of \eqref{eq:Ofull} in the two-point function of the mass operator with $\cO_J$. If any one of these is omitted or changed, the identity no longer holds. The check is therefore sensitive to each ingredient and is not satisfied trivially.

We should point out the limitations of this check. As for the two structures above, the vanishing of $x_1$ follows from the bare theory once the counterterms are known, so that it tests the derivation of \eqref{eq:localsep} with its normalisation and the counterterms, and not a statement which is independent of~\cite{Jack:2013sha}. Further, we did not compare the three-loop coefficients \eqref{eq:threeloopcoeff} with the literature. One four-loop integral was evaluated with Forcer~\cite{Ruijl:2017cxj}, and the first moment was again computed for chains only.

\paragraph{Interpretation.} For structures to which the current and the mass operator do not contribute, the identity can be understood from the bare theory in $d$ dimensions. The bare operator has dimension $d-\vep$, so that the bare stability matrix is proportional to the unit matrix and \eqref{eq:sep} reduces to $g_0^Kh^0_{K;IJ}=0$. This states that the first moments of all integrated vertices about the midpoint add up to zero. With \eqref{eq:Fh0} and the relation $Z^K{}_I\beta_\vep^I=-\vep\,g_0^K$ for $\beta_\vep^I=-\vep g^I+\beta^I$ we find the algebraic identity
\begin{equation}
\left(M^TF-FM\right)_{IJ}+2\,\beta_\vep^Kh_{K;IJ} = -2\,\vep\,Z^{K}{}_IZ^{L}{}_J\;g_0^Mh^0_{M;KL} \, ,
\end{equation}
in which the coefficient 2 is twice the factor $\tfrac12$ of the contact terms. At this order the equation $x_1=0$ is then equivalent to the antisymmetric part of the $1/\vep$ term of a relation of~\cite{Jack:2013sha} among the counterterms $\pr^2g^I\pr^2g^J$ and $\pr^2g^I\pr_\mu g^J\pr^\mu g^K$ and the coefficient $E_{IJ}$, given the finiteness of the renormalised correlation functions. Both state that the bare six-loop moment of the structure has no double pole. For the two structures above the vanishing of the total bare moment also follows from a symmetry of the graphs.

We conclude that the test shows that the derivation of \eqref{eq:sep} and its normalisation, including the coefficients 2, 4 and 16 and the contact terms, are consistent with the counterterms and integrals at this order. It does not test a statement which is independent of the relations of~\cite{Jack:2013sha}, and it does not involve the contact coefficient $E_{[IJ]}$ itself, which requires six-loop poles.

\paragraph{Remaining structures.} The three remaining structures receive no contribution from the current and the mass operator. They require five-loop integrals which do not reduce to one-loop integrals, and for two of them the first moment of a three-point graph which is not a chain. We did not evaluate these three structures.

\subsection{Two dimensions}
\label{app:twod}

In two dimensions Friedan and Konechny proved a gradient formula which contains an antisymmetric tensor $b_{ij}$~\cite{Friedan:2009ik}, and Behr and Konechny computed this tensor at first order in conformal perturbation theory~\cite{Behr:2013vta}. In this section we express $b_{ij}$ through the two-dimensional analogues of the quantities of sections~\ref{sec:ward} and~\ref{sec:form}. We find that $b_{ij}$ is a term of the first type of section~\ref{sec:removable}, see \eqref{eq:twodb} below, and that \eqref{eq:twodb} agrees with the first-order result of~\cite{Behr:2013vta}. We should point out that this is a comparison with the literature. It uses the same correlation functions as section~\ref{app:firstorder} and is not independent of the check performed there, and it is not a test of Theorem~\ref{thm:main}, which concerns four dimensions.

\paragraph{Known results.} We write $g^i$ and $\cO_i$ for the couplings and operators which are called $\lambda^i$ and $\phi_i$ in~\cite{Friedan:2009ik,Behr:2013vta}. The scale of these references is a length, so that their beta function is $\beta_{\rm FK}^i=-\beta^i$ in terms of ours. The gradient formula, eq.~(2.13) of~\cite{Friedan:2009ik}, reads
\begin{equation}
\pr_i\mathcal{C}+\left(g_{ij}+\Delta g_{ij}\right)\beta_{\rm FK}^j+b_{ij}\,\beta_{\rm FK}^j = 0 \, ,
\qquad b_{ij} = \pr_iw_j-\pr_jw_i \, ,
\label{eq:twodFK}
\end{equation}
where $\mathcal{C}$ is the $c$-function of Zamolodchikov, $g_{ij}$ is his metric and $\Delta g_{ij}$ is a correction which is built from currents and which is absent in the cases that we consider. The one-form is given in eqs.~(2.12) and~(3.5) there by
\begin{equation}
w_i = \int d^2x\;G_\Lambda(x)\,\vev{\cO_i(x)\,\Theta(0)} \, , \qquad
G_\Lambda(x) = 3\pi\,|x|^2\,\theta(1-\Lambda|x|) \, ,
\label{eq:twodw}
\end{equation}
where $\Theta$ is the trace of the energy-momentum tensor and $1/\Lambda$ is a fixed distance. The two-point function in \eqref{eq:twodw} is a distribution, and its contact terms contribute. With local couplings and in flat space the trace is assumed in~\cite{Osborn:1991gm} to be of the form, in the absence of currents, eq.~(2.7) of~\cite{Friedan:2009ik},
\begin{equation}
\Theta-\beta_{\rm FK}^i\,\cO_i = \pr^\mu\!\left(W_i\,\pr_\mu g^i\right)
+\tfrac12\,G_{ij}\,\pr_\mu g^i\,\pr^\mu g^j \, ,
\label{eq:twodtrace}
\end{equation}
where both sides act on the generating functional. The one-form $W_i(g)$ is, up to its normalisation, the one-form which multiplies $\pr_\mu\sigma\,\pr^\mu g^i$ in the Weyl variation of~\cite{Osborn:1991gm}, eq.~(2.4) there. It is the analogue of the one-form $W_I$ of section~\ref{sec:consistency}, and it does not vanish in flat space. Its curl plays the role of $E_{[IJ]}$.

\paragraph{Ward identity.} The derivation of section~\ref{sec:ward} applies in two dimensions without currents and without a mass operator. The generating functional with a local scale factor is invariant under diffeomorphisms~\cite{Friedan:2009ik}, and for a conformal Killing vector this gives \eqref{eq:ward} with $\cA=-\int d^2x\,\sigma\,(\Theta-\beta_{\rm FK}^i\cO_i)$. With \eqref{eq:twodtrace} we obtain
\begin{equation}
\cA[\sigma;g] = \int d^2x\left(\pr_\mu\sigma\,W_i\,\pr^\mu g^i
-\tfrac12\,\sigma\,G_{ij}\,\pr_\mu g^i\,\pr^\mu g^j\right) .
\end{equation}
After two derivatives with respect to the couplings the term with $G_{ij}$ is proportional to $b\cdot c$ and enters only the Callan--Symanzik equation. The term with $W_i$ gives $2\,(\pr_iW_j-\pr_jW_i)\,b\cdot\pr\,\delta^2(u)$, and we use $u_\mu\pr^2\delta^2(u)=-2\,\pr_\mu\delta^2(u)$. The identity \eqref{eq:localfull} thus becomes
\begin{equation}
-u^\mu\left(M^TF-FM\right)_{ij}-2\,\beta^km^\mu_{k;ij} = E_{[ij]}\;u^\mu\,\pr^2\delta^2(u) \, ,
\qquad E_{[ij]} = -\left(\pr_iW_j-\pr_jW_i\right) ,
\label{eq:twodward}
\end{equation}
with $M^k{}_i=\pr_i\beta^k$ and the first moment \eqref{eq:moment}. At separated points this is \eqref{eq:sep}. We pair \eqref{eq:twodward} with the two-dimensional analogue of \eqref{eq:testfn},
\begin{equation}
\psi_\mu(u) = \tfrac14\,u_\mu\,\theta(1-\Lambda|u|) \, , \qquad
\varphi(u) = u^\mu\psi_\mu(u) = \tfrac14\,|u|^2\,\theta(1-\Lambda|u|) = \frac{G_\Lambda(u)}{12\pi} \, ,
\label{eq:twodtest}
\end{equation}
for which $\pr^2\varphi(0)=1$. The sharp cutoff is admissible here, because the distributions are smooth functions away from the origin. We obtain the analogue of \eqref{eq:formfull},
\begin{equation}
E_{[ij]} = \left(M^Tc-cM\right)_{ij}+\beta^k\tilde\tau_{k;ij} \, , \qquad
c_{ij} = -\vev{F_{ij},\varphi} \, , \qquad \tilde\tau_{k;ij} = -2\,\vev{m^\mu_{k;ij},\psi_\mu} \, .
\label{eq:twodform}
\end{equation}

\paragraph{Antisymmetric tensor.} We now insert \eqref{eq:twodtrace} into \eqref{eq:twodw}. A derivative of \eqref{eq:twodtrace} with respect to $g^i(x)$ at constant couplings gives $\vev{\cO_i(x)\,\Theta(0)}=\beta_{\rm FK}^jF_{ij}(x)+W_i\,\pr^2\delta^2(x)$, up to a term proportional to $\delta^2(x)$, which does not contribute because $G_\Lambda(0)=0$. With \eqref{eq:twodtest} we find
\begin{equation}
w_i = 12\pi\left(W_i+c_{ij}\,\beta^j\right) .
\label{eq:twodwW}
\end{equation}
Thus the one-form of~\cite{Friedan:2009ik} is the sum of the contact one-form $W_i$ and of $c_{ij}\beta^j$, where $c_{ij}$ is the pairing of the two-point function with $\varphi$. In the notation of~\cite{Friedan:2009ik} this pairing is $c^\Lambda_{ij}=-12\pi\,c_{ij}$, eq.~(9.28) there. We take the curl of \eqref{eq:twodwW} and use \eqref{eq:curlC} and \eqref{eq:twodform}, which gives
\begin{equation}
b_{ij} = 12\pi\left[-E_{[ij]}+\pr_i\!\left(c_{jk}\beta^k\right)-\pr_j\!\left(c_{ik}\beta^k\right)\right]
= 12\pi\,\beta^k\left(\pr_ic_{jk}-\pr_jc_{ik}-\tilde\tau_{k;ij}\right) .
\label{eq:twodb}
\end{equation}
The first equality does not use the Ward identity. It corresponds to the statement of~\cite{Friedan:2009ik}, eqs.~(9.23) and~(9.27) to~(9.29), that their gradient formula is related by a dressing transformation with $c^\Lambda_{ij}$ to a formula in which the antisymmetric tensor is the curl of $W_i$. The second equality uses \eqref{eq:twodform}. By \eqref{eq:curlC} the curl of $c_{ik}\beta^k$ is the sum of the term $M^Tc-cM$ of the third type and of a term of the first type. By \eqref{eq:twodform} the coefficient $E_{[ij]}$ contains the same term of the third type, and it cancels in the difference. Hence $b_{ij}$ is a term of the first type, and by \eqref{eq:type1} it can be replaced in \eqref{eq:twodFK} by a symmetric tensor without a change of $\mathcal{C}$. We should point out that at first order this statement has no content for generic dimensions. If all $\epsilon_k\neq0$, every antisymmetric tensor which is linear in the couplings is of the form $\beta^k\tau_{k;ij}$. The coefficient then contains factors $1/\epsilon_k$, and the symmetric tensor which is obtained from \eqref{eq:type1} is singular for $\epsilon_k\to0$, whereas $b_{ij}$ is not. The statement acquires content when the coefficient is regular in the nearly marginal limit, which we did not examine. Note also that the decomposition into the three types is not unique. For the matrix $c_{ij}$ of \eqref{eq:twodform} the Ward identity converts the term of the third type into $E_{[ij]}$ and a term of the first type.

\paragraph{First order.} We consider the perturbations of section~\ref{app:firstorder} for $d=2$, with $0<\epsilon_i<1$ and without resonances. This is the case of section~4.5 of~\cite{Behr:2013vta}, where the exponents are denoted by $\epsilon_i$ as well and $\beta_{\rm FK}^i=\epsilon_ig^i$, eq.~(4.91) there. There are no contact terms for dimensional reasons~\cite{Behr:2013vta}, so that $W_i=0$ and $E_{[ij]}=0$. The first-order two-point function is $F_{ij}=P_{ijk}\,g^k\,|u|^{\epsilon_i+\epsilon_j+\epsilon_k-4}$, where $P_{ijk}$ is $C_{ijk}$ times the coefficient of $I_0$ in \eqref{eq:I0I1} for $d=2$. This is eq.~(4.93) of~\cite{Behr:2013vta}. The last index of $P_{ijk}$ refers to the integrated operator. As in section~\ref{app:firstorder}, the integrals over the position of this operator converge only in a part of this range and are defined by analytic continuation otherwise. We set $\Lambda=1$ and use the normalisation $F_{ij}=\delta_{ij}|u|^{2\epsilon_i-4}$ at zeroth order. From \eqref{eq:twodform} and \eqref{eq:hfirst} we find
\begin{equation}
c_{ij} = -\frac{\pi}{4\,\epsilon_i}\,\delta_{ij}-\frac{\pi}{2}\,\frac{P_{ijk}\,g^k}{\epsilon_i+\epsilon_j+\epsilon_k} \, , \qquad
\tilde\tau_{k;ij} = -\frac{\pi}{2}\,\frac{(\epsilon_j-\epsilon_i)\,P_{ijk}}{\epsilon_k\,(\epsilon_i+\epsilon_j+\epsilon_k)} \, ,
\label{eq:twodfirst}
\end{equation}
through first and zeroth order respectively. With $\beta^k=-\epsilon_kg^k$ the second form of \eqref{eq:twodb} becomes
\begin{equation}
b_{ij} = \frac{6\pi^2\,g^k}{\epsilon_i+\epsilon_j+\epsilon_k}\left[P_{ijk}\,(\epsilon_i-\epsilon_j)+\epsilon_k\left(P_{kji}-P_{kij}\right)\right] ,
\label{eq:twodBK}
\end{equation}
which is eq.~(4.96) of~\cite{Behr:2013vta}, including the normalisation. The first term in the bracket comes from the first moment of the three-point function and the second from the derivative of $c_{ij}$. We verified the algebra which leads from \eqref{eq:twodfirst} to \eqref{eq:twodBK} by computer algebra for three couplings with general $\epsilon_i$ and $P_{ijk}$. We also confirmed \eqref{eq:I0I1} for $d=2$ by a numerical integration, for which we find agreement to $10^{-7}$ for $(\epsilon_i,\epsilon_j,\epsilon_k)=(0.30,0.25,0.35)$.

We should point out what this agreement does and does not show. In~\cite{Behr:2013vta} the result \eqref{eq:twodBK} is obtained as the curl of \eqref{eq:twodwW} with $W_i=0$, so that the first form of \eqref{eq:twodb} agrees with it by construction. The second form requires in addition that $M^Tc-cM$ equals $-\beta^k\tilde\tau_{k;ij}$ at this order. This is \eqref{eq:sep} integrated with $\varphi$, and it holds by the cancellation found in section~\ref{app:firstorder}. The relation $E_{[ij]}=-(\pr_iW_j-\pr_jW_i)$ is not tested, since both sides vanish at this order. Beyond first order \eqref{eq:twodb} is a formal statement. We did not examine the convergence of the first moment at large distances in two dimensions, and conformal perturbation theory with relevant operators has infrared divergences at higher orders~\cite{Behr:2013vta}. For nearly marginal perturbations it is found in~\cite{Behr:2013vta}, below eq.~(4.82), that $b_{ij}$ is of second order in the couplings. This is the order of $\beta^k$ and seems consistent with \eqref{eq:twodb}. However, the models considered there contain currents, for which we did not derive the two-dimensional identity.

Finally we comment on the dressing transformation of~\cite{Friedan:2009ik}, eq.~(9.27), with $c^\Lambda_{ij}$. It removes $b_{ij}$ whenever the curl of $W_i$ vanishes, and in particular at first order. However, the dressed formula is the formula of~\cite{Friedan:2009ik}, eq.~(9.23), which is built from contact terms. In the present case there are none, so that the dressed metric $g_{ij}-(\cL_{\beta_{\rm FK}}c^\Lambda)_{ij}$ vanishes and the dressed potential is the central charge of the unperturbed theory. The replacement of $b_{ij}$ by a symmetric tensor through \eqref{eq:type1} is different, because it leaves $\mathcal{C}$ unchanged and adds to $g_{ij}$ a symmetric tensor of first order in the couplings.

\paragraph{Summary.} Let us summarise what this comparison shows. In two dimensions the $c$-theorem of Zamolodchikov states that $\mathcal{C}$ decreases along the flow, and the gradient formula with his metric alone is known to hold at leading order near a fixed point, but not beyond~\cite{Friedan:2009ik}. The general formula \eqref{eq:twodFK} therefore contains the antisymmetric tensor $b_{ij}$, as the formula of~\cite{Osborn:1991gm,Jack:2013sha} does in four dimensions. We found three things. First, $b_{ij}$ is the sum of the curl of a one-form of contact terms, which is the analogue of our anomaly coefficient, and of the curl of $c_{ij}\beta^j$, where $c_{ij}$ is the two-point function paired with a test function, as in four dimensions. Second, with the Ward identity $b_{ij}$ is a term of the first type, so that it can be replaced by a symmetric contribution to the metric without a change of the potential. At first order and for generic dimensions this is true of every antisymmetric tensor, and the symmetric contribution is singular in the nearly marginal limit. Third, at first order our expression agrees with the result of~\cite{Behr:2013vta}, and the replacement keeps the $c$-function as the potential, whereas the dressing transformation of~\cite{Friedan:2009ik} leaves a trivial formula at this order. The comparison has clear limits. The agreement at first order follows in part by construction and otherwise from the check of section~\ref{app:firstorder}. The contact terms vanish at this order and are not tested, and beyond first order the relation is formal. It is a comparison of two formulations, and not a test of Theorem~\ref{thm:main}.

\end{document}